\documentclass[prx,reprint,groupedaddress,superscriptaddress]{revtex4-1}

\usepackage{amsmath} \DeclareMathOperator{\Tr}{Tr} \DeclareMathOperator{\col}{col}
\usepackage{braket}
\usepackage{graphicx}
\usepackage{units}
\usepackage[colorlinks=true,citecolor=blue,urlcolor=blue,linkcolor=blue]{hyperref}
\usepackage{bbold}
\usepackage{xcolor}
\usepackage{multirow}

\usepackage{booktabs}
\usepackage{array}
\usepackage[normalem]{ulem}
\usepackage{comment}

\begin{document}

\title{Autonomous dissipative Maxwell's demon in a diamond spin qutrit}

\author{S. Hern\'{a}ndez-G\'{o}mez}\email{hernandez@lens.unifi.it}
\affiliation{European Laboratory for Non-linear Spectroscopy (LENS), Universit\`a di Firenze, I-50019 Sesto Fiorentino, Italy}
\affiliation{Dipartimento di Fisica e Astronomia, Universit\`a di Firenze, I-50019, Sesto Fiorentino, Italy}
\affiliation{Istituto Nazionale di Ottica del Consiglio Nazionale delle Ricerche (CNR-INO), I-50019 Sesto Fiorentino, Italy}
\author{S. Gherardini}
\affiliation{European Laboratory for Non-linear Spectroscopy (LENS), Universit\`a di Firenze, I-50019 Sesto Fiorentino, Italy}
\affiliation{Istituto Nazionale di Ottica del Consiglio Nazionale delle Ricerche (CNR-INO), I-50019 Sesto Fiorentino, Italy}
\affiliation{Scuola Internazionale Superiore di Studi Avanzati (SISSA), I-34136 Trieste, Italy}
\author{N. Staudenmaier}
\altaffiliation{Current address: Institute for Quantum Optics, Ulm University, D-89081, Germany}
\affiliation{European Laboratory for Non-linear Spectroscopy (LENS), Universit\`a di Firenze, I-50019 Sesto Fiorentino, Italy}
\affiliation{Dipartimento di Fisica e Astronomia, Universit\`a di Firenze, I-50019, Sesto Fiorentino, Italy}
\author{F. Poggiali}
\altaffiliation{Current address: Universit\"at Rostock, Institut f\"ur Physik - AG Quantentechnologie, Albert-Einstein-Str. 23, D-18059 Rostock, Germany}
\affiliation{European Laboratory for Non-linear Spectroscopy (LENS), Universit\`a di Firenze, I-50019 Sesto Fiorentino, Italy}
\affiliation{Dipartimento di Fisica e Astronomia, Universit\`a di Firenze, I-50019, Sesto Fiorentino, Italy}
\author{M. Campisi}
\affiliation{NEST, Istituto Nanoscienze-CNR and Scuola Normale Superiore, I-56127 Pisa, Italy}
\affiliation{Dipartimento di Fisica e Astronomia, Universit\`a di Firenze, I-50019, Sesto Fiorentino, Italy}
\affiliation{INFN - Sezione di Pisa, I-56127 Pisa, Italy}
\author{A. Trombettoni}
\affiliation{Scuola Internazionale Superiore di Studi Avanzati (SISSA), I-34136 Trieste, Italy}
\affiliation{CNR-IOM DEMOCRITOS Simulation Center, I-34136 Trieste, Italy}
\author{F. S. Cataliotti}
\affiliation{European Laboratory for Non-linear Spectroscopy (LENS), Universit\`a di Firenze, I-50019 Sesto Fiorentino, Italy}
\affiliation{Istituto Nazionale di Ottica del Consiglio Nazionale delle Ricerche (CNR-INO), I-50019 Sesto Fiorentino, Italy}
\author{P. Cappellaro}
\affiliation{Department of Nuclear Science and Engineering, Department of Physics, Massachusetts Institute of Technology, Cambridge, MA 02139}
\author{N. Fabbri}\email{fabbri@lens.unifi.it}
\affiliation{European Laboratory for Non-linear Spectroscopy (LENS), Universit\`a di Firenze, I-50019 Sesto Fiorentino, Italy}
\affiliation{Istituto Nazionale di Ottica del Consiglio Nazionale delle Ricerche (CNR-INO), I-50019 Sesto Fiorentino, Italy}

\begin{abstract}
Engineered dynamical maps combining coherent and dissipative transformations of quantum states with quantum measurements, have demonstrated a number of technological applications, and promise to be a crucial tool in quantum thermodynamic processes. 
Here, we exploit the control on the  effective open spin qutrit dynamics of an NV center, to experimentally realize an autonomous feedback process (Maxwell demon) with tunable dissipative strength. The feedback is enabled by random measurement events that condition the subsequent dissipative evolution of the qutrit. The efficacy of the autonomous Maxwell demon is quantified by experimentally characterizing the fluctuations of the energy exchanged by the system with the environment by means of a generalized Sagawa-Ueda-Tasaki relation for dissipative dynamics. This opens the way to the implementation of a new class of Maxwell demons, which could be useful for quantum sensing and quantum thermodynamic devices. 
\end{abstract}

\maketitle

\begin{figure*}
\includegraphics[width=0.75\textwidth]{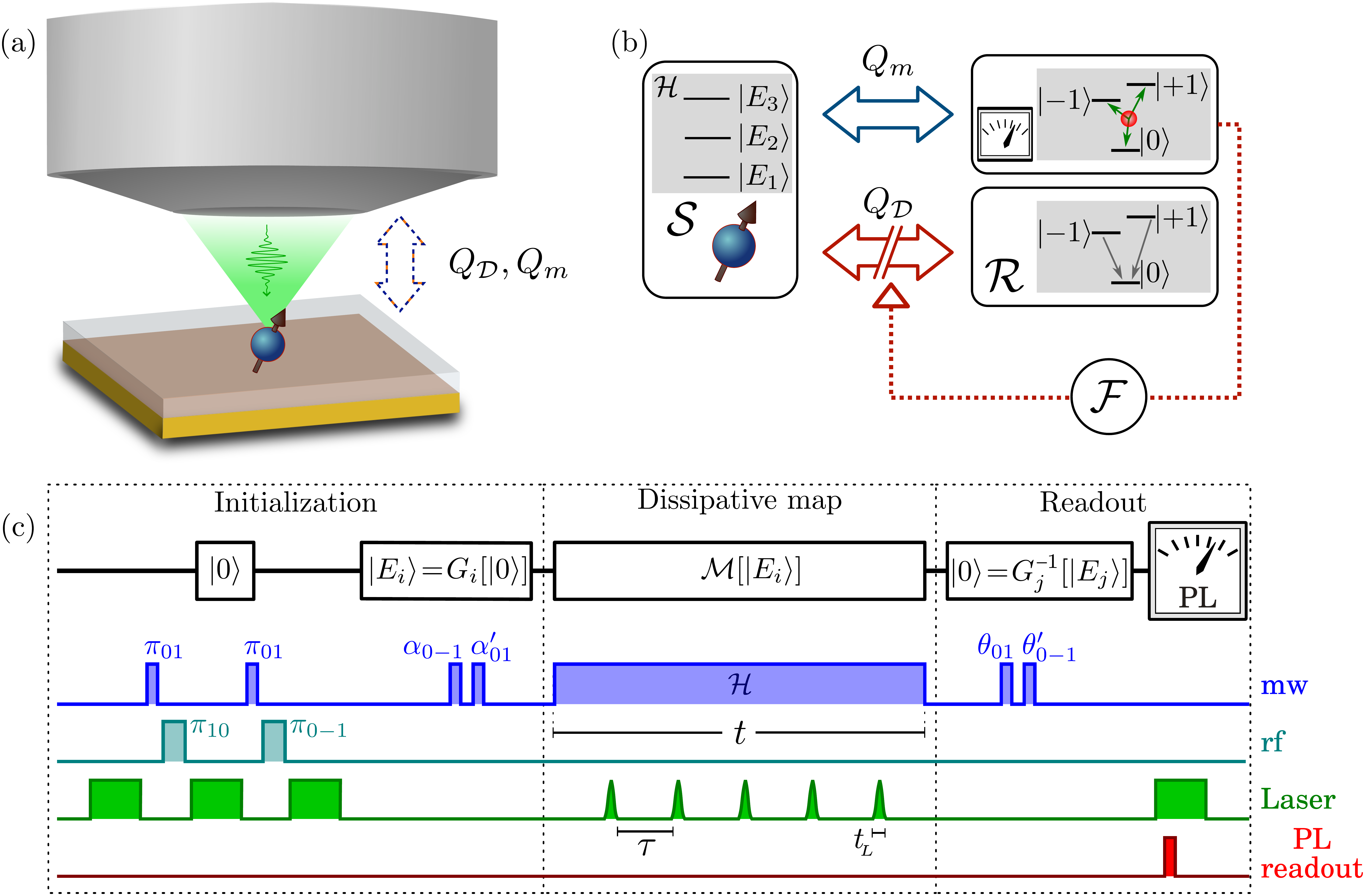}
\caption{
\textbf{(a,b)} Scheme of the spin system $\mathcal{S}$ in the presence of a green laser. Upon interaction with a laser pulse, the spin is subject to a quantum measurement ($Q_m$) of $S_z$, and dissipation ($Q_\mathcal{D}$) towards the $m_S=0$ spin projection. This irreversible dissipation is analogous to put the system in contact with an out-of-equilibrium reservoir~$\mathcal{R}$. Since the interaction between the system and the reservoir is conditioned by the application of a quantum measurement, the dissipation acts as an intrinsic feedback mechanism. The unitary part of the dynamics is defined by the Hamiltonian $\mathcal{H}$, with eigenstates $\ket{E_i}$, $i=1,2,3$.
\textbf{(c)} Scheme to measure
energy conditional probabilities. The stochastic-dissipative map $\mathcal{M}$ is a combination of a train of $N_\mathrm{L}$ equidistant short laser pulses and a continuous driving under the Hamiltonian $\mathcal{H}$ that can be either $\mathcal{H}_\mathrm{NV}$ or $\mathcal{H}_\mathrm{mw}$~(see text).
The total time of the experiment $t = N_\mathrm{L}\tau$ is defined in terms of the number of laser pulses and the time $\tau$ between them. The laser pulse duration $t_\mathrm{L}$ is negligible with respect to the continuous driving.
The gates $G_i:\ket{0}\rightarrow\ket{E_i}$ and $G_j^{-1}:\ket{E_j}\rightarrow\ket{0}$ enable to prepare and readout the Hamiltonian eigenstates, respectively, by exploiting the optical properties of the NV center.}
\label{fig:protocol}
\end{figure*}

\section{Introduction}

The Maxwell's demon paradox, introduced by Maxwell in 1867 to discuss the validity of the second law of thermodynamics, has uncovered the relationship between thermodynamics and information, and still flourishes in modern physics~\cite{Maruyama09}. The demon is an intelligent entity that uses the information resulting from the measurement of a system to condition the system dynamics, with results that may be in apparent contrast with the second law. One century later, Landauer and Bennett~\cite{Bennett1982}, the fathers of so-called information thermodynamics, provided the solution of this paradox, by suggesting to consider in the thermodynamic balance also the information stored in the demon memory, which is erased in the process. The modern formulation of Maxwell's demons is embodied by the combination of measurement and feedback control, a typical setting of information thermodynamics~\cite{FunoBook2018,Koski14,Naghiloo2018}. The feedback mechanism can be either operated by an external agent or even performed internally, with no microscopic information exiting the
system~\cite{Mandal2012,Kutvonen2016}. The demon can accomplish different tasks, e.g., acting to perform information heat engines, refrigerators, thermal accelerators, or heaters~\cite{Buffoni2019}. In quantum settings, projective measurements contribute as a purely quantum component to heat exchange~\cite{Elouard17,Gherardini18}, and enable the feedback mechanism that the demon exploits to convert information into usable energy~\cite{Toyabe10,Masuyama18}.
 This feedback mechanism plays a crucial role in the investigation of quantum information thermodynamics and may find applications in information-powered quantum refrigerator or heat engine~\cite{Koski15,Elouard17PRL}, quantum heat transport~\cite{Campisi17}, quantum computation and error correction~\cite{Schindler11}, and metrology~\cite{Hirose16}. Experimental implementations of Maxwell's demons in the quantum regime have been  carried out in NMR setup to compensate entropy production~\cite{CamatiPRL2016}, photonic platform working at the few-photons level~\cite{VidrighinPRL2016}, superconducting QED circuits~\cite{CottetPNAS2017,Naghiloo2018,SongPRA2021}, solid state spins~\cite{Ji22}, and single Rydberg atoms~\cite{Najera-SantosPRR2020}.

A Maxwell demon generally acts via unitary evolutions. 
However, a proper control and design of the system-environment coupling via the combination of coherent and non-unitary operations can be a fundamental resource for quantum information processing~\cite{Verstraete09,Pastawski11} and thermodynamics~\cite{Scarani02,Nandkishore15,Strasberg17}, and more broadly for quantum simulation~\cite{Barreiro11} and sensing~\cite{Do19,Wolski20,Xie20}.
Dissipative operations can be used to produce quantum states of interest such as non-equilibrium steady states, strongly correlated states, or to prepare and stabilize robust phases and entanglement~\cite{Lin13,Barontini15,Biella17,Lu17,Ma19}.
A relevant feature of dissipative dynamics is the appearance of stationary states, non necessarily in thermal equilibrium~\cite{BenattiBook2003,Dutta21}, as a generalization of thermalization processes.
Among dissipative processes, optical pumping exhibits the peculiar properties of leading the quantum system to a dissipated out-of-equilibrium state that does not depend on the system initialization. 
As we will show, repetitive system-environment interactions via optical pumping can be modelled as an autonomous Maxwell demon where dissipation is conditioned on the absorption of light.

Information and energy exchanges of an open quantum system with its environment inherently involve fluctuations.
Quantum measurements, randomizing the system evolution, introduce quantum energy fluctuations, which impact the observable distribution obtained by averaging over many quantum trajectories~\cite{Gisin84,JacobsBook2014}.
These fluctuations can drive the quantum system towards novel---often out-of-equilibrium---dynamical regimes that could not be otherwise achieved.
Quantum fluctuation relations~\cite{Esposito09,Campisi11} provide a powerful framework to characterize such energy fluctuations in thermodynamic processes.
Experimental investigations of quantum fluctuation relations have been recently conducted on different  
platforms, including single trapped ions~\cite{An15,Smith18}, NMR systems~\cite{Batalhao14,Pal19}, atom chip~\cite{Cerisola17}, superconducting qubits~\cite{Zhang18}, Nitrogen-Vacancy (NV) centers in diamond~\cite{HernandezGomez20}, and entangled photon pairs~\cite{RibeiroPRA2020}. These studies cover closed system dynamics~\cite{An15,Smith18,Batalhao14,Cerisola17,Zhang18}, and certain open 
dynamics~\cite{Smith18,Pal19,HernandezGomez20,RibeiroPRA2020} where micro-reversibility may not be satisfied.

Here, we realize  an autonomous dissipative Maxwell demon with a spin qutrit formed by a Nitrogen-Vacancy (NV) center in diamond at room temperature, and we investigate its purely quantum (non-Gibbsian) energy fluctuations through a generalized Sagawa-Ueda-Tasaki (SUT) quantum fluctuation relation.
The intrinsic feedback mechanism acting on a dissipative dynamics is achieved by performing random projective measurements followed by conditioned and tunable optical pumping. 
The resulting dynamics generates non-thermal steady states in the energy basis, independent of the initial state. 
In the case of conditioned  unitary  evolution, the SUT relation establishes a fundamental connection between the thermodynamic properties of non-equilibrium quantum processes and information-theoretic quantities, evaluated by measuring and manipulating the system \cite{Sagawa08,Morikuni11,Funo13,FunoNJP15}. The SUT relation has been generalized to completely positive trace-preserving (CPTP) maps~\cite{Kafri12,Rastegin13,Albash13,Goold15,SongPRA2021}.
Our work formulates such relation in the case of conditioned dissipative dynamics and verifies it experimentally by measuring the energy change statistics of the spin qutrit. 
Interestingly, we find that expressing the SUT relation in the framework of the superoperators' formalism~\cite{Havel03} renders  the computation  of all the system trajectories unnecessary,  drastically reducing the required computational resources. 
This is a relevant feature in protocols involving repeated measurements.
We also measure the mean energy change of the qutrit and compare it with theoretical bounds.
Finally, we find that the knowledge of the stationary state is sufficient
to characterize the demon in terms of its capacity of energy extraction, without requiring any additional information on the map.

The paper is structured as follows:
In Sec.~\ref{sec:Experiment_map} we introduce the experimental platform and the implementation of the intrinsic feedback dissipative mechanism for a spin qutrit.
We discuss the formalism to model the dynamics of the qutrit as an autonomous Maxwell demon in Sec.~\ref{sec:Modeled_map}. 
Sec.~\ref{sec:g-SUTR} contains a discussion about the SUT fluctuation relation and its extension beyond unitary dynamics.
In Sec.~\ref{sec:protocol}  we present the experimental protocol to measure the energy variation of the spin qutrit due to the action of the dissipative-autonomous Maxwell demon. We also present and discuss our experimental results. 
A brief discussion about the energy extraction capability of dissipative demons is presented in Sec.~\ref{sec:asymptotic_energy_extraction}. 
Conclusions and perspectives are summarized in Sec.~\ref{sec:conclusions}.

\section{Feedback-controlled dissipative dynamics: Experiment}
\label{sec:Experiment_map} 

We now describe the intrinsic-feedback dissipative dynamics realized with a diamond spin qutrit.

The negatively-charged NV center, a quantum defect comprising a substitutional nitrogen atom next to a vacancy in the diamond lattice, forms an electronic spin triplet $S=1$ in its orbital ground state.
The intrinsic electron spin-spin interaction separates in energy the state $\ket{m_S=0}$ from the degenerate $\ket{m_S=\pm1}$ (where $\ket{m_S}$ are the eigenstates of the spin operator $S_z = \ket{+1}\!\!\bra{+1}-\ket{-1}\!\!\bra{-1}$ 
along the NV symmetry axis $z$), while an applied magnetic field $B$ aligned along $z$ removes the degeneracy of the electronic spin states $\ket{\pm1}$,  and leads to the formation of a three-level system. Each of the three states $\ket{m_S}$ is further split into  hyperfine sublevels due to coupling to the NV $^{14}$N nuclear spin $I=1$~\cite{Steiner10}, however we restrict our analysis to the hyperfine subspace with nuclear spin projection $m_I = +1$, since the other states are depleted as a part of the initialization procedure and then are  out of resonance in the following experiments, thus not contributing to the spin dynamics.

The spin qutrit is coherently driven by bichromatic on-resonant microwave radiation, with frequency components $\omega_{\pm1}$.
The spin dynamics under the continuous double driving is described by the Hamiltonian
 $H(t)=\mathcal{H}_{\mathrm{NV}} +\omega\left[\cos(\omega_{+1}t)\ket{+1}\!\!\bra{0}+\cos(\omega_{-1}t)\ket{-1}\!\!\bra{0}+h.c.\right]$, where $\omega$ is the driving Rabi frequency. 
When $\omega$ vanishes, this reduces to the intrinsic spin-1 Hamiltonian
\begin{equation} \label{eq:mathcalH_NV}
\mathcal{H}_{\mathrm{NV}} = \Delta S_z^2 + \gamma_e B S_z \,,
\end{equation}
where $\Delta \simeq 2.87$~GHz is the zero-field-splitting, and $\gamma_e$ is the electron gyromagnetic ratio ($\hbar=1$ here and throughout this paper) . When the microwave (mw) driving is on, with $\omega_{\pm1} =\Delta \pm \gamma_e B$, in the mw rotating frame and after applying the rotating-wave approximation the spin Hamiltonian simplifies as follows:\begin{equation} \label{eq:mathcalH}
\mathcal{H}_\mathrm{mw} =\omega S_x.
\end{equation}
We performed different experiments while using each of the two Hamiltonians [Eq.~\eqref{eq:mathcalH_NV} and \eqref{eq:mathcalH}] to determine the unitary part of dissipative maps, and the energy basis in which fluctuations are evaluated.

On top of the unitary evolution, the system is intermittently opened by means of its interaction with a train of short laser pulses. The pulse length $t_\mathrm{L}$ is negligible compared to the characteristic timescale of the spin dynamics ($t_\mathrm{L} \ll 2\pi/\omega$). 
Although the laser pulses are equidistant, photon absorption events follow a binomial random distribution in time, due to the
finite photon absorption probability ($p_\mathrm{a}<1$).
While a long laser pulse would produce a complete optical pumping in the $\ket {0}$ state~\cite{Doherty13}, the interaction with a short laser pulse---as used here---
has the following effect: 
If photons are not absorbed, the state of the system is unperturbed; if a photon is absorbed, the spin performs an optical transition to a short-lifetime  orbital excited triplet state, loosing any coherence in the $S_z$-basis during the process~\cite{Wolters13},  then it decays back to the orbital ground triplet state $\{\ket{\pm1},\ket{0}\}$. 
The decay occurs through a direct spin-preserving radiative channel, or through spin-nonpreserving nonradiative paths involving intermediate metastable states. The different decay rates of radiative and nonradiative paths result in an optical-pump of the spin towards the state $\ket{0}$. 
Considering the reduced 3-level system formed by the spin levels of the orbital ground state, the photon absorption induces a loss of coherence and an \textit{irreversible dissipation}, originated by optical pumping.
Such a dissipation, conditioned by the absorption  of a photon, constitutes the basis of the 
intrinsic-feedback dissipative dynamics, as schematized in Fig.~\ref{fig:protocol}(a-b). 
The mathematical model of these phenomena is described in Sec.~\ref{sec:Modeled_map}.

\section{Modelled dynamics: Autonomous-dissipative Maxwell demon}
\label{sec:Modeled_map}

Here, we model the qutrit dynamics in terms of a Lindbladian master equation describing an autonomous-dissipative Maxwell's demon.
For this we use the formalism described in Ref.~\cite{Havel03} for superoperators represented as $N^2\times N^2$ matrices, where $N=3$ is the dimension of the Hilbert space of the three-level system (3LS). 
According to this formalism, a given density matrix $\rho$ under unitary evolution $U\equiv e^{-i\tau\mathcal{H}}$ is transformed into
\begin{equation}
\col[U \rho U^\dagger ] = \boldsymbol{U} \col[\rho]
\end{equation}
where $\col[\rho]$ denotes the vectorization of $\rho$, obtained by stacking the columns of $\rho$ to form a `column' vector, and $\boldsymbol{U} \equiv \exp(-i \tau ( \mathcal{H}\otimes \mathbb{1}_{3\times 3}  - \mathbb{1}_{3\times 3} \otimes \mathcal{H}^* ))$ with $\otimes$ being the Kronecker product.

On the other hand, the interaction between the 3LS and a short laser pulse can be described by a positive operator-valued measure~(POVM) followed by a dissipation operator conditioned on the POVM result. 
Specifically, the interaction with a single short laser pulse transforms a density matrix $\rho$ into 
\begin{equation}
\col[\rho(t_\mathrm{L})] = \boldsymbol{\mathcal{A}} \col[\rho]
\end{equation}
where the superoperator $\boldsymbol{\mathcal{A}}$ models the mean effect of the single short laser pulse. Taking into account all the possible outcomes of the 3LS-laser interaction, $\boldsymbol{\mathcal{A}}$ is written as
\begin{equation} \label{eq:mathcalA_bis}
\boldsymbol{\mathcal{A}} \equiv \sum_{j=1}^{4} \boldsymbol{\mathcal{D}}_j \boldsymbol{m}_j
\end{equation}
where $\boldsymbol{m}_j \equiv m_j \otimes m_j$ is one of the measurement super-operator associated with the POVM~$m_j=\{m_1, m_2 ,m_3 ,m_4 \}$ with
\begin{subequations}
 \label{eq:povm1}
\begin{align} 
m_1 &\equiv \sqrt{p_\mathrm{a}}\ket{-1}\!\!\bra{-1} \\
m_2 &\equiv \sqrt{p_\mathrm{a}}\ket{0}\!\!\bra{0} \\
m_3 &\equiv \sqrt{p_\mathrm{a}}\ket{+1}\!\!\bra{+1} \\
m_4 &\equiv \sqrt{(1-p_\mathrm{a})}\mathbb{1}_{3\times 3},
\end{align}
\end{subequations}
such that $\sum_{j=1}^{4} m_j m_j^\dagger = \mathbb{1}_{3\times 3}$,
and $\boldsymbol{\mathcal{D}}_j$ represents the action of a superoperator conditioned to the result of the POVM:
\begin{equation}\label{eq:Super_operator_S_j}
\boldsymbol{\mathcal{D}}_j =
\begin{cases}
    \pmb{\mathbb{1}}_{9\times 9} , & \text{if } j = 4 \\
	\boldsymbol{\mathcal{L}} , & \text{otherwise}
\end{cases}
\end{equation}
with
\begin{align}
\boldsymbol{\mathcal{L}} \equiv \exp \left( t_\mathrm{L}  \sum_{\ell=0}^1 L^*_\ell \otimes L_\ell - \frac{1}{2} \mathbb{1}_{3\times 3}\otimes L^\dagger_\ell L_\ell \right. & \notag \\
\left. - \frac{1}{2}(L^\dagger_\ell L_\ell)^* \otimes  \mathbb{1}_{3\times 3} \right) & \label{eq:lindbladian_operator_def}
\end{align}
where $(\cdot)^*$~denotes complex conjugate, and $L_\ell$ are the Lindblad jump operators $\{ L_0,L_1 \} \equiv \{ \sqrt{\Gamma}\ket{0}\!\!\bra{+1} , \sqrt{\Gamma}\ket{0}\!\!\bra{-1} \}$, describing the dissipation towards the state $\ket{0}$.
In Eq.~\eqref{eq:mathcalA_bis}, the term for $j=4$ corresponds to the case where the laser pulse is not absorbed, while the other three terms model the absorption of a single laser pulse. The Lindblad dissipative super-operator $\boldsymbol{\mathcal{L}}$ is defined in terms of the product between the effective decay rate and the laser duration $\Gamma t_\mathrm{L}$, which dictates the strength of the dissipation that brings the system towards $\ket{0}$; the dissipation probability is $p_\mathrm{d}\equiv(1-e^{-\Gamma t_\mathrm{L} })$. The explicit expression of $\boldsymbol{\mathcal{L}}$ can be found in the Supplemental Material at [URL will be inserted by publisher].
Using short laser pulses with $t_\mathrm{L}=41$~ns, we experimentally characterized the strength of this decay rate resulting in a value such that $\Gamma t_\mathrm{L} \simeq 1/2$. 
Given the effective nature of this model, the value of $\Gamma$ might vary for different NV centers and under different experimental conditions. 
Notice that for a long laser pulse ($t_\mathrm{L} \Gamma \gg 1$) any given state $\rho$ is transformed into $ \boldsymbol{\mathcal{L}} \col[\rho] = \col[\ket{0}\!\!\bra{0}]$, which is consistent with the usual protocol employed to optically initialize the electronic spin state. 
Note  that, in the hypothetical limit where $p_\mathrm{d}=0$, then no dissipation will occur ($\boldsymbol{\mathcal{L}} = \pmb{\mathbb{1}}_{9\times 9}$) and consequently $\boldsymbol{\mathcal{D}}_j = \pmb{\mathbb{1}}_{9\times 9}$.

Here, the action of the dissipation operator is conditioned on the result of the previously-applied POVM. 
Thus, the combined effect of unitary evolution, POVMs, and dissipation results in an \emph{autonomous Maxwell's demon}, whose action is defined by the specific photodynamics of the NV center~[see Fig.~\ref{fig:protocol}(b)]. Overall, the effect of this feedback-controlled dissipative map $\mathcal{M}$, after $N_\mathrm{L}$ laser pulses, is thus modeled as $\mathcal{M}[\rho] \rightarrow \boldsymbol{\mathcal{B}}^{N_\mathrm{L}} \col[\rho]$, where $\boldsymbol{\mathcal{B}}$ is the superoperator that describes a single block of the dynamics formed by unitary evolution followed by the interaction of the system with a short laser pulse 
\begin{equation}\label{eq:supOp_B_singleblock}
\boldsymbol{\mathcal{B}} \equiv \boldsymbol{\mathcal{A}} \,\boldsymbol{U} .
\end{equation} 
The study of energy variation fluctuations induced by this autonomous Maxwell demon 
is a central issue of this article.

\section{Dissipative Sagawa-Ueda-Tasaki relation}
\label{sec:g-SUTR}

In this context, quantum fluctuation relations (QFR)~\cite{Campisi11} provide a powerful framework to characterize fluctuations of energy by inspecting its characteristic function. The quantum fluctuation relation for dynamics under measurements and feedback control, also known as the quantum Sagawa-Ueda-Tasaki relation, was originally proposed for protocols where specific unitary operations are applied to the quantum system depending on the outcomes of a sequence of projective measurements~\cite{Morikuni11,Funo13}. 
Later on it was extended to the more general case of non-unitary dynamics described by completely positive trace-preserving (CPTP) maps, subject to the result of the POVMs---which generalize the projective measurements~\cite{Kafri12,Rastegin13,Albash13,Goold15,SongPRA2021}. 
Here, we detail how to formulate such extension using the superoperator formalism, instead of the usual Kraus operators formalism.

A two-point measurement (TPM) scheme~\cite{Talkner07} is used to characterize the statistics of the energy variation: An initial energy measurement projects an initial thermal state $\rho^{\mathrm{th}}$ into one of the Hamiltonian eigenstates, which then evolves under the map we are studying, and a final energy measurement allows us to extract the energy difference. By repeating this process several times 
it is possible to reconstruct probability distribution of the energy variation  \begin{equation}\label{eq:probsDeltaE}
P_{\Delta E} \equiv {\rm Prob}(\Delta E) = \sum_{i,j} \delta(\Delta E - \Delta E_{j,i})P_{j|i}P_{i} ,
\end{equation}
where $\Delta E_{j,i} \equiv E_j - E_i$, $P_i$ is the probability
to obtain $E_i$ as a result of the first energy measurement of $\rho^{\mathrm{th}}$, and $P_{j|i}$ is the conditional probability to obtain $E_j$ as a result of the second energy measurement at the end of the TPM scheme. 
Once that the statistics of $\Delta E$ are known, the characteristic function $\mathcal{G}(\eta) \equiv \langle e^{iu\Delta E} \rangle_{u=i\eta}$ of $P_{\Delta E}$ can be experimentally computed as
\begin{equation}\label{eq:LHS_f}
\mathcal{G}(\eta) = \sum_{i,j} e^{-\eta\Delta E_{j,i}} P_{j|i} P_i\,.
\end{equation}

The general protocol is the following.
An $n$-dimensional quantum system (in our experiments, $n = 3$) evolves under the Hamiltonian $H_0$; then, a POVM is performed on the system. The quantum measurement is defined by a set of positive semidefinite operators $\Pi_1,...\Pi_{n'}$ such that $\sum_{k=1}^{n'}\Pi_k=\mathbb{1}_{n\times n}$. 
According to a feedback mechanism, the measurement outcome $k$ determines the CPTP map $\Phi_k$ under which the system continues to evolve. Since $\Phi_k$ is a CPTP map, we can define the superoperator propagator $\boldsymbol{\Phi}_k$ such that the evolution of a generic density matrix $\rho$ is described as $\col[\Phi_k[\rho]] = \boldsymbol{\Phi}_k \col{[\rho]}$~\cite{Havel03}. 
Hence, the complete feedback map $\mathcal{M}_{\Phi}$ transforms the density matrix $\rho$ into
\begin{equation} \label{eq:qSUT_generic_map}
\col[\mathcal{M}_{\Phi}[\rho]] = \sum_{k=1}^{n'} \boldsymbol{\Phi}_k \boldsymbol{\Pi}_k \boldsymbol{U}_0 \col[\rho],
\end{equation}
where $\boldsymbol{U}_0 = \exp(-i t ( H_0\otimes \mathbb{1}_{n\times n}  - \mathbb{1}_{n\times n} \otimes H_0^* ))$ is the superoperator that describes the unitary evolution before the POVM, and $\boldsymbol{\Pi}_k \equiv \Pi_k\otimes \Pi_k$ represents the action of the measurement operators on the quantum system.

As a result, the characteristic function of the energy variation is equal to a parameter $\gamma$ that represents the efficacy of the feedback mechanism:
\begin{equation} \label{eq:qSUT_generic}
\mathcal{G}(\beta) = \gamma \equiv \sum_{k=1}^{n'} \Tr_{n\times n}\left[\boldsymbol{\Pi}_k^\dagger \boldsymbol{\Phi}_k^\dagger \,\col[\rho^{\mathrm{th}}_{\mathrm{f}}]\right],
\end{equation}
where $\beta$ is the inverse temperature of the initial thermal state, $\rho^{\mathrm{th}}_{\mathrm{f}}$ is 
a thermal state at the final time of the TPM  scheme, the symbol $\dagger$ denotes conjugate-transpose, and $\Tr_{n\times n}[\col[\cdot]] \equiv \Tr[\cdot]$. 
The efficacy of the feedback mechanism, as measured by $-\ln \gamma$, determines a lower bound on the energy variation of the system, as we will discuss in Sec.~\ref{sec:asymptotic_energy_extraction}. 
The mathematical proof of Eq.~\eqref{eq:qSUT_generic} can be found in Appendix~\ref{app:proof_GSUTR}~(see \textit{Corollary~1}). The proof is based on the fact that the map $\mathcal{M}_{\Phi}$ is itself a CPTP map, meaning that Eq.~\eqref{eq:qSUT_generic} is a particular case of the general quantum fluctuation relation for CPTP maps~\cite{Kafri12,Rastegin13,Albash13,Rastegin14,Goold15,SongPRA2021}. 
It is worth observing that Eq.~\eqref{eq:qSUT_generic} reduces to the original quantum SUT relation~\cite{Morikuni11} in the particular case where the intermediate quantum measurements are projective and $\Phi_k$ are unitary evolution operators.
Moreover, if $\Phi_k$ are unital CPTP maps, then $\boldsymbol{\Phi}_k^\dagger$ is trace preserving~\cite{Havel03}, hence $\gamma=1$~\cite{Kafri12,Albash13} for any time-independent Hamiltonian. In contrast, for non-unital maps where microreversibility is not satisfied, the value of $\gamma$ can be different from $1$ and, in general, involves non trace-preserving operators $\boldsymbol{\Phi}_k^\dagger$.

Equations~\eqref{eq:qSUT_generic_map} and \eqref{eq:qSUT_generic} refer to a feedback operation on the quantum system enabled by applying a single POVM measurement, and they can be simplified by defining $\boldsymbol{\mathcal{B}}_\Phi \equiv  \sum_{k=1}^{n'} \boldsymbol{\Phi}_k \boldsymbol{\Pi}_k$ that leads to $\gamma = \Tr_{n\times n}[ \boldsymbol{\mathcal{B}}_\Phi^\dagger \,\col[\rho^{\mathrm{th}}]]$. Therefore, extending the protocol to  the scenario where measurements and feedback are applied repeatedly is quite straightforward: After $N$ repetitions $\gamma= \Tr_{n\times n}[ (\boldsymbol{\mathcal{B}}_\Phi^\dagger)^N \,\col[\rho^{\mathrm{th}}]]$. 
Expressing $\gamma$ in this way significantly simplifies its computation, because it removes the requirement to calculate every possible quantum trajectory originated by the system dynamics; compare for example with Refs.~\cite{Morikuni11,Campisi17}. This advantage may be significant since the number of trajectories scales exponentially with the number of measurements. 
In addition, for any CPTP dissipative map, whereby the system asymptotically reaches the single steady state 
$\col[\rho^\infty] \equiv \lim_{N\rightarrow\infty}(\boldsymbol{\mathcal{B}}_\Phi)^N \col[\rho]$, 
for a generic initial state $\rho$, we can write the asymptotic value of $\gamma$ as (see also \textit{Corollary~2} in Appendix~\ref{app:proof_GSUTR})
\begin{equation}\label{eq:gamma_infty}
\gamma^{\infty}\equiv \lim_{t\rightarrow\infty}\gamma = n\langle\rho^\infty, \rho^\mathrm{th}\rangle_{\mathrm{hs}} 
\end{equation}
where $\langle \rho_1,\rho_2 \rangle_{\mathrm{hs}}\equiv \Tr[\rho_1^\dagger \rho_2]$ denotes the Hilbert-Schmidt inner product, and $n$ is the dimension of the quantum system. Remarkably, the quantity in Eq.~\eqref{eq:gamma_infty} can be measured experimentally, even for non-unital maps.

\begin{figure*}[t!]
\includegraphics[width=0.925\textwidth]{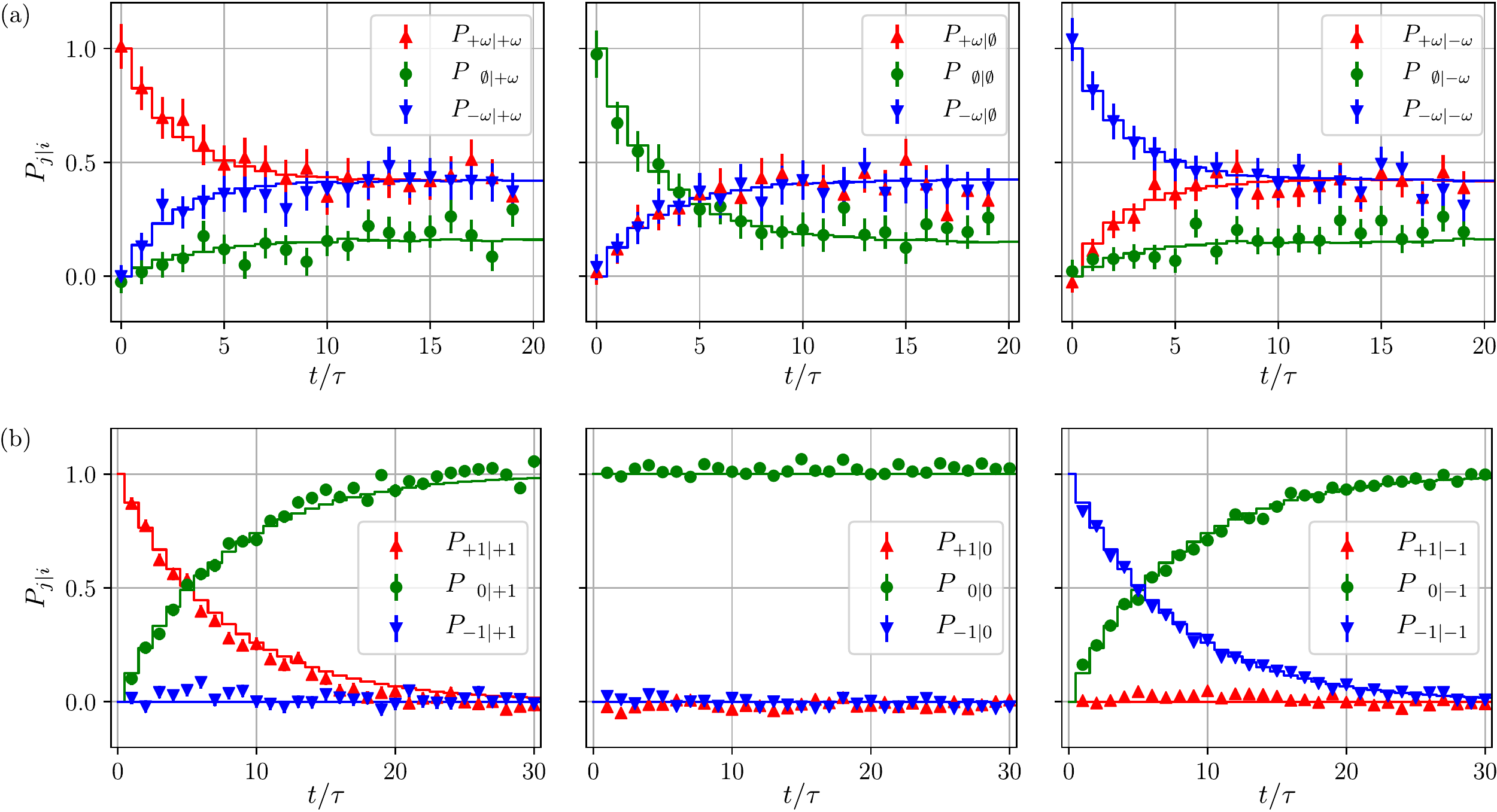}
\caption{
Dynamics of the spin qutrit under the autonomous-dissipative Maxwell demon. 
The probability $P_{j|i}$ of measuring the state $\ket{E_j}$ after applying the map $\mathcal{M}$ to the state $\ket{E_i}$ is shown as a function of the number of laser pulses $N_\mathrm{L} = t /\tau$, where $\ket{E_i}$ and $\ket{E_j}$ represent each of the eigenstates of the $3LS$ Hamiltonian
\textbf{(a)} $\mathcal{H}_\mathrm{mw}$ and \textbf{(b)} $\mathcal{H}_\mathrm{NV}$. In both cases, the duration of the laser pulses is $t_\mathrm{L}=41$~ns, and the time between pulses is $\tau=424$~ns.
Each panel corresponds to a different initial eigenstate: \textbf{(a)} $\{\ket{+\omega},\ket{\emptyset},\ket{-\omega}\}$ and \textbf{(b)} $\{\ket{+1},\ket{0},\ket{-1}\}$.
The markers with errorbars represent the experimental data, and each solid line corresponds to the calculation performed with the theoretical model of the map $\mathcal{M}$, as described in Sec.~\ref{sec:Modeled_map}. Experimental error bars are mainly due to photon shot noise. 
Notice that these measurement stem from a comparison between the PL of the two reference states and the PL of the NV after the studied dynamics, which may result in values that are slightly outside the $[0,1]$ interval.
}
\label{fig:conditional_probs}
\end{figure*}

\section{Dynamics and thermodynamics of the demon}\label{sec:protocol}

We have characterized the dynamics and the quantum (non-thermal) energy fluctuations of the 3-level spin system induced by non-unital  
quantum dissipative maps in two independent experiments:
\begin{enumerate}
\item Coherent double driving and short laser pulses. The unitary part of the map $\mathcal{M}$ is ruled by the Hamiltonian $\mathcal{H}_\mathrm{mw}$ defined in Eq.~\eqref{eq:mathcalH}, with eigenstates $\ket{E_1} = \ket{- \omega}$, $\ket{E_2} = \ket{\emptyset}$, and $\ket{E_3} = \ket{+\omega}$, where $\ket{\pm \omega} \equiv \frac{1}{2}\left(\ket{-1} \pm \sqrt{2}\ket{0} + \ket{1}\right)$, and $\ket{\emptyset} \equiv \frac{1}{\sqrt{2}}\left(\ket{-1} - \ket{1}\right)$.
\item Undriven spin, subject to short laser pulses. The spin Hamiltonian is $\mathcal{H}_\mathrm{NV}$ with eigenstates $\ket{E_1} = \ket{0}$, $\ket{E_2} = \ket{- 1}$, and $\ket{E_3} = \ket{+ 1}$.
\end{enumerate}

The effect of the map $\mathcal{M}$ on the spin energy is characterized by measuring the energy jump probabilities, i.e., the conditional probabilities associated with the energy variation in a given time interval.
The scheme used to measure conditional probabilities, as depicted in Fig.~\ref{fig:protocol}(c) consists in  the following steps:

a) Initialize the system into one of the Hamiltonian eigenstates, say $\ket{E_i}$;

b) Evolve the system under the map $\mathcal{M}$ up to  time  $t$;

c) Read out the probability of the spin to be in the Hamiltonian eigenstate $\ket{E_j}$ at final time $t$;

d) Repeat the procedure for each initial and final Hamiltonian eigenstates.

\subsection{Spin initialization and readout}

The NV spin is initially prepared in the spin qutrit Hamitonian eigenstate $\ket{E_i}$. 
The starting point of the initialization is a thermal spin mixture 
$\overline{\rho}=\sum_{m_S,m_I}\ket{m_S,m_I}\!\!\bra{m_S,m_I}$ within the $9\times9$ space described by the hyperfine manifold within orbital ground state. The quantum gate $G_0$ prepares the  hyperfine state $\ket{m_S,m_I}\equiv\ket{0,1}$ (see Fig.~\ref{fig:protocol}). 
From now on we drop the hyperfine spin to simplify the notation, $\ket{m_S,1}=\ket{m_S}$. The details about the nuclear spin initialization can be found in the Supplemental Material at [URL will be inserted by publisher].
Pure electron spin states in the energy basis ($\ket{E_i}$) are then prepared by applying opportune two-level-system quantum gates ($G_i$, in Fig.~\ref{fig:protocol}) realized by means of nuclear spin selective monochromatic microwave pulses resonant with the electronic transitions $\ket{0}\rightarrow \ket{+1}$ or $\ket{0}\rightarrow \ket{-1}$ as described in Appendix~\ref{app:prep_and_readout_gates}.

After a spin evolution under the dissipative map $\mathcal{M}$, the spin state readout is performed by exploiting the difference in the NV center photoluminescence~(PL) intensity, among state $\ket{0}$ and states $\ket{\pm1}$, upon illumination with green laser light.
To measure the probability of the spin state to be equal to each of the three Hamiltonian eigenstates $\ket{E_j}$ at the end of the protocol, that eigenstate is projected into the $S_z$ eigenstate $\ket{0}$ (quantum gate $G_j^{-1}$ in Fig.~\ref{fig:protocol}), then the PL intensity is  recorded.
The description of the gates $G_i$ and $G_j^{-1}$ for each of the initial and final eigenstates can be found in Appendix~\ref{app:prep_and_readout_gates}.
Since the readout process is destructive, each experiment is composed by three different runs, where the probability of each of the three eigenstates is recorded. 
Due to limited diamond PL collection efficiency and to photon shot noise, each experiment is  repeated $\sim10^6$ times.

Notice that the duration of each realization of the experiment is much shorter than the nuclear spin lifetime so that the three-dimensional hyperfine spin subspace is well defined for the whole experiment duration.

\begin{figure*}[t!]
\includegraphics[width=0.9\textwidth]{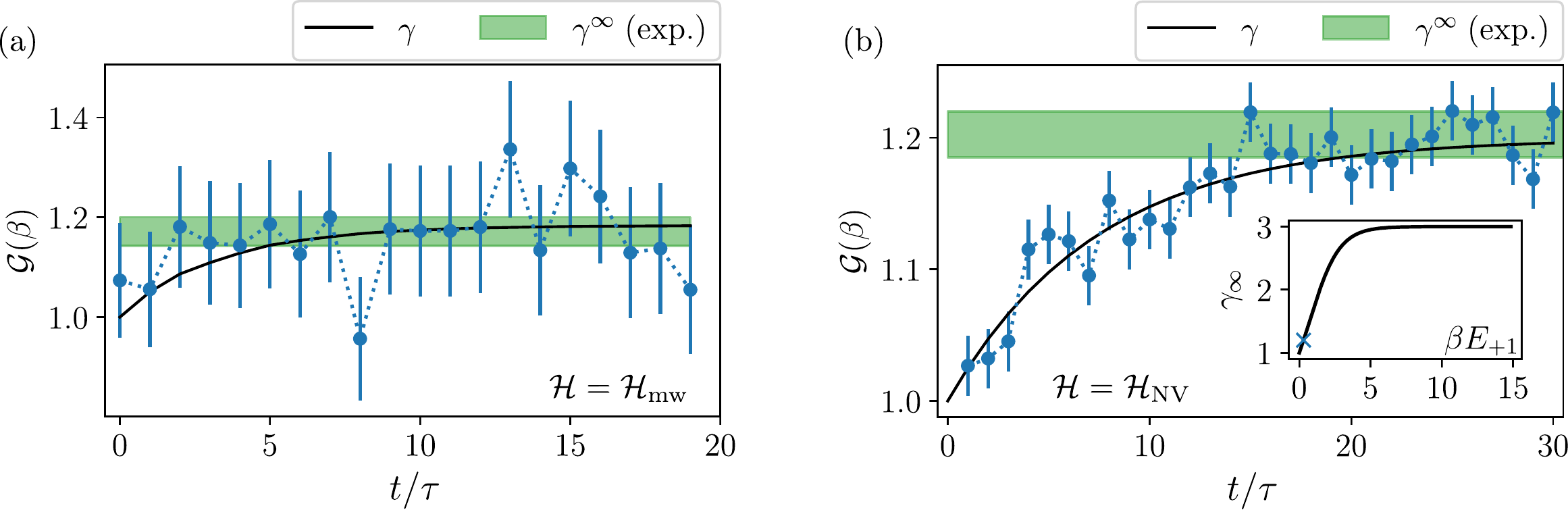}
\caption{
Experimental verification of the generalized quantum Sagawa-Ueda-Tasaki relation \eqref{eq:qSUT_generic} for a 3LS controlled by an autonomous-dissipative Maxwell's demon. The energy variation statistics are shown in terms of the total time of the experiment divided by the time between laser pulses, i.e. in terms of the number of laser pulses $N_\mathrm{L} = t/\tau$.
\textbf{(a)}~The continuous coherent driving of the 3LS is defined by the Hamiltonian $\mathcal{H}_\mathrm{mw}$ and the value of the inverse temperature of the initial state is $\beta = 3/E_{+\omega}$.
\textbf{(b)}~Same as (a) but for a Hamiltonian $\mathcal{H}_\mathrm{NV}$, and an inverse temperature of the initial state $\beta = 0.297/E_{+1}$.
Blue circles represent the measured values of the characteristic function $\mathcal{G}(\beta)$, the black continuous line denotes the theoretical estimation of $\gamma$~[Eq.~\eqref{eq:gamma_def}],
and the green area marks the experimental value of $\gamma^\infty$~[Eq.~\eqref{eq:gamma_infty}], the asymptotic value of $\gamma$.
Inset: theoretical value of $\gamma^\infty$ as a function of the initial inverse temperature $\beta$. The blue cross is associated with the value of $\beta$ used for the experimental data in (b).
}
\label{fig:FR_results}
\end{figure*}

\subsection{Demon dynamics}
\label{subsec:conditional_probabilities}

We then reconstruct the dynamics of the spin qutrit under the autonomous-dissipative Maxwell's demon. 
The conditional probabilities $P_{j|i}$ associated with energy jumps from the initial  ($\ket{E_i}$) to the final ($\ket{E_j}$) eigenstates are shown in Fig.~\ref{fig:conditional_probs}, as a function of the number of laser pulses applied before performing the readout.
The experimental data are shown together with the theoretical model of the dynamics, described in Sec.~\ref{sec:Modeled_map}.
The excellent agreement between experiment and theory leads us to conclude that the dynamics of the system is very well described by an autonomous feedback mechanism. 
The results show that the conditional probabilities $P_{j|i}$ tend to a single constant value in the long-time regime (large $N_\mathrm{L}$).
In other words, the spin state asymptotically approaches a \emph{steady state in the energy basis} (SSE) that does not depend on the initial state, thus confirming the dissipative nature of the map $\mathcal{M}$. In the case
of $\mathcal{H} = \mathcal{H}_\mathrm{NV}$~[Fig.~\ref{fig:conditional_probs}(b)], the asymptotic state is
$\rho^{\infty}_\mathrm{NV} = \sum_{\ell=-1}^{+1} p^{\infty}_{\ell} \ket{\ell}\!\!\bra{\ell}$,
with populations, obtained from the experimental data, $p^{\infty}_{+1} = (-0.01 \pm 0.01)$,
$p^{\infty}_{0} = (1.01 \pm 0.01)$, and $p^{\infty}_{-1} = (0.00 \pm 0.01)$. In such case the protocol is asymptotically  equivalent to an initialization procedure into $\rho^{\infty}_\mathrm{NV} =\ket{0}\!\!\bra{0}$. On the other hand, if $\mathcal{H} = \mathcal{H}_\mathrm{mw}$~[Fig.~\ref{fig:conditional_probs}(a)], the asymptotic state significantly differ from $\ket{0} = \frac{1}{\sqrt{2}}\left( \ket{+\omega} - \ket{-\omega}\right)$.
Although the interaction with each laser pulse pushes the system towards $\ket{0}$,
the unitary evolution modifies the density operator populations in the $S_z$ basis, thus changing the SSE at large times. In such case, the asymptotic state is $\rho^{\infty}_\mathrm{mw} = \sum_{\ell=-\omega}^{+\omega} p^{\infty}_{\ell} \ket{\ell}\!\!\bra{\ell}$,
with populations, obtained from the experimental data,  $p^{\infty}_{+\omega} = (0.41 \pm 0.01)$, $p^{\infty}_{\emptyset} = (0.20 \pm 0.02)$, and $p^{\infty}_{-\omega} = (0.40 \pm 0.01)$, as shown in Fig.~\ref{fig:conditional_probs}(a).

\subsection{Energy variation distribution}

Based on the formalism established in Sec.~\ref{sec:g-SUTR}, here we study the statistics of the energy exchange fluctuations of the demon and we calculate the feedback efficacy for the specific map $\mathcal{M}$.

For an initial thermal state, measuring the conditional probabilities $P_{j|i}$ gives access to the probability distribution of the energy variation $P_{\Delta E}$, as defined in Eq.~\eqref{eq:probsDeltaE}. 
We remind that in the usual two-point measurement (TPM) scheme~\cite{Talkner07}, the probability $P_i$ is measured by performing an energy projective measurement
on the initial state. Here, we initialize the spin into each of the eigenstates of the Hamiltonian [Sec.~\ref{sec:protocol}], and we obtain mixed (equilibrium) states as statistical mixture of the eigenstates with probabilities $P_{i}$ as weight factors. Our scheme
gives equivalent results to a TPM scheme, owing to the large number of experimental realizations~\cite{HernandezGomez20,Cimini20}, while overcoming the difficulties to prepare an initial thermal state. Moreover, our scheme removes  possible experimental errors inherent in the first energy measurement, and allows to use one single set of measurements to study different initial states.

Once that we have measured the energy variation distribution $P_{\Delta E}$, we can experimentally compute the values of the characteristic function $\mathcal{G}(\eta)$ as in Eq.~\eqref{eq:LHS_f}. In addition, as pointed out in Sec.~\ref{sec:g-SUTR}, the value of $\gamma^\infty$ can be independently computed from the initial and asymptotic states of the map~[Eq.~\eqref{eq:gamma_infty}]. 
The measurements of $\mathcal{G}(\beta)$ and $\gamma^\infty$ are shown in Fig.~\ref{fig:FR_results}.
The agreement between the measured values of these two parameters 
represents the experimental verification of the generalized SUT fluctuation relation~[Eq.~\eqref{eq:qSUT_generic}], in the SSE regime, for an open three-level system under quantum (non-thermalizing) dissipative dynamics conditioned by POVM quantum measurements.
Let us observe that the theoretical model of the map $\mathcal{M}$~[Sec.~\ref{sec:Modeled_map}] allows us to calculate the efficacy $\gamma$ also in the transient regime, as
\begin{equation} \label{eq:gamma_def}
\gamma = \Tr_{3\times3}\left[ (\boldsymbol{\mathcal{B}}^\dagger)^{N_\mathrm{L}} \,\col[\rho^{\mathrm{th}}]\right] 
\end{equation}
with $\boldsymbol{\mathcal{B}}$ defined in Eq.~(\ref{eq:supOp_B_singleblock}). Here, it is worth noting
that the `backwards' superoperator $\boldsymbol{\mathcal{B}}^\dagger$ is not trace preserving, a clear sign of the non-reversibility associated with the dissipative process~\cite{Campisi17}.
In the transient regime, the values of $\gamma$~[Eq.~\eqref{eq:gamma_def}] were compared with the experimental values of the characteristic function $\mathcal{G}(\beta)$, as a function of the number of laser pulses. 
As shown in Fig.~\ref{fig:FR_results}(a-b), there is an excellent agreement between the two quantities.

In the specific case of $\mathcal{H}=\mathcal{H}_\mathrm{NV}$, where the Hamiltonian commutes with the POVM operators and with the dissipative operator, we can derive an analytic expression for the efficacy $\gamma$ defined in Eq.~\eqref{eq:gamma_def} as
\begin{equation}\label{eq:gamma_H_NV}
\gamma = \mu^{N_\mathrm{L}}  + 3( 1-\mu^{N_\mathrm{L}} )e^{\beta F}
\end{equation}
where $F \equiv -\beta^{-1} \ln Z$ is the initial free energy of the system, with $Z \equiv \sum_{k=-1}^1e^{-\beta E_i}$, and $\mu \equiv  1 - p_\mathrm{d} p_\mathrm{a}\in[0,1]$ 
defines the probability for the system to not be subjected to feedback. We also recall that $p_\mathrm{a}$ denotes the laser pulse absorption probability, and $p_\mathrm{d}\equiv (1 - e^{-t_\mathrm{L} \Gamma})$ is the dissipation probability for the interaction with a single laser pulse [Sec~\ref{sec:Modeled_map}]. 
The derivation of Eq.~\eqref{eq:gamma_H_NV} is given in the Supplemental Material at [URL will be inserted by publisher].
As one would expect, $\gamma = 1$ if $p_\mathrm{a}=0$ (closed quantum system), or in case of pure projective measurements without dissipation $e^{-t_\mathrm{L} \Gamma} =1$ (no feedback). In addition, since $Z<3$ for $\beta \neq 0$, Eq.\,(\ref{eq:gamma_H_NV}) implies that $\gamma > 1$, a necessary condition for energy extraction~\cite{Campisi17}~(see also Sec.~\ref{sec:asymptotic_energy_extraction}). 
As final remark, notice that $\gamma$ in Eq.~\eqref{eq:gamma_H_NV} is defined in terms of macroscopic quantities -- it does not depend on the trajectories followed by the system. In the inset of Fig.~\ref{fig:FR_results}(b), we show the behavior of the asymptotic value of $\gamma^\infty = 3 \, e^{\beta F}$ obtained from Eq.~\eqref{eq:gamma_H_NV} in the SSE regime, as a function of the inverse initial temperature $\beta$. This result is in agreement with Eq.~\eqref{eq:gamma_infty} for the asymptotic state $\rho^{\infty}_\mathrm{NV} = \ket{0}\!\!\bra{0}$.

\section{Extractable energy in autonomous-dissipative demons}\label{sec:asymptotic_energy_extraction}

In this section, we characterize the Maxwell's demon in terms of its capability of energy extraction. 

The mean energy variation originated by the demon is defined as $\langle\Delta E\rangle \equiv \int \Delta E P_{\Delta E}\,d\Delta E$, which can be rewritten as $\langle\Delta E\rangle = \sum_{i,j}(E_j - E_i)P_{j|i} P_i$ using Eq.~\eqref{eq:probsDeltaE}. 
Negative values of $\langle\Delta E\rangle$ denote the extraction of energy from the system. 
The maximal amount of extractable energy ($-\langle\Delta E\rangle$) is expected to be  bounded from 
above~\cite{Morikuni11, Campisi17} by two fundamental quantities, $\ln\gamma$ and 
the classical \emph{mutual information} $\mathcal{I}$ that 
measures the degree of correlation between the actual outcomes of the POVMs and the outcomes recorded by the demon. 
In the case of an autonomous demon -- where no microscopic information needs to be read out by an external agent during the feedback process -- no errors are associated with the intermediate measurements. In this error-free limit the mutual information $\mathcal{I}$ reduces to  the Shannon entropy~\cite{Morikuni11}, \emph{i.e.}, the measurement of the degree of correlation between the trajectories of the system.  
For a quantum dynamics under the action of measurements and feedback,  each sequence of measurements $(k_1, \ldots, k_{N_L})$ corresponds to a different quantum trajectory. Thus, formally one has that in the error-free limit 
\begin{eqnarray}\label{eq:mutual_info}
    &\mathcal{I} \longrightarrow \langle S_{k_1, \ldots, k_{N_L}}\rangle = & \nonumber \\
    & \displaystyle{ -\sum_{k_1, \ldots, k_{N_L}} p(k_1, \ldots, k_{N_L})\ln p(k_1, \ldots, k_{N_L}) }&
\end{eqnarray}
where $\langle S_{k_1, \ldots, k_{N_L}}\rangle$ is the Shannon entropy, and $p(k_1, \ldots, k_{N_L})$ is the probability for the system to follow the $(k_1, \ldots, k_{N_L})$ trajectory. 
This probability is defined as 
\begin{align}
p(k_1, \ldots, k_{N_L}) = \Tr_{n\times n}[ &  \boldsymbol{m}_{k_{N_L}} \boldsymbol{U} \boldsymbol{\mathcal{D}}_{k_{N_L-1}} \boldsymbol{m}_{k_{N_L-1}} \boldsymbol{U} \cdots \notag\\
& \cdots\boldsymbol{\mathcal{D}}_{k_1} \boldsymbol{m}_{k_1} \boldsymbol{U} \,\col[\rho^{\mathrm{th}}] ],
\end{align} where each $k_i$ can take the values $\{1,2,3,4\}$ that indicate one of the results of the $i-$th POVM, as described by Eqs.~\eqref{eq:povm1}.

\begin{figure}
\includegraphics[width=0.925\columnwidth]{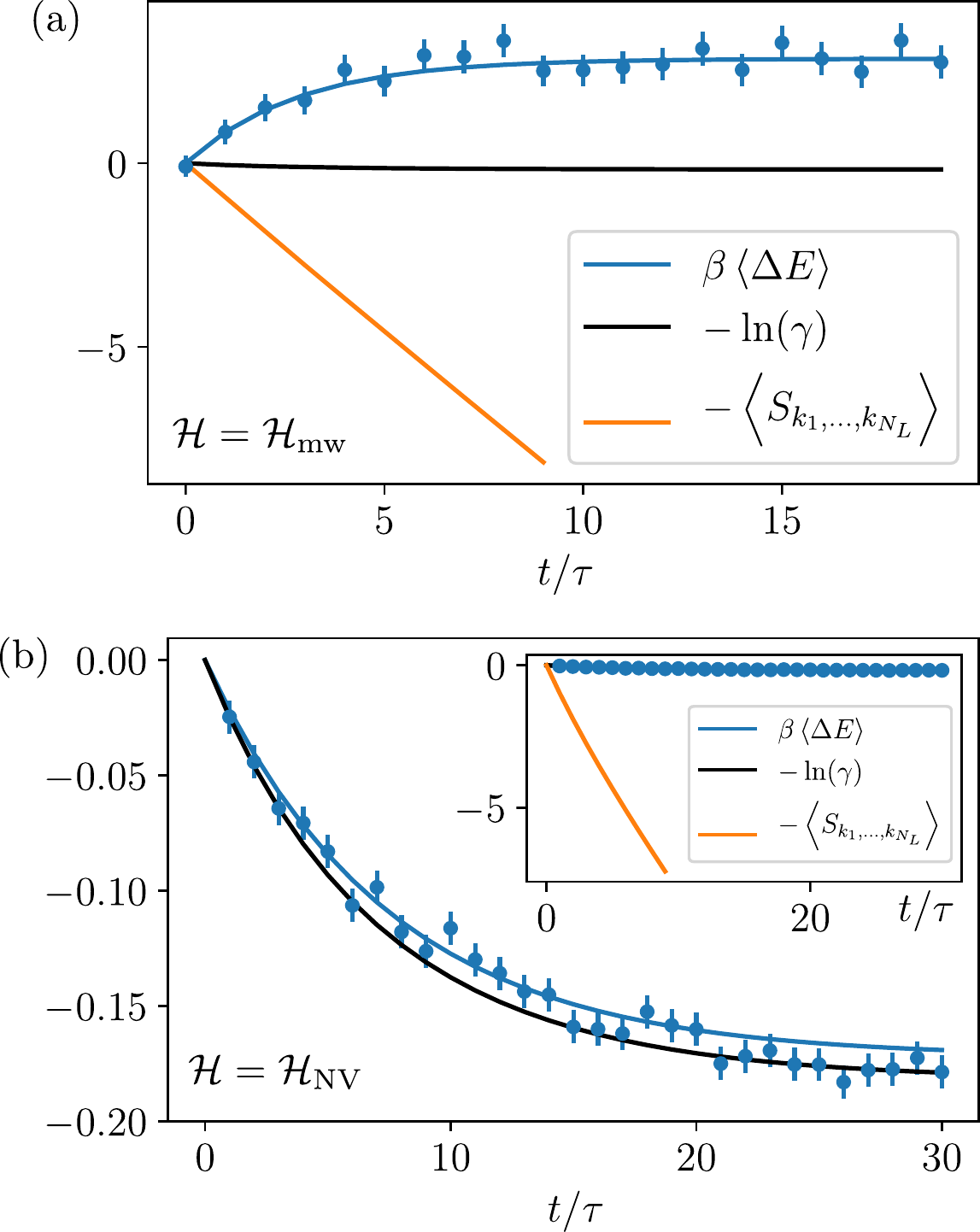}
\caption{
Experimental values of $\beta\langle \Delta E\rangle$ as a function of the number of laser pulses, for a 3LS with Hamiltonian operators $\mathcal{H}_\mathrm{mw}$ (a) and $\mathcal{H}_\mathrm{NV}$ (b). Negative values of $\beta\langle \Delta E\rangle$ (with $\beta>0$) may allow for energy extraction from the 3LS. The solid blue line corresponds to the values of $\beta\langle\Delta E\rangle$ as predicted by theory, while the theoretical bounds $-\ln\gamma$ and $-\langle S_{k_1,\dots,k_{N_L}}\rangle$ are denoted by solid black and orange lines, respectively. The inset in (b) 
shows at the same time the tightness of $-\ln\gamma$ and the discrepancy of $-\langle S_{k_1,\dots,k_{N_L}}\rangle$, both with respect to the mean energy variation $\beta \langle \Delta E\rangle$.
}
\label{fig:DeltaE_logGamma_MutualInfo}
\end{figure}

In Fig.~\ref{fig:DeltaE_logGamma_MutualInfo} we compare the experimental values of $\beta\langle\Delta E\rangle$ (with fixed inverse temperature $\beta > 0$) with the  
calculation of $-\ln\gamma$ and $-\mathcal{I}$, as a function of the number of laser pulses. 
Note that the number of trajectories grows as $4^{N_\mathrm{L}}$, therefore the exact estimation of $\mathcal{I}$ becomes impractical for large $N_\mathrm{L}$. In Fig.~\ref{fig:DeltaE_logGamma_MutualInfo} we show the exact values of $-\mathcal{I}$ for $N_\mathrm{L}<10$ (up to $\sim2.6\times10^5$ trajectories). In contrast, $\langle\Delta E\rangle$ and $\gamma$ do not require the computation of single trajectories, since they are simply determined in terms of $\boldsymbol{\mathcal{B}}^{N_\mathrm{L}}$ and $(\boldsymbol{\mathcal{B}}^\dagger)^{N_\mathrm{L}}$ respectively. 
In the case where $\mathcal{H}=\mathcal{H}_\mathrm{mw}$, energy extraction does not occur (indeed, $\beta\langle\Delta E\rangle>0$).
In contrast, in the case $\mathcal{H}=\mathcal{H}_\mathrm{NV}$, $\beta\langle\Delta E\rangle<0$ for any time $t>0$. Moreover, one can observe that even experimentally the inequality
\begin{equation}
    \beta\langle\Delta E\rangle \geq \max\left\{-\ln\gamma, -\langle S_{k_1, \ldots, k_{N_L}}\rangle \right\}
\end{equation}
is always validated for any value of $t$. However, in both experimental scenarios, the tightest bound is provided by $\ln\gamma$.

\begin{figure}[t!]
\includegraphics[width=0.85\columnwidth]{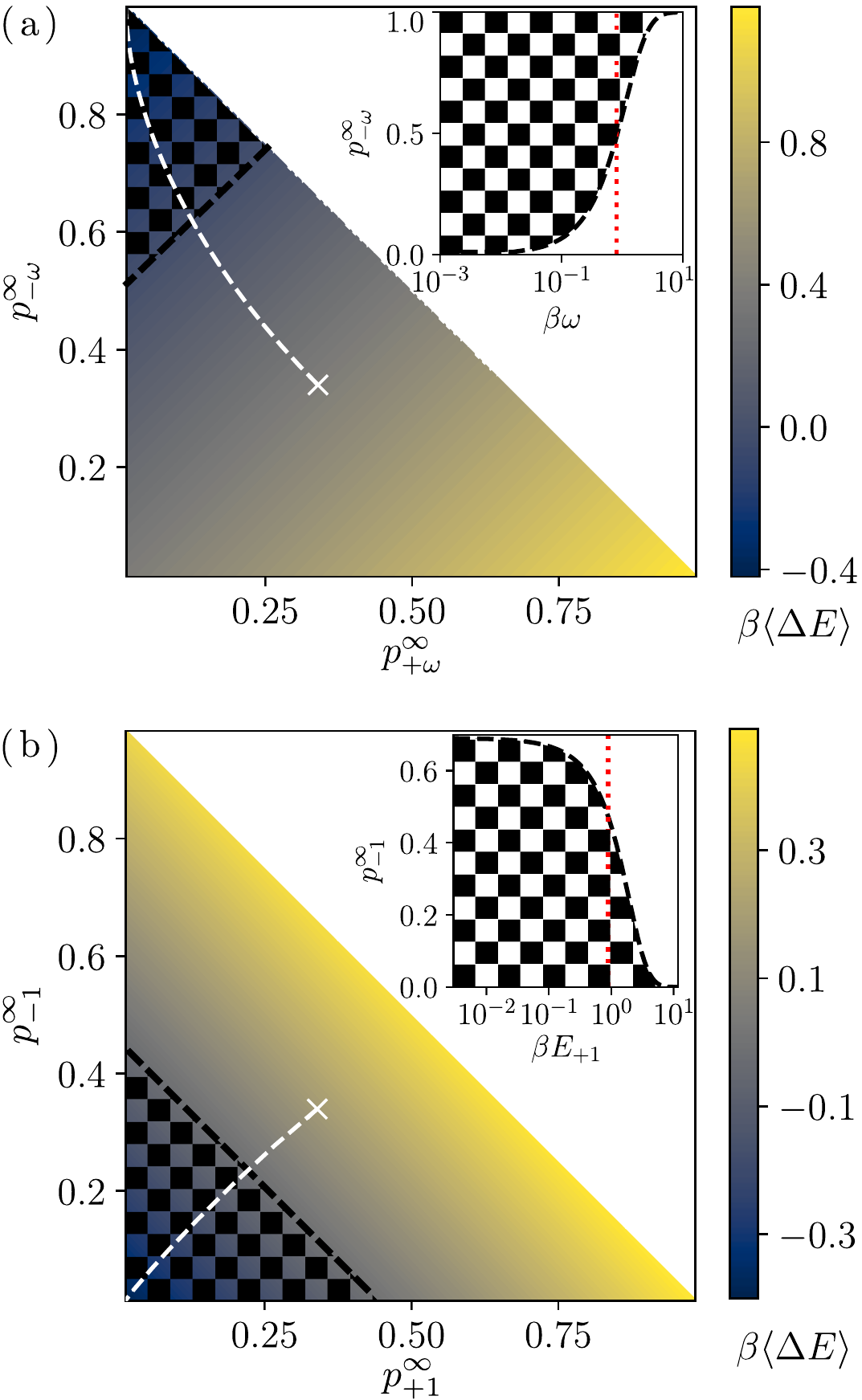}
\caption{
Values of $\beta \langle \Delta E\rangle$ as a function of the asymptotic state populations, for $\mathcal{H}=\mathcal{H}_\mathrm{mw}$~(a) and for $\mathcal{H}=\mathcal{H}_\mathrm{NV}$~(b). In these plots, each point represents a different asymptotic state of the generic quantum dissipative map $\Phi_\mathrm{d}$. The white cross represents a completely mixed asymptotic state for a 3LS, corresponding to the only case in which the map is unital. The subspace of thermalizing dynamics is indicated by the white dashed line, denoting that the asymptotic state is a thermal state with inverse temperature $\beta_{\infty}$. Instead, the black line corresponds to the subspace of states for which $\beta \langle \Delta E\rangle = 0$. The squared-area corresponds to the cases where the energy of the system is reduced, $\beta \langle \Delta E\rangle < 0$, i.e. the asymptotic states for which it is possible to extract energy from the system. 
Note that the inverse temperature of the initial state is fixed for each plot: (a)~$\beta = 0.8/E_{+\omega}$ and (b)~$\beta = 0.891/E_{+1}$. 
Insets: By fixing  $p_{+\omega}^\infty = 0$ (a) (respectively, $p_{+1}^\infty = 0$ (b)) we are able to show the region where $\beta \langle \Delta E\rangle <0 $ as a function of both, the asymptotic state populations and the initial inverse temperature $\beta$. Notice that $\beta$ (x-axis) is scaled with the highest Hamiltonian eigenvalue. The red dashed line corresponds to the value of $\beta$ in figures (a) and (b).
}
\label{fig:epsilon_vs_asymptoticPopuls}
\end{figure}

Let us now analyse the task of energy extraction in  stationary conditions.
Implementing the intrinsic feedback mechanism of the demon by means of dissipative operations implies that the quantum system asymptotically reaches a SSE, i.e., a state for which, on average, the open system does not exchange energy with the external environment, despite the active presence of interaction dynamics (indeed, the quantum system is not closed). 
In the SSE regime, the energy jump conditional probabilities are independent of the initial state, and this entails that $\lim_{t\rightarrow\infty}P_{j|i}(t) = p_{j}^\infty$ as shown in Sec.~\ref{subsec:conditional_probabilities}. Therefore, in such a regime the mean energy variation is time independent and it only depends on the difference between the mean energy of the asymptotic and the initial states: $\langle\Delta E\rangle = \langle E\rangle_\infty - \langle E\rangle_0$, where $\langle E\rangle_\infty \equiv \Tr[\rho^\infty H] = \sum_{j} p_{j}^\infty E_j$ and $\langle E\rangle_0 \equiv \Tr[\rho^{\mathrm{th}} H]$. 
This requires only the knowledge of the initial state, and the measurement of the stationary state $\rho^{\infty}$ induced by dissipation (or more precisely, the measurement of the stationary populations of the system, since the TPM scheme projects the final state into the energy basis of the 3LS). 
Notice that, although the SSE is independent of the initial state, it does depend on the parameters that determine the feedback-controlled quantum map. For example, for the demon that we realize, in the limit of vanishing dissipation  ($p_{\mathrm{d}}\rightarrow 0$, i.e., unital dynamics) the SSE would be a completely mixed state, with no possibility to extract energy.
More broadly, the sign and the value of the energy change might be arbitrarily tuned by a careful choice of the setup parameters. 

In Fig.~\ref{fig:epsilon_vs_asymptoticPopuls} we provide numerical simulations showing the values of $\beta\langle\Delta E\rangle$ ($\beta > 0$) as a function of two populations of the asymptotic state for a 3LS. 
Notice that each point in these plots indicate a different asymptotic state, hence a different dissipative Maxwell's demon. Depending on the specific Hamiltonian, a set of different asymptotic states may allow for energy extraction from the 3LS, as indicated by the triangular shaped squared-area. The black dashed line indicates the limit of such region, i.e., the SSE for which $\beta\langle\Delta E\rangle=0$. 
The slope of this dashed line depends on the specific Hamiltonian, and the \textit{y-intercept} depends on the initial inverse temperature $\beta$. 
In the inset of Fig.~\ref{fig:epsilon_vs_asymptoticPopuls}, we show the value of the $y$-intercept for which the energy variation of the quantum system is zero (black dashed line) as a function of $\beta$. Thus, the squared-area represent the values of the asymptotic populations for which the energy variation of the system is negative. 
As expected, in the limit of low temperature ($\beta\rightarrow \infty)$ the energy extraction from the system is impossible, and the lowest energy Hamiltonian eigenstate is the only possible SSE for which the mean energy variation is zero. In contrast, when the initial inverse temperature $\beta$ approaches zero (infinite temperature limit), the number of asymptotic states that would permit the extraction of energy is maximized.

It is worth observing that for the considered example in Fig.~\ref{fig:epsilon_vs_asymptoticPopuls} (but also for any dissipative quantum map with a unique fixed point), the non-unitality of the underlying quantum process is a \textit{necessary condition} for energy extraction. When the Maxwell's demon is responsible for unital dynamics, the value of the characteristic function $\mathcal{G}(\beta)$ [Eq.~\eqref{eq:qSUT_generic}] must be equal to $1$, for any time value and thus even asymptotically~\cite{Rastegin13,Guarnieri17,Campisi17}. Notice though that $\gamma = 1$ does not necessarily imply that the system exhibits unital dynamics, as discussed in [Appendix~\ref{app:unitality}].
On the other hand, energy extraction is only possible when $\gamma> 1$~\cite{Campisi17}, as a consequence of Eq.~\eqref{eq:qSUT_generic} and Jensen's inequality. Therefore, energy extraction implies $\gamma > 1$, which in turn implies non-unital dynamics. Finally, from Fig.~\ref{fig:epsilon_vs_asymptoticPopuls}, it is apparent that energy extraction is possible for thermal but also for non-thermal asymptotic states.

\section{Conclusions}
\label{sec:conclusions}

In this work, we  use the electronic spin qutrit associated to an NV center in diamond at room temperature to realize an autonomous-dissipative quantum Maxwell demon.  
The interaction of the NV spin qutrit with short laser pulses, is effectively described as an intrinsic feedback process, where dissipative operations (optical pumping) are applied conditioned on the result of a POVM.
The demon can be effectively considered as being \emph{autonomous}, since the feedback mechanism is inherent in the laser-induced photodynamics of the NV spin. Hence no external agent exchanges information with the system.

We have theoretically and experimentally quantified the efficacy of the demon by measuring purely quantum (non-Gibbsian) energy fluctuations, which we have described by means of an appropriate extension of the Sagawa-Ueda-Tasaki formalism for non-unitary, and even non-unital, feedback processes. 
For non-unital dynamics it is not always possible to measure the efficacy, but the dissipative feature (stemming from  optical pumping) of the autonomous demon allows us to measure its asymptotic value.

Finally, we also characterized the demon capability of extracting energy from the system by directly measuring the mean energy variation. We have found that the efficacy is indeed a tighter bound of the mean energy variation compared with the mutual information. 

Our results pave the way for the use of NV centers in diamond to further investigate open quantum system dynamics and thermodynamics. 
In particular, by applying cyclic interactions with the non-thermal reservoir, it has been conjectured the possibility to create a non-Gibbsian quantum heat engine~\cite{Gardas15}, where quantum correlations affect the total amount of heat during the interaction processes. 
More broadly, this work opens the possibility for reinterpreting dissipative phenomena, such as optical pumping, as Maxwell demons to shed light on energy-information relations.
In addition, the proposed experimental scheme to measure energy variation statistics can be adjusted to one-time measurements schemes~\cite{Gherardini20x,Micadei20,Sone20} or to quasi-probability measurements~\cite{Levy20} with the aim to investigate the role of coherence in energy exchange mechanisms, with the final goal of understanding the effects of genuine quantum features in thermodynamic variables. 
Moreover, other forms of quantum fluctuation relations based on observables that do not commute with the system Hamiltonian may be measured to explore quantum synchronization~\cite{Roulet18}, and the relation of the latter with quantum mutual information and with entanglement between the quantum system and its dissipative environment. 
Such kind of studies could also be exploited to realize multipartite entangled systems~\cite{Abobeih19} formed by single NV electronic spins and nuclear spins inside the diamond. Indeed, the high degree of control and long coherence time for such complex spin systems would represent a very useful test-bed for the relation of quantum information and quantum thermodynamics at the nanoscale.

\section*{Acknowledgements}

We gratefully thank Massimo Inguscio for enlightening discussions. We acknowledge financial support from the MISTI Global Seed Funds MIT-FVG Collaboration Grant and from the European Union's 2020 Research and Innovation Program - Qombs Project (FET Flagship on Quantum Technologies grant No 820419). S.G. also acknowledges The Blanceflor Foundation for financial support through the project ``The theRmodynamics behInd thE meaSuremenT postulate of quantum mEchanics (TRIESTE)''.

\appendix

\section{Formal derivation of the dissipative SUT relation}\label{app:proof_GSUTR}

In this Appendix, we first demonstrate the validity of the general QFR, for an $n$-level quantum system under a completely-positive trace-preserving (CPTP) map $\Phi$.
Then we demonstrate the validity of Eq.~\eqref{eq:qSUT_generic}, as a corollary of the previous proof.
Finally, we demonstrate the validity of the dissipative SUT relation for a dissipative map with a unique fixed point. 
We recall that the validity of the general QFR for CPTP map has been proved before~\cite{Kafri12,Rastegin13,Albash13,Goold15}.

Assuming a two-point-measurement scheme, the system energy is measured at the beginning of the protocol, then the system evolves under a CPTP map, and finally its energy is measured again.
The Hamiltonian of the system can be decomposed in terms of the energy eigenstates that define the projectors $\mathcal{P}_i \equiv \ket{E_i}\!\!\bra{E_i}$.
In agreement with the superoperator formalism~\cite{Havel03} used in Sec.~\ref{sec:Modeled_map}, the state after an ideal energy measurement is given by $\col[\mathcal{P}_i \rho \mathcal{P}_i ] = \boldsymbol{\mathcal{P}}_i \col[\rho]$, where $\boldsymbol{\mathcal{P}}_i \equiv \mathcal{P}_i \otimes \mathcal{P}_i$.
Therefore, the joint probability to obtain $E_i$ in the first energy measurement, and $E_f$ in the final one, is written as
\begin{equation}
P_{f,i} = \Tr_{n\times n}[\boldsymbol{\mathcal{P}}_f \, \boldsymbol{\mathcal{J}} \,\boldsymbol{\mathcal{P}}_i \,\col[\rho^{\mathrm{th}}] ]
\end{equation}
where $\rho^{\mathrm{th}}$ is the initial thermal state, $\Tr_{n\times n}[\col[\cdot]] \equiv \Tr[\cdot]$, and $\boldsymbol{\mathcal{J}}$ is the superoperator propagator associated with the CPTP map~$\Phi$. 
The characteristic function $\mathcal{G}(\beta)\equiv \langle e^{-\beta \Delta E} \rangle$ of the energy variation distribution can be then written as
\begin{equation}
\mathcal{G}(\beta) = \sum_{i,f=1}^n \;P_{f,i} e^{-\beta (E_f-E_i)} .
\label{eq:lhs_traj1}
\end{equation}

By expressing the initial thermal state as $\rho^{\mathrm{th}} \equiv \sum_{k=1}^n \mathcal{P}_k e^{-\beta E_k}/Z$ with $Z \equiv \sum_{k=1}^ne^{-\beta E_k}$, we obtain the following:
\begin{align}
\mathcal{G}(\beta) & = \sum_{i,f=1}^n \;P_{f,i} e^{-\beta (E_f-E_i)} \notag \\
& = \sum_{i,f,k=1}^n \;\Tr_{n\times n}[\boldsymbol{\mathcal{P}}_f \, \boldsymbol{\mathcal{J}} \,\boldsymbol{\mathcal{P}}_i \, \col[\mathcal{P}_k] ] e^{-\beta (E_f-E_i+E_k)}/Z \notag \\
& = \sum_{f,k=1}^n \;\Tr_{n\times n}[\boldsymbol{\mathcal{P}}_f \, \boldsymbol{\mathcal{J}} \,\col[\mathcal{P}_k] ] e^{-\beta (E_f)}/Z \label{eq:proof_GSUTR_step1}
\end{align}
where we have used the equality
$\boldsymbol{\mathcal{P}}_i \,\col[\mathcal{P}_k] = \col[ \mathcal{P}_i \,\mathcal{P}_k \,\mathcal{P}_i] = \col[ \mathcal{P}_k ] \delta_{k, i}$ with $ \delta_{k, i}$ the Kronecker delta.
On the other hand, we know that $\Tr_{n\times n}[\boldsymbol{\mathcal{P}}_f \, \col[\rho] ] = (\col[\mathcal{P}_f])^\dagger\col[\rho]$, for any given density matrix $\rho$.
Hence, from Eq.~\eqref{eq:proof_GSUTR_step1} we get
\begin{align}
\mathcal{G}(\beta) & = \sum_{f,k=1}^n \;  (\col[\mathcal{P}_f])^\dagger \, \boldsymbol{\mathcal{J}} \,\col[\mathcal{P}_k]   e^{-\beta (E_f)}/Z \notag \\
& = \sum_{f,k=1}^n \; \left( \boldsymbol{\mathcal{J}} \,\col[\mathcal{P}_k] \right)^\dagger \,\col[\mathcal{P}_f]   e^{-\beta (E_f)}/Z \notag \\
& = \sum_{f,k=1}^n \; (\col[\mathcal{P}_k])^\dagger \, \boldsymbol{\mathcal{J}}^\dagger \,\col[\mathcal{P}_f]   e^{-\beta (E_f)}/Z \notag \\
& = \sum_{k=1}^n \; \Tr_{n\times n}[\boldsymbol{\mathcal{P}}_k \, \boldsymbol{\mathcal{J}}^\dagger \,\col[\rho^{\mathrm{th}}_\mathrm{f}] ] \label{eq:proof_GSUTR_step2_bis} \\
& = \Tr_{n\times n}[ \boldsymbol{\mathcal{J}}^\dagger \,\col[\rho^{\mathrm{th}}_\mathrm{f}] ], \label{eq:proof_GSUTR_step2}
\end{align}
where $\rho^{\mathrm{th}}_\mathrm{f}$ denotes the thermal state at inverse temperature $\beta$ 
taking the system Hamiltonian at the time instant in which the second energy measurement of the TPM scheme is performed. In the case of a time invariant Hamiltonian (as in the experiment of the main text) $\rho^{\mathrm{th}}_\mathrm{f}$ coincides with the initial thermal state. Equation~\eqref{eq:proof_GSUTR_step2} concludes the 
proof.

\textbf{Corollary 1:}
Let us assume that the quantum system is under the feedback map $\mathcal{M}_{\Phi}$ described in Sec.~\ref{sec:g-SUTR}. Given the fact that $\mathcal{M}_{\Phi}$ is formed by a combination of a POVM followed by CPTP maps, it is easy to prove that the map $\mathcal{M}_{\Phi}$ is itself a CPTP map, such that $\col[\mathcal{M}_{\Phi}[\rho]] = \boldsymbol{\mathcal{J}}_\Phi \col[\rho] $, with $\boldsymbol{\mathcal{J}}_\Phi \equiv \sum_{k=1}^{n'} \boldsymbol{\Phi}_k \boldsymbol{\Pi}_k \boldsymbol{U}_0$. Therefore, using Eq.~\eqref{eq:proof_GSUTR_step2} we obtain that  $\mathcal{G}(\beta) = \gamma = \Tr_{n\times n}[ \sum_{k=1}^{n'}  \boldsymbol{\Pi}_k^\dagger \boldsymbol{\Phi}_k^\dagger \,\col[\rho^{\mathrm{th}}] ] $, hence proving the validity of Eq.~\eqref{eq:qSUT_generic}.

\textbf{Corollary 2:} 
Assuming that the CPTP map is a dissipative map $\Phi_\mathrm{d}$ with a unique fixed point $\rho^{\infty}$, then we can write $\lim_{t\rightarrow\infty} \boldsymbol{\mathcal{J}}\col[\mathcal{P}_k] = \col[\rho^{\infty}]$, for every value of $k\in\{1,\dots,n\}$. Hence, from Eq.~\eqref{eq:proof_GSUTR_step2_bis} one gets $
\gamma^{\infty}\equiv \lim_{t\rightarrow\infty}\gamma = n\Tr[\rho^\infty \rho^\mathrm{th}]$.

In the particular case of the dissipative map $\mathcal{M}$ used in during our experiments, $\boldsymbol{\mathcal{J}} = \boldsymbol{\mathcal{B}}^{N_\mathrm{L}}$, which describes the effect of the dissipative map after $N_\mathrm{L}$ laser pulses.

\section{Hamiltonian eigenstate preparation and final readout gates}
\label{app:prep_and_readout_gates}

As shown in Fig.~\ref{fig:protocol}, the preparation of the Hamiltonian initial eigenstate requires the application of the quantum gate $G_i^{(i)}:\ket{0}\rightarrow\ket{E_i}$, while the readout of the eigenstate $\ket{E_j}$ requires a second quantum gate, i.e., $G_j^{-1} :\ket{E_j}\rightarrow\ket{0}$. In this section we describe these gates for each of the possible states $\ket{E_i}$ and $\ket{E_j}$.

There are two possibilities for preparing the Hamiltonian eigenstates. One is with a double-driving microwave (MW) gate driving transitions between $\ket{0}$ and $\ket{\pm1}$, and the other is with two MW pulses applied subsequently to transfer parts of the population from $\ket{0}$ to $\ket{-1}$ and $\ket{+1}$ separately. In our experimental setup we have opted for the latter method due to easier handling of the MW operations. To induce the transition $ \ket{0} \rightarrow \ket{\emptyset} \equiv \frac{1}{\sqrt{2}}\left(\ket{-1} - \ket{+1}\right)$, the population in $\ket{0}$ has to be transferred in equal parts to $\ket{\pm1}$ where both parts have an opposite phase. This is achieved by applying a $\pi/2$-pulse that transfers half of the population to $\ket{-1}$, and subsequently applying a $\pi$-pulse to transfer the remaining population in $\ket{0}$ to $\ket{+1}$. To obtain the correct phase between the $\ket{\pm1}$, it is required that the phase of both pulses is $\pi/2$ (or $-\pi/2$), as one can verify by calculation.
Also the preparation of $\ket{\pm\omega} \equiv \frac{1}{2}\left(\ket{-1} \pm \sqrt{2}\ket{0} + \ket{+1}\right)$ works in a very similar way. A $\pi/3$-pulse has to be applied to transfer one quarter of the population to $\ket{-1}$ and, then, an $\arccos(1/3)$-pulse transfers another one quarter of population from $\ket{0}$ to $\ket{+1}$. Calculation shows that the microwave phases have to be $\mp\pi/2$ and $\pm\pi/2$, respectively for the first and second MW, if we aim to prepare $\ket{\pm\omega}$.

The second quantum gate $G_j^{-1} \equiv G_j^{\dagger}$ applies the reversed process with respect to the preparation one. Thus $G_2^{(j)}$ is obtained by performing the operations of the state preparation in reversed order and assigning to the implemented MW pulses an opposite phase.

\section{Efficacy of the demon \& unitality witness of quantum dissipative maps}\label{app:unitality}

As mentioned in Sec.~\ref{sec:asymptotic_energy_extraction}, $\gamma=1$ is a consequence of unital dynamics. The opposite is not necessarily true though.  Here we show how $\gamma=1$ does not necessarily imply that the system is under unital dynamics. In order to do this, we will focus on the case of $\gamma^\infty$~[Eq.~\eqref{eq:gamma_infty}], as it is a much simpler quantity than $\gamma$.

Let us take a generic $n$-dimensional quantum system. 
Notice that under the assumption that $\Phi_\mathrm{d}$ is a dissipative map with a unique fixed point $\rho^{\infty}$, which is reached asymptotically and does not depend on the initial state, then the formal definition of unitality, $\Phi_\mathrm{d}[\frac{\mathbb{1}}{n}] = \frac{\mathbb{1}}{n}$, is equivalent to 
\begin{equation}\label{eq:non_unital_ineq_gamma}
  \Phi_\mathrm{d}\;\mathrm{unital}
  \Leftrightarrow \rho^\infty  = \frac{\mathbb{1}}{n}  .
\end{equation}
On the other hand, we can write the asymptotic state after the second energy measurement, $\rho^\infty$, according to its spectral decomposition in the $H(t=0)$ basis, i.e., $\rho^\infty = \sum_{k=0}^{n-1}\ket{E_k}\!\!\bra{E_k} p_k^\infty + \xi$, where $\xi$ contains any coherent terms that may appear in the case of  $[H(t=0),H(t)]\neq 0$. 
Using Eq.~\eqref{eq:gamma_infty} it can be  shown that
\begin{align} \label{eq:gamma1_counterexample}
  \gamma^\infty = 1 \Leftrightarrow 
  \sum_{k=1}^{n-1}\left(p_k^\infty - \frac{1}{n}\right)\left(e^{-\beta E_0}-e^{-\beta E_k} \right) = 0\,.
\end{align}
Clearly, the state $\rho^\infty = \frac{\mathbb{1}}{n}$ is a solution of equation~\eqref{eq:gamma1_counterexample}, but it is not the only one. 
There is a whole family of solutions given by the condition
\begin{equation}
p_{n-1}^\infty = \frac{1}{n} - \sum_{k=1}^{n-2}\left(p_k^\infty - \frac{1}{n}\right)\left(\frac{e^{-\beta E_0}-e^{-\beta E_k}}{e^{-\beta E_0}-e^{-\beta E_{n-1}}} \right) .
\end{equation}
Notice though that if $\rho^\infty$ is constrained to be a thermal state  (thermalizing dynamics), then the only solution is  $\rho^\infty = \frac{\mathbb{1}}{n}$.

\bibliography{OQS-Biblio}

\begin{thebibliography}{79}%
\makeatletter
\providecommand \@ifxundefined [1]{%
 \@ifx{#1\undefined}
}%
\providecommand \@ifnum [1]{%
 \ifnum #1\expandafter \@firstoftwo
 \else \expandafter \@secondoftwo
 \fi
}%
\providecommand \@ifx [1]{%
 \ifx #1\expandafter \@firstoftwo
 \else \expandafter \@secondoftwo
 \fi
}%
\providecommand \natexlab [1]{#1}%
\providecommand \enquote  [1]{``#1''}%
\providecommand \bibnamefont  [1]{#1}%
\providecommand \bibfnamefont [1]{#1}%
\providecommand \citenamefont [1]{#1}%
\providecommand \href@noop [0]{\@secondoftwo}%
\providecommand \href [0]{\begingroup \@sanitize@url \@href}%
\providecommand \@href[1]{\@@startlink{#1}\@@href}%
\providecommand \@@href[1]{\endgroup#1\@@endlink}%
\providecommand \@sanitize@url [0]{\catcode `\\12\catcode `\$12\catcode
  `\&12\catcode `\#12\catcode `\^12\catcode `\_12\catcode `\%12\relax}%
\providecommand \@@startlink[1]{}%
\providecommand \@@endlink[0]{}%
\providecommand \url  [0]{\begingroup\@sanitize@url \@url }%
\providecommand \@url [1]{\endgroup\@href {#1}{\urlprefix }}%
\providecommand \urlprefix  [0]{URL }%
\providecommand \Eprint [0]{\href }%
\providecommand \doibase [0]{http://dx.doi.org/}%
\providecommand \selectlanguage [0]{\@gobble}%
\providecommand \bibinfo  [0]{\@secondoftwo}%
\providecommand \bibfield  [0]{\@secondoftwo}%
\providecommand \translation [1]{[#1]}%
\providecommand \BibitemOpen [0]{}%
\providecommand \bibitemStop [0]{}%
\providecommand \bibitemNoStop [0]{.\EOS\space}%
\providecommand \EOS [0]{\spacefactor3000\relax}%
\providecommand \BibitemShut  [1]{\csname bibitem#1\endcsname}%
\let\auto@bib@innerbib\@empty
\bibitem [{\citenamefont {Maruyama}\ \emph {et~al.}(2009)\citenamefont
  {Maruyama}, \citenamefont {Nori},\ and\ \citenamefont {Vedral}}]{Maruyama09}%
  \BibitemOpen
  \bibfield  {author} {\bibinfo {author} {\bibfnamefont {K.}~\bibnamefont
  {Maruyama}}, \bibinfo {author} {\bibfnamefont {F.}~\bibnamefont {Nori}}, \
  and\ \bibinfo {author} {\bibfnamefont {V.}~\bibnamefont {Vedral}},\ }\href
  {\doibase 10.1103/RevModPhys.81.1} {\bibfield  {journal} {\bibinfo  {journal}
  {Rev. Mod. Phys.}\ }\textbf {\bibinfo {volume} {81}},\ \bibinfo {pages} {1}
  (\bibinfo {year} {2009})}\BibitemShut {NoStop}%
\bibitem [{\citenamefont {Bennett}(1982)}]{Bennett1982}%
  \BibitemOpen
  \bibfield  {author} {\bibinfo {author} {\bibfnamefont {C.}~\bibnamefont
  {Bennett}},\ }\href {\doibase 10.1007/BF02084158} {\bibfield  {journal}
  {\bibinfo  {journal} {Int. J. Theor. Phys.}\ }\textbf {\bibinfo {volume}
  {21}},\ \bibinfo {pages} {905} (\bibinfo {year} {1982})}\BibitemShut
  {NoStop}%
\bibitem [{\citenamefont {Funo}\ \emph {et~al.}(2018)\citenamefont {Funo},
  \citenamefont {Ueda},\ and\ \citenamefont {Sagawa}}]{FunoBook2018}%
  \BibitemOpen
  \bibfield  {author} {\bibinfo {author} {\bibfnamefont {K.}~\bibnamefont
  {Funo}}, \bibinfo {author} {\bibfnamefont {M.}~\bibnamefont {Ueda}}, \ and\
  \bibinfo {author} {\bibfnamefont {T.}~\bibnamefont {Sagawa}},\ }\href@noop {}
  {\emph {\bibinfo {title} {Thermodynamics in the quantum regime - Fundamental
  Aspects and New Directions}}}\ (\bibinfo  {publisher} {Springer International
  Publishing},\ \bibinfo {year} {2018})\ pp.\ \bibinfo {pages}
  {249--273}\BibitemShut {NoStop}%
\bibitem [{\citenamefont {Koski}\ \emph {et~al.}(2014)\citenamefont {Koski},
  \citenamefont {Maisi}, \citenamefont {Sagawa},\ and\ \citenamefont
  {Pekola}}]{Koski14}%
  \BibitemOpen
  \bibfield  {author} {\bibinfo {author} {\bibfnamefont {J.~V.}\ \bibnamefont
  {Koski}}, \bibinfo {author} {\bibfnamefont {V.~F.}\ \bibnamefont {Maisi}},
  \bibinfo {author} {\bibfnamefont {T.}~\bibnamefont {Sagawa}}, \ and\ \bibinfo
  {author} {\bibfnamefont {J.~P.}\ \bibnamefont {Pekola}},\ }\href {\doibase
  10.1103/PhysRevLett.113.030601} {\bibfield  {journal} {\bibinfo  {journal}
  {Phys. Rev. Lett.}\ }\textbf {\bibinfo {volume} {113}},\ \bibinfo {pages}
  {030601} (\bibinfo {year} {2014})}\BibitemShut {NoStop}%
\bibitem [{\citenamefont {Naghiloo}\ \emph {et~al.}(2018)\citenamefont
  {Naghiloo}, \citenamefont {Alonso}, \citenamefont {Romito}, \citenamefont
  {Lutz},\ and\ \citenamefont {Murch}}]{Naghiloo2018}%
  \BibitemOpen
  \bibfield  {author} {\bibinfo {author} {\bibfnamefont {M.}~\bibnamefont
  {Naghiloo}}, \bibinfo {author} {\bibfnamefont {J.~J.}\ \bibnamefont
  {Alonso}}, \bibinfo {author} {\bibfnamefont {A.}~\bibnamefont {Romito}},
  \bibinfo {author} {\bibfnamefont {E.}~\bibnamefont {Lutz}}, \ and\ \bibinfo
  {author} {\bibfnamefont {K.~W.}\ \bibnamefont {Murch}},\ }\href {\doibase
  10.1103/PhysRevLett.121.030604} {\bibfield  {journal} {\bibinfo  {journal}
  {Phys. Rev. Lett.}\ }\textbf {\bibinfo {volume} {121}},\ \bibinfo {pages}
  {030604} (\bibinfo {year} {2018})}\BibitemShut {NoStop}%
\bibitem [{\citenamefont {Mandal}\ and\ \citenamefont
  {Jarzynski}(2012)}]{Mandal2012}%
  \BibitemOpen
  \bibfield  {author} {\bibinfo {author} {\bibfnamefont {D.}~\bibnamefont
  {Mandal}}\ and\ \bibinfo {author} {\bibfnamefont {C.}~\bibnamefont
  {Jarzynski}},\ }\href {\doibase 10.1073/pnas.1204263109} {\bibfield
  {journal} {\bibinfo  {journal} {Proc. Nat. Acad. Sc.}\ }\textbf {\bibinfo
  {volume} {109}},\ \bibinfo {pages} {1} (\bibinfo {year} {2012})}\BibitemShut
  {NoStop}%
\bibitem [{\citenamefont {Kutvonen}\ \emph {et~al.}(2016)\citenamefont
  {Kutvonen}, \citenamefont {Koski},\ and\ \citenamefont
  {Ala-Nissila}}]{Kutvonen2016}%
  \BibitemOpen
  \bibfield  {author} {\bibinfo {author} {\bibfnamefont {A.}~\bibnamefont
  {Kutvonen}}, \bibinfo {author} {\bibfnamefont {J.}~\bibnamefont {Koski}}, \
  and\ \bibinfo {author} {\bibfnamefont {T.}~\bibnamefont {Ala-Nissila}},\
  }\href {\doibase 10.1038/srep21126} {\bibfield  {journal} {\bibinfo
  {journal} {Scientific Reports}\ ,\ \bibinfo {pages} {1}} (\bibinfo {year}
  {2016})}\BibitemShut {NoStop}%
\bibitem [{\citenamefont {Buffoni}\ \emph {et~al.}(2019)\citenamefont
  {Buffoni}, \citenamefont {Solfanelli}, \citenamefont {Verrucchi},
  \citenamefont {Cuccoli},\ and\ \citenamefont {Campisi}}]{Buffoni2019}%
  \BibitemOpen
  \bibfield  {author} {\bibinfo {author} {\bibfnamefont {L.}~\bibnamefont
  {Buffoni}}, \bibinfo {author} {\bibfnamefont {A.}~\bibnamefont {Solfanelli}},
  \bibinfo {author} {\bibfnamefont {P.}~\bibnamefont {Verrucchi}}, \bibinfo
  {author} {\bibfnamefont {A.}~\bibnamefont {Cuccoli}}, \ and\ \bibinfo
  {author} {\bibfnamefont {M.}~\bibnamefont {Campisi}},\ }\href {\doibase
  10.1103/PhysRevLett.122.070603} {\bibfield  {journal} {\bibinfo  {journal}
  {Phys. Rev. Lett.}\ }\textbf {\bibinfo {volume} {122}},\ \bibinfo {pages}
  {070603} (\bibinfo {year} {2019})}\BibitemShut {NoStop}%
\bibitem [{\citenamefont {Elouard}\ \emph
  {et~al.}(2017{\natexlab{a}})\citenamefont {Elouard}, \citenamefont
  {Herrera-Mart{\'\i}}, \citenamefont {Clusel},\ and\ \citenamefont
  {Auffeves}}]{Elouard17}%
  \BibitemOpen
  \bibfield  {author} {\bibinfo {author} {\bibfnamefont {C.}~\bibnamefont
  {Elouard}}, \bibinfo {author} {\bibfnamefont {D.~A.}\ \bibnamefont
  {Herrera-Mart{\'\i}}}, \bibinfo {author} {\bibfnamefont {M.}~\bibnamefont
  {Clusel}}, \ and\ \bibinfo {author} {\bibfnamefont {A.}~\bibnamefont
  {Auffeves}},\ }\href {\doibase 10.1038/s41534-017-0008-4} {\bibfield
  {journal} {\bibinfo  {journal} {npj Quantum Information}\ }\textbf {\bibinfo
  {volume} {3}} (\bibinfo {year} {2017}{\natexlab{a}}),\
  10.1038/s41534-017-0008-4}\BibitemShut {NoStop}%
\bibitem [{\citenamefont {Gherardini}\ \emph {et~al.}(2018)\citenamefont
  {Gherardini}, \citenamefont {Buffoni}, \citenamefont {M{\"u}ller},
  \citenamefont {Caruso}, \citenamefont {Campisi}, \citenamefont
  {Trombettoni},\ and\ \citenamefont {Ruffo}}]{Gherardini18}%
  \BibitemOpen
  \bibfield  {author} {\bibinfo {author} {\bibfnamefont {S.}~\bibnamefont
  {Gherardini}}, \bibinfo {author} {\bibfnamefont {L.}~\bibnamefont {Buffoni}},
  \bibinfo {author} {\bibfnamefont {M.~M.}\ \bibnamefont {M{\"u}ller}},
  \bibinfo {author} {\bibfnamefont {F.}~\bibnamefont {Caruso}}, \bibinfo
  {author} {\bibfnamefont {M.}~\bibnamefont {Campisi}}, \bibinfo {author}
  {\bibfnamefont {A.}~\bibnamefont {Trombettoni}}, \ and\ \bibinfo {author}
  {\bibfnamefont {S.}~\bibnamefont {Ruffo}},\ }\href {\doibase
  10.1103/PhysRevE.98.032108} {\bibfield  {journal} {\bibinfo  {journal} {Phys.
  Rev. E}\ }\textbf {\bibinfo {volume} {98}},\ \bibinfo {pages} {032108}
  (\bibinfo {year} {2018})}\BibitemShut {NoStop}%
\bibitem [{\citenamefont {Toyabe}\ \emph {et~al.}(2010)\citenamefont {Toyabe},
  \citenamefont {Sagawa}, \citenamefont {Ueda}, \citenamefont {Muneyuki},\ and\
  \citenamefont {Sano}}]{Toyabe10}%
  \BibitemOpen
  \bibfield  {author} {\bibinfo {author} {\bibfnamefont {S.}~\bibnamefont
  {Toyabe}}, \bibinfo {author} {\bibfnamefont {T.}~\bibnamefont {Sagawa}},
  \bibinfo {author} {\bibfnamefont {M.}~\bibnamefont {Ueda}}, \bibinfo {author}
  {\bibfnamefont {E.}~\bibnamefont {Muneyuki}}, \ and\ \bibinfo {author}
  {\bibfnamefont {M.}~\bibnamefont {Sano}},\ }\href {\doibase
  10.1038/nphys1821} {\bibfield  {journal} {\bibinfo  {journal} {Nat. Phys.}\
  }\textbf {\bibinfo {volume} {6}},\ \bibinfo {pages} {988} (\bibinfo {year}
  {2010})}\BibitemShut {NoStop}%
\bibitem [{\citenamefont {Masuyama}\ \emph {et~al.}(2018)\citenamefont
  {Masuyama}, \citenamefont {Funo}, \citenamefont {Murashita}, \citenamefont
  {Noguchi}, \citenamefont {Kono}, \citenamefont {Tabuchi}, \citenamefont
  {Yamazaki}, \citenamefont {Ueda},\ and\ \citenamefont
  {Nakamura}}]{Masuyama18}%
  \BibitemOpen
  \bibfield  {author} {\bibinfo {author} {\bibfnamefont {Y.}~\bibnamefont
  {Masuyama}}, \bibinfo {author} {\bibfnamefont {K.}~\bibnamefont {Funo}},
  \bibinfo {author} {\bibfnamefont {Y.}~\bibnamefont {Murashita}}, \bibinfo
  {author} {\bibfnamefont {A.}~\bibnamefont {Noguchi}}, \bibinfo {author}
  {\bibfnamefont {S.}~\bibnamefont {Kono}}, \bibinfo {author} {\bibfnamefont
  {Y.}~\bibnamefont {Tabuchi}}, \bibinfo {author} {\bibfnamefont
  {R.}~\bibnamefont {Yamazaki}}, \bibinfo {author} {\bibfnamefont
  {M.}~\bibnamefont {Ueda}}, \ and\ \bibinfo {author} {\bibfnamefont
  {Y.}~\bibnamefont {Nakamura}},\ }\href {\doibase 10.1038/s41467-018-03686-y}
  {\bibfield  {journal} {\bibinfo  {journal} {Nature Communications}\ }\textbf
  {\bibinfo {volume} {9}},\ \bibinfo {pages} {1291} (\bibinfo {year}
  {2018})}\BibitemShut {NoStop}%
\bibitem [{\citenamefont {Koski}\ \emph {et~al.}(2015)\citenamefont {Koski},
  \citenamefont {Kutvonen}, \citenamefont {Khaymovich}, \citenamefont
  {Ala-Nissila},\ and\ \citenamefont {Pekola}}]{Koski15}%
  \BibitemOpen
  \bibfield  {author} {\bibinfo {author} {\bibfnamefont {J.~V.}\ \bibnamefont
  {Koski}}, \bibinfo {author} {\bibfnamefont {A.}~\bibnamefont {Kutvonen}},
  \bibinfo {author} {\bibfnamefont {I.~M.}\ \bibnamefont {Khaymovich}},
  \bibinfo {author} {\bibfnamefont {T.}~\bibnamefont {Ala-Nissila}}, \ and\
  \bibinfo {author} {\bibfnamefont {J.~P.}\ \bibnamefont {Pekola}},\ }\href
  {\doibase 10.1103/PhysRevLett.115.260602} {\bibfield  {journal} {\bibinfo
  {journal} {Phys. Rev. Lett.}\ }\textbf {\bibinfo {volume} {115}},\ \bibinfo
  {pages} {260602} (\bibinfo {year} {2015})}\BibitemShut {NoStop}%
\bibitem [{\citenamefont {Elouard}\ \emph
  {et~al.}(2017{\natexlab{b}})\citenamefont {Elouard}, \citenamefont
  {Herrera-Mart\'{\i}}, \citenamefont {Huard},\ and\ \citenamefont
  {Auff\`eves}}]{Elouard17PRL}%
  \BibitemOpen
  \bibfield  {author} {\bibinfo {author} {\bibfnamefont {C.}~\bibnamefont
  {Elouard}}, \bibinfo {author} {\bibfnamefont {D.}~\bibnamefont
  {Herrera-Mart\'{\i}}}, \bibinfo {author} {\bibfnamefont {B.}~\bibnamefont
  {Huard}}, \ and\ \bibinfo {author} {\bibfnamefont {A.}~\bibnamefont
  {Auff\`eves}},\ }\href {\doibase 10.1103/PhysRevLett.118.260603} {\bibfield
  {journal} {\bibinfo  {journal} {Phys. Rev. Lett.}\ }\textbf {\bibinfo
  {volume} {118}},\ \bibinfo {pages} {260603} (\bibinfo {year}
  {2017}{\natexlab{b}})}\BibitemShut {NoStop}%
\bibitem [{\citenamefont {Campisi}\ \emph {et~al.}(2017)\citenamefont
  {Campisi}, \citenamefont {Pekola},\ and\ \citenamefont {Fazio}}]{Campisi17}%
  \BibitemOpen
  \bibfield  {author} {\bibinfo {author} {\bibfnamefont {M.}~\bibnamefont
  {Campisi}}, \bibinfo {author} {\bibfnamefont {J.}~\bibnamefont {Pekola}}, \
  and\ \bibinfo {author} {\bibfnamefont {R.}~\bibnamefont {Fazio}},\ }\href
  {\doibase 10.1088/1367-2630/aa6acb} {\bibfield  {journal} {\bibinfo
  {journal} {New Journal of Physics}\ }\textbf {\bibinfo {volume} {19}},\
  \bibinfo {pages} {053027} (\bibinfo {year} {2017})}\BibitemShut {NoStop}%
\bibitem [{\citenamefont {Schindler}\ \emph {et~al.}(2011)\citenamefont
  {Schindler}, \citenamefont {Barreiro}, \citenamefont {Monz}, \citenamefont
  {Nebendahl}, \citenamefont {Nigg}, \citenamefont {Chwalla}, \citenamefont
  {Hennrich},\ and\ \citenamefont {Blatt}}]{Schindler11}%
  \BibitemOpen
  \bibfield  {author} {\bibinfo {author} {\bibfnamefont {P.}~\bibnamefont
  {Schindler}}, \bibinfo {author} {\bibfnamefont {J.~T.}\ \bibnamefont
  {Barreiro}}, \bibinfo {author} {\bibfnamefont {T.}~\bibnamefont {Monz}},
  \bibinfo {author} {\bibfnamefont {V.}~\bibnamefont {Nebendahl}}, \bibinfo
  {author} {\bibfnamefont {D.}~\bibnamefont {Nigg}}, \bibinfo {author}
  {\bibfnamefont {M.}~\bibnamefont {Chwalla}}, \bibinfo {author} {\bibfnamefont
  {M.}~\bibnamefont {Hennrich}}, \ and\ \bibinfo {author} {\bibfnamefont
  {R.}~\bibnamefont {Blatt}},\ }\href {\doibase 10.1126/science.1203329}
  {\bibfield  {journal} {\bibinfo  {journal} {Science}\ }\textbf {\bibinfo
  {volume} {332}},\ \bibinfo {pages} {1059} (\bibinfo {year}
  {2011})}\BibitemShut {NoStop}%
\bibitem [{\citenamefont {Hirose}\ and\ \citenamefont
  {Cappellaro}(2016)}]{Hirose16}%
  \BibitemOpen
  \bibfield  {author} {\bibinfo {author} {\bibfnamefont {M.}~\bibnamefont
  {Hirose}}\ and\ \bibinfo {author} {\bibfnamefont {P.}~\bibnamefont
  {Cappellaro}},\ }\href {\doibase 10.1038/nature17404} {\bibfield  {journal}
  {\bibinfo  {journal} {Nature}\ }\textbf {\bibinfo {volume} {532}},\ \bibinfo
  {pages} {77} (\bibinfo {year} {2016})}\BibitemShut {NoStop}%
\bibitem [{\citenamefont {Camati}\ \emph {et~al.}(2016)\citenamefont {Camati},
  \citenamefont {Peterson}, \citenamefont {Batalh\~ao}, \citenamefont
  {Micadei}, \citenamefont {Souza}, \citenamefont {Sarthour}, \citenamefont
  {Oliveira},\ and\ \citenamefont {Serra}}]{CamatiPRL2016}%
  \BibitemOpen
  \bibfield  {author} {\bibinfo {author} {\bibfnamefont {P.~A.}\ \bibnamefont
  {Camati}}, \bibinfo {author} {\bibfnamefont {J.~P.~S.}\ \bibnamefont
  {Peterson}}, \bibinfo {author} {\bibfnamefont {T.~B.}\ \bibnamefont
  {Batalh\~ao}}, \bibinfo {author} {\bibfnamefont {K.}~\bibnamefont {Micadei}},
  \bibinfo {author} {\bibfnamefont {A.~M.}\ \bibnamefont {Souza}}, \bibinfo
  {author} {\bibfnamefont {R.~S.}\ \bibnamefont {Sarthour}}, \bibinfo {author}
  {\bibfnamefont {I.~S.}\ \bibnamefont {Oliveira}}, \ and\ \bibinfo {author}
  {\bibfnamefont {R.~M.}\ \bibnamefont {Serra}},\ }\href {\doibase
  10.1103/PhysRevLett.117.240502} {\bibfield  {journal} {\bibinfo  {journal}
  {Phys. Rev. Lett.}\ }\textbf {\bibinfo {volume} {117}},\ \bibinfo {pages}
  {240502} (\bibinfo {year} {2016})}\BibitemShut {NoStop}%
\bibitem [{\citenamefont {Vidrighin}\ \emph {et~al.}(2016)\citenamefont
  {Vidrighin}, \citenamefont {Dahlsten}, \citenamefont {Barbieri},
  \citenamefont {Kim}, \citenamefont {Vedral},\ and\ \citenamefont
  {Walmsley}}]{VidrighinPRL2016}%
  \BibitemOpen
  \bibfield  {author} {\bibinfo {author} {\bibfnamefont {M.~D.}\ \bibnamefont
  {Vidrighin}}, \bibinfo {author} {\bibfnamefont {O.}~\bibnamefont {Dahlsten}},
  \bibinfo {author} {\bibfnamefont {M.}~\bibnamefont {Barbieri}}, \bibinfo
  {author} {\bibfnamefont {M.~S.}\ \bibnamefont {Kim}}, \bibinfo {author}
  {\bibfnamefont {V.}~\bibnamefont {Vedral}}, \ and\ \bibinfo {author}
  {\bibfnamefont {I.~A.}\ \bibnamefont {Walmsley}},\ }\href {\doibase
  10.1103/PhysRevLett.116.050401} {\bibfield  {journal} {\bibinfo  {journal}
  {Phys. Rev. Lett.}\ }\textbf {\bibinfo {volume} {116}},\ \bibinfo {pages}
  {050401} (\bibinfo {year} {2016})}\BibitemShut {NoStop}%
\bibitem [{\citenamefont {Cottet}\ \emph {et~al.}(2017)\citenamefont {Cottet},
  \citenamefont {Jezouin}, \citenamefont {Bretheau}, \citenamefont
  {Campagne-Ibarcq}, \citenamefont {Ficheux}, \citenamefont {Anders},
  \citenamefont {Auff\`eves}, \citenamefont {Azouit}, \citenamefont {Rouchon},\
  and\ \citenamefont {Huard}}]{CottetPNAS2017}%
  \BibitemOpen
  \bibfield  {author} {\bibinfo {author} {\bibfnamefont {N.}~\bibnamefont
  {Cottet}}, \bibinfo {author} {\bibfnamefont {S.}~\bibnamefont {Jezouin}},
  \bibinfo {author} {\bibfnamefont {L.}~\bibnamefont {Bretheau}}, \bibinfo
  {author} {\bibfnamefont {P.}~\bibnamefont {Campagne-Ibarcq}}, \bibinfo
  {author} {\bibfnamefont {Q.}~\bibnamefont {Ficheux}}, \bibinfo {author}
  {\bibfnamefont {J.}~\bibnamefont {Anders}}, \bibinfo {author} {\bibfnamefont
  {A.}~\bibnamefont {Auff\`eves}}, \bibinfo {author} {\bibfnamefont
  {R.}~\bibnamefont {Azouit}}, \bibinfo {author} {\bibfnamefont
  {P.}~\bibnamefont {Rouchon}}, \ and\ \bibinfo {author} {\bibfnamefont
  {B.}~\bibnamefont {Huard}},\ }\href {\doibase 10.1073/pnas.1704827114}
  {\bibfield  {journal} {\bibinfo  {journal} {PNAS}\ }\textbf {\bibinfo
  {volume} {114}},\ \bibinfo {pages} {7561} (\bibinfo {year}
  {2017})}\BibitemShut {NoStop}%
\bibitem [{\citenamefont {Song}\ \emph {et~al.}(2021)\citenamefont {Song},
  \citenamefont {Naghiloo},\ and\ \citenamefont {Murch}}]{SongPRA2021}%
  \BibitemOpen
  \bibfield  {author} {\bibinfo {author} {\bibfnamefont {X.}~\bibnamefont
  {Song}}, \bibinfo {author} {\bibfnamefont {M.}~\bibnamefont {Naghiloo}}, \
  and\ \bibinfo {author} {\bibfnamefont {K.}~\bibnamefont {Murch}},\ }\href
  {\doibase 10.1103/PhysRevA.104.022211} {\bibfield  {journal} {\bibinfo
  {journal} {Phys. Rev. A}\ }\textbf {\bibinfo {volume} {104}},\ \bibinfo
  {pages} {022211} (\bibinfo {year} {2021})}\BibitemShut {NoStop}%
\bibitem [{\citenamefont {Ji}\ \emph {et~al.}(2022)\citenamefont {Ji},
  \citenamefont {Chai}, \citenamefont {Wang}, \citenamefont {Guo},
  \citenamefont {Rong}, \citenamefont {Shi}, \citenamefont {Ren}, \citenamefont
  {Wang},\ and\ \citenamefont {Du}}]{Ji22}%
  \BibitemOpen
  \bibfield  {author} {\bibinfo {author} {\bibfnamefont {W.}~\bibnamefont
  {Ji}}, \bibinfo {author} {\bibfnamefont {Z.}~\bibnamefont {Chai}}, \bibinfo
  {author} {\bibfnamefont {M.}~\bibnamefont {Wang}}, \bibinfo {author}
  {\bibfnamefont {Y.}~\bibnamefont {Guo}}, \bibinfo {author} {\bibfnamefont
  {X.}~\bibnamefont {Rong}}, \bibinfo {author} {\bibfnamefont {F.}~\bibnamefont
  {Shi}}, \bibinfo {author} {\bibfnamefont {C.}~\bibnamefont {Ren}}, \bibinfo
  {author} {\bibfnamefont {Y.}~\bibnamefont {Wang}}, \ and\ \bibinfo {author}
  {\bibfnamefont {J.}~\bibnamefont {Du}},\ }\href {\doibase
  10.1103/PhysRevLett.128.090602} {\bibfield  {journal} {\bibinfo  {journal}
  {Phys. Rev. Lett.}\ }\textbf {\bibinfo {volume} {128}},\ \bibinfo {pages}
  {090602} (\bibinfo {year} {2022})}\BibitemShut {NoStop}%
\bibitem [{\citenamefont {Najera-Santos}\ \emph {et~al.}(2020)\citenamefont
  {Najera-Santos}, \citenamefont {Camati}, \citenamefont {M\'etillon},
  \citenamefont {Brune}, \citenamefont {Raimond}, \citenamefont {Auff\`eves},\
  and\ \citenamefont {Dotsenko}}]{Najera-SantosPRR2020}%
  \BibitemOpen
  \bibfield  {author} {\bibinfo {author} {\bibfnamefont {B.-L.}\ \bibnamefont
  {Najera-Santos}}, \bibinfo {author} {\bibfnamefont {P.~A.}\ \bibnamefont
  {Camati}}, \bibinfo {author} {\bibfnamefont {V.}~\bibnamefont {M\'etillon}},
  \bibinfo {author} {\bibfnamefont {M.}~\bibnamefont {Brune}}, \bibinfo
  {author} {\bibfnamefont {J.-M.}\ \bibnamefont {Raimond}}, \bibinfo {author}
  {\bibfnamefont {A.}~\bibnamefont {Auff\`eves}}, \ and\ \bibinfo {author}
  {\bibfnamefont {I.}~\bibnamefont {Dotsenko}},\ }\href {\doibase
  10.1103/PhysRevResearch.2.032025} {\bibfield  {journal} {\bibinfo  {journal}
  {Phys. Rev. Research}\ }\textbf {\bibinfo {volume} {2}},\ \bibinfo {pages}
  {032025} (\bibinfo {year} {2020})}\BibitemShut {NoStop}%
\bibitem [{\citenamefont {Verstraete}\ \emph {et~al.}(2009)\citenamefont
  {Verstraete}, \citenamefont {Wolf},\ and\ \citenamefont
  {Ignacio~Cirac}}]{Verstraete09}%
  \BibitemOpen
  \bibfield  {author} {\bibinfo {author} {\bibfnamefont {F.}~\bibnamefont
  {Verstraete}}, \bibinfo {author} {\bibfnamefont {M.~M.}\ \bibnamefont
  {Wolf}}, \ and\ \bibinfo {author} {\bibfnamefont {J.}~\bibnamefont
  {Ignacio~Cirac}},\ }\href {\doibase 10.1038/nphys1342} {\bibfield  {journal}
  {\bibinfo  {journal} {Nature Physics}\ }\textbf {\bibinfo {volume} {5}},\
  \bibinfo {pages} {633} (\bibinfo {year} {2009})}\BibitemShut {NoStop}%
\bibitem [{\citenamefont {Pastawski}\ \emph {et~al.}(2011)\citenamefont
  {Pastawski}, \citenamefont {Clemente},\ and\ \citenamefont
  {Cirac}}]{Pastawski11}%
  \BibitemOpen
  \bibfield  {author} {\bibinfo {author} {\bibfnamefont {F.}~\bibnamefont
  {Pastawski}}, \bibinfo {author} {\bibfnamefont {L.}~\bibnamefont {Clemente}},
  \ and\ \bibinfo {author} {\bibfnamefont {J.~I.}\ \bibnamefont {Cirac}},\
  }\href {\doibase 10.1103/PhysRevA.83.012304} {\bibfield  {journal} {\bibinfo
  {journal} {Phys. Rev. A}\ }\textbf {\bibinfo {volume} {83}},\ \bibinfo
  {pages} {012304} (\bibinfo {year} {2011})}\BibitemShut {NoStop}%
\bibitem [{\citenamefont {Scarani}\ \emph {et~al.}(2002)\citenamefont
  {Scarani}, \citenamefont {Ziman}, \citenamefont {\ifmmode \check{S}\else
  \v{S}\fi{}telmachovi\ifmmode~\check{c}\else \v{c}\fi{}}, \citenamefont
  {Gisin},\ and\ \citenamefont {Bu\ifmmode~\check{z}\else
  \v{z}\fi{}ek}}]{Scarani02}%
  \BibitemOpen
  \bibfield  {author} {\bibinfo {author} {\bibfnamefont {V.}~\bibnamefont
  {Scarani}}, \bibinfo {author} {\bibfnamefont {M.}~\bibnamefont {Ziman}},
  \bibinfo {author} {\bibfnamefont {P.}~\bibnamefont {\ifmmode \check{S}\else
  \v{S}\fi{}telmachovi\ifmmode~\check{c}\else \v{c}\fi{}}}, \bibinfo {author}
  {\bibfnamefont {N.}~\bibnamefont {Gisin}}, \ and\ \bibinfo {author}
  {\bibfnamefont {V.}~\bibnamefont {Bu\ifmmode~\check{z}\else \v{z}\fi{}ek}},\
  }\href {\doibase 10.1103/PhysRevLett.88.097905} {\bibfield  {journal}
  {\bibinfo  {journal} {Phys. Rev. Lett.}\ }\textbf {\bibinfo {volume} {88}},\
  \bibinfo {pages} {097905} (\bibinfo {year} {2002})}\BibitemShut {NoStop}%
\bibitem [{\citenamefont {Nandkishore}\ and\ \citenamefont
  {Huse}(2015)}]{Nandkishore15}%
  \BibitemOpen
  \bibfield  {author} {\bibinfo {author} {\bibfnamefont {R.}~\bibnamefont
  {Nandkishore}}\ and\ \bibinfo {author} {\bibfnamefont {D.~A.}\ \bibnamefont
  {Huse}},\ }\href {\doibase 10.1146/annurev-conmatphys-031214-014726}
  {\bibfield  {journal} {\bibinfo  {journal} {Annual Review of Condensed Matter
  Physics}\ }\textbf {\bibinfo {volume} {6}},\ \bibinfo {pages} {15} (\bibinfo
  {year} {2015})}\BibitemShut {NoStop}%
\bibitem [{\citenamefont {Strasberg}\ \emph {et~al.}(2017)\citenamefont
  {Strasberg}, \citenamefont {Schaller}, \citenamefont {Brandes},\ and\
  \citenamefont {Esposito}}]{Strasberg17}%
  \BibitemOpen
  \bibfield  {author} {\bibinfo {author} {\bibfnamefont {P.}~\bibnamefont
  {Strasberg}}, \bibinfo {author} {\bibfnamefont {G.}~\bibnamefont {Schaller}},
  \bibinfo {author} {\bibfnamefont {T.}~\bibnamefont {Brandes}}, \ and\
  \bibinfo {author} {\bibfnamefont {M.}~\bibnamefont {Esposito}},\ }\href
  {\doibase 10.1103/PhysRevX.7.021003} {\bibfield  {journal} {\bibinfo
  {journal} {Phys. Rev. X}\ }\textbf {\bibinfo {volume} {7}},\ \bibinfo {pages}
  {021003} (\bibinfo {year} {2017})}\BibitemShut {NoStop}%
\bibitem [{\citenamefont {Barreiro}\ \emph {et~al.}(2011)\citenamefont
  {Barreiro}, \citenamefont {Müller}, \citenamefont {Schindler}, \citenamefont
  {Nigg}, \citenamefont {Monz}, \citenamefont {Chwalla}, \citenamefont
  {Hennrich}, \citenamefont {Roos}, \citenamefont {Zoller},\ and\ \citenamefont
  {Blatt}}]{Barreiro11}%
  \BibitemOpen
  \bibfield  {author} {\bibinfo {author} {\bibfnamefont {J.~T.}\ \bibnamefont
  {Barreiro}}, \bibinfo {author} {\bibfnamefont {M.}~\bibnamefont {Müller}},
  \bibinfo {author} {\bibfnamefont {P.}~\bibnamefont {Schindler}}, \bibinfo
  {author} {\bibfnamefont {D.}~\bibnamefont {Nigg}}, \bibinfo {author}
  {\bibfnamefont {T.}~\bibnamefont {Monz}}, \bibinfo {author} {\bibfnamefont
  {M.}~\bibnamefont {Chwalla}}, \bibinfo {author} {\bibfnamefont
  {M.}~\bibnamefont {Hennrich}}, \bibinfo {author} {\bibfnamefont {C.~F.}\
  \bibnamefont {Roos}}, \bibinfo {author} {\bibfnamefont {P.}~\bibnamefont
  {Zoller}}, \ and\ \bibinfo {author} {\bibfnamefont {R.}~\bibnamefont
  {Blatt}},\ }\href {\doibase 10.1038/nature09801} {\bibfield  {journal}
  {\bibinfo  {journal} {Nature}\ }\textbf {\bibinfo {volume} {470}},\ \bibinfo
  {pages} {486} (\bibinfo {year} {2011})}\BibitemShut {NoStop}%
\bibitem [{\citenamefont {Do}\ \emph {et~al.}(2019)\citenamefont {Do},
  \citenamefont {Lovecchio}, \citenamefont {Mastroserio}, \citenamefont
  {Fabbri}, \citenamefont {Cataliotti}, \citenamefont {Gherardini},
  \citenamefont {M{\"u}ller}, \citenamefont {Dalla~Pozza},\ and\ \citenamefont
  {Caruso}}]{Do19}%
  \BibitemOpen
  \bibfield  {author} {\bibinfo {author} {\bibfnamefont {H.-V.}\ \bibnamefont
  {Do}}, \bibinfo {author} {\bibfnamefont {C.}~\bibnamefont {Lovecchio}},
  \bibinfo {author} {\bibfnamefont {I.}~\bibnamefont {Mastroserio}}, \bibinfo
  {author} {\bibfnamefont {N.}~\bibnamefont {Fabbri}}, \bibinfo {author}
  {\bibfnamefont {F.~S.}\ \bibnamefont {Cataliotti}}, \bibinfo {author}
  {\bibfnamefont {S.}~\bibnamefont {Gherardini}}, \bibinfo {author}
  {\bibfnamefont {M.~M.}\ \bibnamefont {M{\"u}ller}}, \bibinfo {author}
  {\bibfnamefont {N.}~\bibnamefont {Dalla~Pozza}}, \ and\ \bibinfo {author}
  {\bibfnamefont {F.}~\bibnamefont {Caruso}},\ }\href {\doibase
  10.1088/1367-2630/ab5740} {\bibfield  {journal} {\bibinfo  {journal} {New
  Journal of Physics}\ }\textbf {\bibinfo {volume} {21}},\ \bibinfo {pages}
  {113056} (\bibinfo {year} {2019})}\BibitemShut {NoStop}%
\bibitem [{\citenamefont {Wolski}\ \emph {et~al.}(2020)\citenamefont {Wolski},
  \citenamefont {Lachance-Quirion}, \citenamefont {Tabuchi}, \citenamefont
  {Kono}, \citenamefont {Noguchi}, \citenamefont {Usami},\ and\ \citenamefont
  {Nakamura}}]{Wolski20}%
  \BibitemOpen
  \bibfield  {author} {\bibinfo {author} {\bibfnamefont {S.~P.}\ \bibnamefont
  {Wolski}}, \bibinfo {author} {\bibfnamefont {D.}~\bibnamefont
  {Lachance-Quirion}}, \bibinfo {author} {\bibfnamefont {Y.}~\bibnamefont
  {Tabuchi}}, \bibinfo {author} {\bibfnamefont {S.}~\bibnamefont {Kono}},
  \bibinfo {author} {\bibfnamefont {A.}~\bibnamefont {Noguchi}}, \bibinfo
  {author} {\bibfnamefont {K.}~\bibnamefont {Usami}}, \ and\ \bibinfo {author}
  {\bibfnamefont {Y.}~\bibnamefont {Nakamura}},\ }\href {\doibase
  10.1103/PhysRevLett.125.117701} {\bibfield  {journal} {\bibinfo  {journal}
  {Phys. Rev. Lett.}\ }\textbf {\bibinfo {volume} {125}},\ \bibinfo {pages}
  {117701} (\bibinfo {year} {2020})}\BibitemShut {NoStop}%
\bibitem [{\citenamefont {Xie}\ \emph {et~al.}(2020)\citenamefont {Xie},
  \citenamefont {Geng}, \citenamefont {Yu}, \citenamefont {Rong}, \citenamefont
  {Wang},\ and\ \citenamefont {Du}}]{Xie20}%
  \BibitemOpen
  \bibfield  {author} {\bibinfo {author} {\bibfnamefont {Y.}~\bibnamefont
  {Xie}}, \bibinfo {author} {\bibfnamefont {J.}~\bibnamefont {Geng}}, \bibinfo
  {author} {\bibfnamefont {H.}~\bibnamefont {Yu}}, \bibinfo {author}
  {\bibfnamefont {X.}~\bibnamefont {Rong}}, \bibinfo {author} {\bibfnamefont
  {Y.}~\bibnamefont {Wang}}, \ and\ \bibinfo {author} {\bibfnamefont
  {J.}~\bibnamefont {Du}},\ }\href {\doibase 10.1103/PhysRevApplied.14.014013}
  {\bibfield  {journal} {\bibinfo  {journal} {Phys. Rev. Applied}\ }\textbf
  {\bibinfo {volume} {14}},\ \bibinfo {pages} {014013} (\bibinfo {year}
  {2020})}\BibitemShut {NoStop}%
\bibitem [{\citenamefont {Lin}\ \emph {et~al.}(2013)\citenamefont {Lin},
  \citenamefont {Gaebler}, \citenamefont {Reiter}, \citenamefont {Tan},
  \citenamefont {Bowler}, \citenamefont {Sørensen}, \citenamefont
  {Leibfried},\ and\ \citenamefont {Wineland}}]{Lin13}%
  \BibitemOpen
  \bibfield  {author} {\bibinfo {author} {\bibfnamefont {Y.}~\bibnamefont
  {Lin}}, \bibinfo {author} {\bibfnamefont {J.~P.}\ \bibnamefont {Gaebler}},
  \bibinfo {author} {\bibfnamefont {F.}~\bibnamefont {Reiter}}, \bibinfo
  {author} {\bibfnamefont {T.~R.}\ \bibnamefont {Tan}}, \bibinfo {author}
  {\bibfnamefont {R.}~\bibnamefont {Bowler}}, \bibinfo {author} {\bibfnamefont
  {A.~S.}\ \bibnamefont {Sørensen}}, \bibinfo {author} {\bibfnamefont
  {D.}~\bibnamefont {Leibfried}}, \ and\ \bibinfo {author} {\bibfnamefont
  {D.~J.}\ \bibnamefont {Wineland}},\ }\href {\doibase 10.1038/nature12801}
  {\bibfield  {journal} {\bibinfo  {journal} {Nature}\ }\textbf {\bibinfo
  {volume} {504}},\ \bibinfo {pages} {415} (\bibinfo {year}
  {2013})}\BibitemShut {NoStop}%
\bibitem [{\citenamefont {Barontini}\ \emph {et~al.}(2015)\citenamefont
  {Barontini}, \citenamefont {Hohmann}, \citenamefont {Haas}, \citenamefont
  {Est{\`e}ve},\ and\ \citenamefont {Reichel}}]{Barontini15}%
  \BibitemOpen
  \bibfield  {author} {\bibinfo {author} {\bibfnamefont {G.}~\bibnamefont
  {Barontini}}, \bibinfo {author} {\bibfnamefont {L.}~\bibnamefont {Hohmann}},
  \bibinfo {author} {\bibfnamefont {F.}~\bibnamefont {Haas}}, \bibinfo {author}
  {\bibfnamefont {J.}~\bibnamefont {Est{\`e}ve}}, \ and\ \bibinfo {author}
  {\bibfnamefont {J.}~\bibnamefont {Reichel}},\ }\href {\doibase
  10.1126/science.aaa0754} {\bibfield  {journal} {\bibinfo  {journal}
  {Science}\ }\textbf {\bibinfo {volume} {349}},\ \bibinfo {pages} {1317}
  (\bibinfo {year} {2015})}\BibitemShut {NoStop}%
\bibitem [{\citenamefont {Biella}\ \emph {et~al.}(2017)\citenamefont {Biella},
  \citenamefont {Storme}, \citenamefont {Lebreuilly}, \citenamefont {Rossini},
  \citenamefont {Fazio}, \citenamefont {Carusotto},\ and\ \citenamefont
  {Ciuti}}]{Biella17}%
  \BibitemOpen
  \bibfield  {author} {\bibinfo {author} {\bibfnamefont {A.}~\bibnamefont
  {Biella}}, \bibinfo {author} {\bibfnamefont {F.}~\bibnamefont {Storme}},
  \bibinfo {author} {\bibfnamefont {J.}~\bibnamefont {Lebreuilly}}, \bibinfo
  {author} {\bibfnamefont {D.}~\bibnamefont {Rossini}}, \bibinfo {author}
  {\bibfnamefont {R.}~\bibnamefont {Fazio}}, \bibinfo {author} {\bibfnamefont
  {I.}~\bibnamefont {Carusotto}}, \ and\ \bibinfo {author} {\bibfnamefont
  {C.}~\bibnamefont {Ciuti}},\ }\href {\doibase 10.1103/PhysRevA.96.023839}
  {\bibfield  {journal} {\bibinfo  {journal} {Phys. Rev. A}\ }\textbf {\bibinfo
  {volume} {96}},\ \bibinfo {pages} {023839} (\bibinfo {year}
  {2017})}\BibitemShut {NoStop}%
\bibitem [{\citenamefont {Lu}\ \emph {et~al.}(2017)\citenamefont {Lu},
  \citenamefont {Chakram}, \citenamefont {Leung}, \citenamefont {Earnest},
  \citenamefont {Naik}, \citenamefont {Huang}, \citenamefont {Groszkowski},
  \citenamefont {Kapit}, \citenamefont {Koch},\ and\ \citenamefont
  {Schuster}}]{Lu17}%
  \BibitemOpen
  \bibfield  {author} {\bibinfo {author} {\bibfnamefont {Y.}~\bibnamefont
  {Lu}}, \bibinfo {author} {\bibfnamefont {S.}~\bibnamefont {Chakram}},
  \bibinfo {author} {\bibfnamefont {N.}~\bibnamefont {Leung}}, \bibinfo
  {author} {\bibfnamefont {N.}~\bibnamefont {Earnest}}, \bibinfo {author}
  {\bibfnamefont {R.~K.}\ \bibnamefont {Naik}}, \bibinfo {author}
  {\bibfnamefont {Z.}~\bibnamefont {Huang}}, \bibinfo {author} {\bibfnamefont
  {P.}~\bibnamefont {Groszkowski}}, \bibinfo {author} {\bibfnamefont
  {E.}~\bibnamefont {Kapit}}, \bibinfo {author} {\bibfnamefont
  {J.}~\bibnamefont {Koch}}, \ and\ \bibinfo {author} {\bibfnamefont {D.~I.}\
  \bibnamefont {Schuster}},\ }\href {\doibase 10.1103/PhysRevLett.119.150502}
  {\bibfield  {journal} {\bibinfo  {journal} {Phys. Rev. Lett.}\ }\textbf
  {\bibinfo {volume} {119}},\ \bibinfo {pages} {150502} (\bibinfo {year}
  {2017})}\BibitemShut {NoStop}%
\bibitem [{\citenamefont {Ma}\ \emph {et~al.}(2019)\citenamefont {Ma},
  \citenamefont {Saxberg}, \citenamefont {Owens}, \citenamefont {Leung},
  \citenamefont {Lu}, \citenamefont {Simon},\ and\ \citenamefont
  {Schuster}}]{Ma19}%
  \BibitemOpen
  \bibfield  {author} {\bibinfo {author} {\bibfnamefont {R.}~\bibnamefont
  {Ma}}, \bibinfo {author} {\bibfnamefont {B.}~\bibnamefont {Saxberg}},
  \bibinfo {author} {\bibfnamefont {C.}~\bibnamefont {Owens}}, \bibinfo
  {author} {\bibfnamefont {N.}~\bibnamefont {Leung}}, \bibinfo {author}
  {\bibfnamefont {Y.}~\bibnamefont {Lu}}, \bibinfo {author} {\bibfnamefont
  {J.}~\bibnamefont {Simon}}, \ and\ \bibinfo {author} {\bibfnamefont {D.~I.}\
  \bibnamefont {Schuster}},\ }\href {\doibase 10.1038/s41586-019-0897-9}
  {\bibfield  {journal} {\bibinfo  {journal} {Nature}\ }\textbf {\bibinfo
  {volume} {566}},\ \bibinfo {pages} {51} (\bibinfo {year} {2019})}\BibitemShut
  {NoStop}%
\bibitem [{\citenamefont {Benatti}\ and\ \citenamefont
  {Floreanini}(2003)}]{BenattiBook2003}%
  \BibitemOpen
  \bibfield  {author} {\bibinfo {author} {\bibfnamefont {F.}~\bibnamefont
  {Benatti}}\ and\ \bibinfo {author} {\bibfnamefont {R.}~\bibnamefont
  {Floreanini}},\ }\href@noop {} {\emph {\bibinfo {title} {Irreversible quantum
  dynamics}}},\ Vol.\ \bibinfo {volume} {622}\ (\bibinfo  {publisher} {Springer
  Science \& Business Media},\ \bibinfo {year} {2003})\BibitemShut {NoStop}%
\bibitem [{\citenamefont {Dutta}\ and\ \citenamefont {Cooper}(2021)}]{Dutta21}%
  \BibitemOpen
  \bibfield  {author} {\bibinfo {author} {\bibfnamefont {S.}~\bibnamefont
  {Dutta}}\ and\ \bibinfo {author} {\bibfnamefont {N.~R.}\ \bibnamefont
  {Cooper}},\ }\href {\doibase 10.1103/PhysRevResearch.3.L012016} {\bibfield
  {journal} {\bibinfo  {journal} {Phys. Rev. Research}\ }\textbf {\bibinfo
  {volume} {3}},\ \bibinfo {pages} {L012016} (\bibinfo {year}
  {2021})}\BibitemShut {NoStop}%
\bibitem [{\citenamefont {Gisin}(1984)}]{Gisin84}%
  \BibitemOpen
  \bibfield  {author} {\bibinfo {author} {\bibfnamefont {N.}~\bibnamefont
  {Gisin}},\ }\href {\doibase 10.1103/PhysRevLett.52.1657} {\bibfield
  {journal} {\bibinfo  {journal} {Phys. Rev. Lett.}\ }\textbf {\bibinfo
  {volume} {52}},\ \bibinfo {pages} {1657} (\bibinfo {year}
  {1984})}\BibitemShut {NoStop}%
\bibitem [{\citenamefont {Jacobs}(2014)}]{JacobsBook2014}%
  \BibitemOpen
  \bibfield  {author} {\bibinfo {author} {\bibfnamefont {K.}~\bibnamefont
  {Jacobs}},\ }\href@noop {} {\emph {\bibinfo {title} {Quantum measurement
  theory and its applications}}}\ (\bibinfo  {publisher} {Cambridge University
  Press},\ \bibinfo {year} {2014})\BibitemShut {NoStop}%
\bibitem [{\citenamefont {Esposito}\ \emph {et~al.}(2009)\citenamefont
  {Esposito}, \citenamefont {Harbola},\ and\ \citenamefont
  {Mukamel}}]{Esposito09}%
  \BibitemOpen
  \bibfield  {author} {\bibinfo {author} {\bibfnamefont {M.}~\bibnamefont
  {Esposito}}, \bibinfo {author} {\bibfnamefont {U.}~\bibnamefont {Harbola}}, \
  and\ \bibinfo {author} {\bibfnamefont {S.}~\bibnamefont {Mukamel}},\ }\href
  {\doibase 10.1103/RevModPhys.81.1665} {\bibfield  {journal} {\bibinfo
  {journal} {Rev. Mod. Phys.}\ }\textbf {\bibinfo {volume} {81}},\ \bibinfo
  {pages} {1665} (\bibinfo {year} {2009})}\BibitemShut {NoStop}%
\bibitem [{\citenamefont {Campisi}\ \emph {et~al.}(2011)\citenamefont
  {Campisi}, \citenamefont {H\"anggi},\ and\ \citenamefont
  {Talkner}}]{Campisi11}%
  \BibitemOpen
  \bibfield  {author} {\bibinfo {author} {\bibfnamefont {M.}~\bibnamefont
  {Campisi}}, \bibinfo {author} {\bibfnamefont {P.}~\bibnamefont {H\"anggi}}, \
  and\ \bibinfo {author} {\bibfnamefont {P.}~\bibnamefont {Talkner}},\ }\href
  {\doibase 10.1103/RevModPhys.83.771} {\bibfield  {journal} {\bibinfo
  {journal} {Rev. Mod. Phys.}\ }\textbf {\bibinfo {volume} {83}},\ \bibinfo
  {pages} {771} (\bibinfo {year} {2011})}\BibitemShut {NoStop}%
\bibitem [{\citenamefont {An}\ \emph {et~al.}(2015)\citenamefont {An},
  \citenamefont {Zhang}, \citenamefont {Um}, \citenamefont {Lv}, \citenamefont
  {Lu}, \citenamefont {Zhang}, \citenamefont {Yin}, \citenamefont {Quan},\ and\
  \citenamefont {Kim}}]{An15}%
  \BibitemOpen
  \bibfield  {author} {\bibinfo {author} {\bibfnamefont {S.}~\bibnamefont
  {An}}, \bibinfo {author} {\bibfnamefont {J.-N.}\ \bibnamefont {Zhang}},
  \bibinfo {author} {\bibfnamefont {M.}~\bibnamefont {Um}}, \bibinfo {author}
  {\bibfnamefont {D.}~\bibnamefont {Lv}}, \bibinfo {author} {\bibfnamefont
  {Y.}~\bibnamefont {Lu}}, \bibinfo {author} {\bibfnamefont {J.}~\bibnamefont
  {Zhang}}, \bibinfo {author} {\bibfnamefont {Z.-Q.}\ \bibnamefont {Yin}},
  \bibinfo {author} {\bibfnamefont {H.~T.}\ \bibnamefont {Quan}}, \ and\
  \bibinfo {author} {\bibfnamefont {K.}~\bibnamefont {Kim}},\ }\href {\doibase
  10.1038/nphys3197} {\bibfield  {journal} {\bibinfo  {journal} {Nat. Phys.}\
  }\textbf {\bibinfo {volume} {11}},\ \bibinfo {pages} {193} (\bibinfo {year}
  {2015})}\BibitemShut {NoStop}%
\bibitem [{\citenamefont {Smith}\ \emph {et~al.}(2018)\citenamefont {Smith},
  \citenamefont {Lu}, \citenamefont {An}, \citenamefont {Zhang}, \citenamefont
  {Zhang}, \citenamefont {Gong}, \citenamefont {Quan}, \citenamefont
  {Jarzynski},\ and\ \citenamefont {Kim}}]{Smith18}%
  \BibitemOpen
  \bibfield  {author} {\bibinfo {author} {\bibfnamefont {A.}~\bibnamefont
  {Smith}}, \bibinfo {author} {\bibfnamefont {Y.}~\bibnamefont {Lu}}, \bibinfo
  {author} {\bibfnamefont {S.}~\bibnamefont {An}}, \bibinfo {author}
  {\bibfnamefont {X.}~\bibnamefont {Zhang}}, \bibinfo {author} {\bibfnamefont
  {J.-N.}\ \bibnamefont {Zhang}}, \bibinfo {author} {\bibfnamefont
  {Z.}~\bibnamefont {Gong}}, \bibinfo {author} {\bibfnamefont {H.~T.}\
  \bibnamefont {Quan}}, \bibinfo {author} {\bibfnamefont {C.}~\bibnamefont
  {Jarzynski}}, \ and\ \bibinfo {author} {\bibfnamefont {K.}~\bibnamefont
  {Kim}},\ }\href {\doibase 10.1088/1367-2630/aa9cd6} {\bibfield  {journal}
  {\bibinfo  {journal} {New J. Phys.}\ }\textbf {\bibinfo {volume} {20}},\
  \bibinfo {pages} {013008} (\bibinfo {year} {2018})}\BibitemShut {NoStop}%
\bibitem [{\citenamefont {Batalh{\~a}o}\ \emph {et~al.}(2014)\citenamefont
  {Batalh{\~a}o}, \citenamefont {Souza}, \citenamefont {Mazzola}, \citenamefont
  {Auccaise}, \citenamefont {Sarthour}, \citenamefont {Oliveira}, \citenamefont
  {Goold}, \citenamefont {De~Chiara}, \citenamefont {Paternostro},\ and\
  \citenamefont {Serra}}]{Batalhao14}%
  \BibitemOpen
  \bibfield  {author} {\bibinfo {author} {\bibfnamefont {T.~B.}\ \bibnamefont
  {Batalh{\~a}o}}, \bibinfo {author} {\bibfnamefont {A.~M.}\ \bibnamefont
  {Souza}}, \bibinfo {author} {\bibfnamefont {L.}~\bibnamefont {Mazzola}},
  \bibinfo {author} {\bibfnamefont {R.}~\bibnamefont {Auccaise}}, \bibinfo
  {author} {\bibfnamefont {R.~S.}\ \bibnamefont {Sarthour}}, \bibinfo {author}
  {\bibfnamefont {I.~S.}\ \bibnamefont {Oliveira}}, \bibinfo {author}
  {\bibfnamefont {J.}~\bibnamefont {Goold}}, \bibinfo {author} {\bibfnamefont
  {G.}~\bibnamefont {De~Chiara}}, \bibinfo {author} {\bibfnamefont
  {M.}~\bibnamefont {Paternostro}}, \ and\ \bibinfo {author} {\bibfnamefont
  {R.~M.}\ \bibnamefont {Serra}},\ }\href
  {https://link.aps.org/doi/10.1103/PhysRevLett.113.140601} {\bibfield
  {journal} {\bibinfo  {journal} {Phys. Rev. Lett.}\ }\textbf {\bibinfo
  {volume} {113}},\ \bibinfo {pages} {140601} (\bibinfo {year}
  {2014})}\BibitemShut {NoStop}%
\bibitem [{\citenamefont {Pal}\ \emph {et~al.}(2019)\citenamefont {Pal},
  \citenamefont {Mahesh},\ and\ \citenamefont {Agarwalla}}]{Pal19}%
  \BibitemOpen
  \bibfield  {author} {\bibinfo {author} {\bibfnamefont {S.}~\bibnamefont
  {Pal}}, \bibinfo {author} {\bibfnamefont {T.~S.}\ \bibnamefont {Mahesh}}, \
  and\ \bibinfo {author} {\bibfnamefont {B.~K.}\ \bibnamefont {Agarwalla}},\
  }\href {\doibase 10.1103/PhysRevA.100.042119} {\bibfield  {journal} {\bibinfo
   {journal} {Phys. Rev. A}\ }\textbf {\bibinfo {volume} {100}},\ \bibinfo
  {pages} {042119} (\bibinfo {year} {2019})}\BibitemShut {NoStop}%
\bibitem [{\citenamefont {Cerisola}\ \emph {et~al.}(2017)\citenamefont
  {Cerisola}, \citenamefont {Margalit}, \citenamefont {Machluf}, \citenamefont
  {Roncaglia}, \citenamefont {Paz},\ and\ \citenamefont {Folman}}]{Cerisola17}%
  \BibitemOpen
  \bibfield  {author} {\bibinfo {author} {\bibfnamefont {F.}~\bibnamefont
  {Cerisola}}, \bibinfo {author} {\bibfnamefont {Y.}~\bibnamefont {Margalit}},
  \bibinfo {author} {\bibfnamefont {S.}~\bibnamefont {Machluf}}, \bibinfo
  {author} {\bibfnamefont {A.}~\bibnamefont {Roncaglia}}, \bibinfo {author}
  {\bibfnamefont {J.}~\bibnamefont {Paz}}, \ and\ \bibinfo {author}
  {\bibfnamefont {R.}~\bibnamefont {Folman}},\ }\href {\doibase
  10.1038/s41467-017-01308-7} {\bibfield  {journal} {\bibinfo  {journal} {Nat.
  Comm.}\ }\textbf {\bibinfo {volume} {8}},\ \bibinfo {pages} {1241} (\bibinfo
  {year} {2017})}\BibitemShut {NoStop}%
\bibitem [{\citenamefont {Zhang}\ \emph {et~al.}(2018)\citenamefont {Zhang},
  \citenamefont {Wang}, \citenamefont {Xiang}, \citenamefont {Jia},
  \citenamefont {Duan}, \citenamefont {Cai}, \citenamefont {Zhan},
  \citenamefont {Zong}, \citenamefont {Wu}, \citenamefont {Sun}, \citenamefont
  {Yin},\ and\ \citenamefont {Guo}}]{Zhang18}%
  \BibitemOpen
  \bibfield  {author} {\bibinfo {author} {\bibfnamefont {Z.}~\bibnamefont
  {Zhang}}, \bibinfo {author} {\bibfnamefont {T.}~\bibnamefont {Wang}},
  \bibinfo {author} {\bibfnamefont {L.}~\bibnamefont {Xiang}}, \bibinfo
  {author} {\bibfnamefont {Z.}~\bibnamefont {Jia}}, \bibinfo {author}
  {\bibfnamefont {P.}~\bibnamefont {Duan}}, \bibinfo {author} {\bibfnamefont
  {W.}~\bibnamefont {Cai}}, \bibinfo {author} {\bibfnamefont {Z.}~\bibnamefont
  {Zhan}}, \bibinfo {author} {\bibfnamefont {Z.}~\bibnamefont {Zong}}, \bibinfo
  {author} {\bibfnamefont {J.}~\bibnamefont {Wu}}, \bibinfo {author}
  {\bibfnamefont {L.}~\bibnamefont {Sun}}, \bibinfo {author} {\bibfnamefont
  {Y.}~\bibnamefont {Yin}}, \ and\ \bibinfo {author} {\bibfnamefont
  {G.}~\bibnamefont {Guo}},\ }\href {\doibase 10.1088/1367-2630/aad4e7}
  {\bibfield  {journal} {\bibinfo  {journal} {New J. Phys.}\ }\textbf {\bibinfo
  {volume} {20}},\ \bibinfo {pages} {085001} (\bibinfo {year}
  {2018})}\BibitemShut {NoStop}%
\bibitem [{\citenamefont {Hern\'andez-G\'omez}\ \emph
  {et~al.}(2020)\citenamefont {Hern\'andez-G\'omez}, \citenamefont
  {Gherardini}, \citenamefont {Poggiali}, \citenamefont {Cataliotti},
  \citenamefont {Trombettoni}, \citenamefont {Cappellaro},\ and\ \citenamefont
  {Fabbri}}]{HernandezGomez20}%
  \BibitemOpen
  \bibfield  {author} {\bibinfo {author} {\bibfnamefont {S.}~\bibnamefont
  {Hern\'andez-G\'omez}}, \bibinfo {author} {\bibfnamefont {S.}~\bibnamefont
  {Gherardini}}, \bibinfo {author} {\bibfnamefont {F.}~\bibnamefont
  {Poggiali}}, \bibinfo {author} {\bibfnamefont {F.~S.}\ \bibnamefont
  {Cataliotti}}, \bibinfo {author} {\bibfnamefont {A.}~\bibnamefont
  {Trombettoni}}, \bibinfo {author} {\bibfnamefont {P.}~\bibnamefont
  {Cappellaro}}, \ and\ \bibinfo {author} {\bibfnamefont {N.}~\bibnamefont
  {Fabbri}},\ }\href {\doibase 10.1103/PhysRevResearch.2.023327} {\bibfield
  {journal} {\bibinfo  {journal} {Phys. Rev. Research}\ }\textbf {\bibinfo
  {volume} {2}},\ \bibinfo {pages} {023327} (\bibinfo {year}
  {2020})}\BibitemShut {NoStop}%
\bibitem [{\citenamefont {Ribeiro}\ \emph {et~al.}(2020)\citenamefont
  {Ribeiro}, \citenamefont {H\"affner}, \citenamefont {Zanin}, \citenamefont
  {da~Silva}, \citenamefont {de~Ara\'ujo}, \citenamefont {Soares},
  \citenamefont {de~Assis}, \citenamefont {C\'eleri},\ and\ \citenamefont
  {Forbes}}]{RibeiroPRA2020}%
  \BibitemOpen
  \bibfield  {author} {\bibinfo {author} {\bibfnamefont {P.~H.~S.}\
  \bibnamefont {Ribeiro}}, \bibinfo {author} {\bibfnamefont {T.}~\bibnamefont
  {H\"affner}}, \bibinfo {author} {\bibfnamefont {G.~L.}\ \bibnamefont
  {Zanin}}, \bibinfo {author} {\bibfnamefont {N.~R.}\ \bibnamefont {da~Silva}},
  \bibinfo {author} {\bibfnamefont {R.~M.}\ \bibnamefont {de~Ara\'ujo}},
  \bibinfo {author} {\bibfnamefont {W.~C.}\ \bibnamefont {Soares}}, \bibinfo
  {author} {\bibfnamefont {R.~J.}\ \bibnamefont {de~Assis}}, \bibinfo {author}
  {\bibfnamefont {L.~C.}\ \bibnamefont {C\'eleri}}, \ and\ \bibinfo {author}
  {\bibfnamefont {A.}~\bibnamefont {Forbes}},\ }\href {\doibase
  10.1103/PhysRevA.101.052113} {\bibfield  {journal} {\bibinfo  {journal}
  {Phys. Rev. A}\ }\textbf {\bibinfo {volume} {101}},\ \bibinfo {pages}
  {052113} (\bibinfo {year} {2020})}\BibitemShut {NoStop}%
\bibitem [{\citenamefont {Sagawa}\ and\ \citenamefont {Ueda}(2008)}]{Sagawa08}%
  \BibitemOpen
  \bibfield  {author} {\bibinfo {author} {\bibfnamefont {T.}~\bibnamefont
  {Sagawa}}\ and\ \bibinfo {author} {\bibfnamefont {M.}~\bibnamefont {Ueda}},\
  }\href {\doibase 10.1103/PhysRevLett.100.080403} {\bibfield  {journal}
  {\bibinfo  {journal} {Phys. Rev. Lett.}\ }\textbf {\bibinfo {volume} {100}},\
  \bibinfo {pages} {080403} (\bibinfo {year} {2008})}\BibitemShut {NoStop}%
\bibitem [{\citenamefont {Morikuni}\ and\ \citenamefont
  {Tasaki}(2011)}]{Morikuni11}%
  \BibitemOpen
  \bibfield  {author} {\bibinfo {author} {\bibfnamefont {Y.}~\bibnamefont
  {Morikuni}}\ and\ \bibinfo {author} {\bibfnamefont {H.}~\bibnamefont
  {Tasaki}},\ }\href {\doibase 10.1007/s10955-011-0153-7} {\bibfield  {journal}
  {\bibinfo  {journal} {Journal of Statistical Physics}\ }\textbf {\bibinfo
  {volume} {143}},\ \bibinfo {pages} {1} (\bibinfo {year} {2011})}\BibitemShut
  {NoStop}%
\bibitem [{\citenamefont {Funo}\ \emph {et~al.}(2013)\citenamefont {Funo},
  \citenamefont {Watanabe},\ and\ \citenamefont {Ueda}}]{Funo13}%
  \BibitemOpen
  \bibfield  {author} {\bibinfo {author} {\bibfnamefont {K.}~\bibnamefont
  {Funo}}, \bibinfo {author} {\bibfnamefont {Y.}~\bibnamefont {Watanabe}}, \
  and\ \bibinfo {author} {\bibfnamefont {M.}~\bibnamefont {Ueda}},\ }\href
  {\doibase 10.1103/PhysRevE.88.052121} {\bibfield  {journal} {\bibinfo
  {journal} {Phys. Rev. E}\ }\textbf {\bibinfo {volume} {88}},\ \bibinfo
  {pages} {052121} (\bibinfo {year} {2013})}\BibitemShut {NoStop}%
\bibitem [{\citenamefont {Funo}\ \emph {et~al.}(2015)\citenamefont {Funo},
  \citenamefont {Murashita},\ and\ \citenamefont {Ueda}}]{FunoNJP15}%
  \BibitemOpen
  \bibfield  {author} {\bibinfo {author} {\bibfnamefont {K.}~\bibnamefont
  {Funo}}, \bibinfo {author} {\bibfnamefont {Y.}~\bibnamefont {Murashita}}, \
  and\ \bibinfo {author} {\bibfnamefont {M.}~\bibnamefont {Ueda}},\ }\href
  {\doibase 10.1088/1367-2630/17/7/075005} {\bibfield  {journal} {\bibinfo
  {journal} {New J. Phys.}\ }\textbf {\bibinfo {volume} {17}},\ \bibinfo
  {pages} {075005} (\bibinfo {year} {2015})}\BibitemShut {NoStop}%
\bibitem [{\citenamefont {Kafri}\ and\ \citenamefont
  {Deffner}(2012)}]{Kafri12}%
  \BibitemOpen
  \bibfield  {author} {\bibinfo {author} {\bibfnamefont {D.}~\bibnamefont
  {Kafri}}\ and\ \bibinfo {author} {\bibfnamefont {S.}~\bibnamefont
  {Deffner}},\ }\href {\doibase 10.1103/PhysRevA.86.044302} {\bibfield
  {journal} {\bibinfo  {journal} {Phys. Rev. A}\ }\textbf {\bibinfo {volume}
  {86}},\ \bibinfo {pages} {044302} (\bibinfo {year} {2012})}\BibitemShut
  {NoStop}%
\bibitem [{\citenamefont {Rastegin}(2013)}]{Rastegin13}%
  \BibitemOpen
  \bibfield  {author} {\bibinfo {author} {\bibfnamefont {A.~E.}\ \bibnamefont
  {Rastegin}},\ }\href {\doibase 10.1088/1742-5468/2013/06/p06016} {\bibfield
  {journal} {\bibinfo  {journal} {J. Stat. Mech.}\ }\textbf {\bibinfo {volume}
  {2013}},\ \bibinfo {pages} {P06016} (\bibinfo {year} {2013})}\BibitemShut
  {NoStop}%
\bibitem [{\citenamefont {Albash}\ \emph {et~al.}(2013)\citenamefont {Albash},
  \citenamefont {Lidar}, \citenamefont {Marvian},\ and\ \citenamefont
  {Zanardi}}]{Albash13}%
  \BibitemOpen
  \bibfield  {author} {\bibinfo {author} {\bibfnamefont {T.}~\bibnamefont
  {Albash}}, \bibinfo {author} {\bibfnamefont {D.~A.}\ \bibnamefont {Lidar}},
  \bibinfo {author} {\bibfnamefont {M.}~\bibnamefont {Marvian}}, \ and\
  \bibinfo {author} {\bibfnamefont {P.}~\bibnamefont {Zanardi}},\ }\href
  {\doibase 10.1103/PhysRevE.88.032146} {\bibfield  {journal} {\bibinfo
  {journal} {Phys. Rev. E}\ }\textbf {\bibinfo {volume} {88}},\ \bibinfo
  {pages} {032146} (\bibinfo {year} {2013})}\BibitemShut {NoStop}%
\bibitem [{\citenamefont {Goold}\ \emph {et~al.}(2015)\citenamefont {Goold},
  \citenamefont {Paternostro},\ and\ \citenamefont {Modi}}]{Goold15}%
  \BibitemOpen
  \bibfield  {author} {\bibinfo {author} {\bibfnamefont {J.}~\bibnamefont
  {Goold}}, \bibinfo {author} {\bibfnamefont {M.}~\bibnamefont {Paternostro}},
  \ and\ \bibinfo {author} {\bibfnamefont {K.}~\bibnamefont {Modi}},\ }\href
  {\doibase 10.1103/PhysRevLett.114.060602} {\bibfield  {journal} {\bibinfo
  {journal} {Phys. Rev. Lett.}\ }\textbf {\bibinfo {volume} {114}},\ \bibinfo
  {pages} {060602} (\bibinfo {year} {2015})}\BibitemShut {NoStop}%
\bibitem [{\citenamefont {Havel}(2003)}]{Havel03}%
  \BibitemOpen
  \bibfield  {author} {\bibinfo {author} {\bibfnamefont {T.~F.}\ \bibnamefont
  {Havel}},\ }\href {\doibase 10.1063/1.1518555} {\bibfield  {journal}
  {\bibinfo  {journal} {Journal of Mathematical Physics}\ }\textbf {\bibinfo
  {volume} {44}},\ \bibinfo {pages} {534} (\bibinfo {year} {2003})}\BibitemShut
  {NoStop}%
\bibitem [{\citenamefont {Steiner}\ \emph {et~al.}(2010)\citenamefont
  {Steiner}, \citenamefont {Neumann}, \citenamefont {Beck}, \citenamefont
  {Jelezko},\ and\ \citenamefont {Wrachtrup}}]{Steiner10}%
  \BibitemOpen
  \bibfield  {author} {\bibinfo {author} {\bibfnamefont {M.}~\bibnamefont
  {Steiner}}, \bibinfo {author} {\bibfnamefont {P.}~\bibnamefont {Neumann}},
  \bibinfo {author} {\bibfnamefont {J.}~\bibnamefont {Beck}}, \bibinfo {author}
  {\bibfnamefont {F.}~\bibnamefont {Jelezko}}, \ and\ \bibinfo {author}
  {\bibfnamefont {J.}~\bibnamefont {Wrachtrup}},\ }\href {\doibase
  10.1103/PhysRevB.81.035205} {\bibfield  {journal} {\bibinfo  {journal} {Phys.
  Rev. B}\ }\textbf {\bibinfo {volume} {81}},\ \bibinfo {pages} {035205}
  (\bibinfo {year} {2010})}\BibitemShut {NoStop}%
\bibitem [{\citenamefont {Doherty}\ \emph {et~al.}(2013)\citenamefont
  {Doherty}, \citenamefont {Manson}, \citenamefont {Delaney}, \citenamefont
  {Jelezko}, \citenamefont {Wrachtrup},\ and\ \citenamefont
  {Hollenberg}}]{Doherty13}%
  \BibitemOpen
  \bibfield  {author} {\bibinfo {author} {\bibfnamefont {M.~W.}\ \bibnamefont
  {Doherty}}, \bibinfo {author} {\bibfnamefont {N.~B.}\ \bibnamefont {Manson}},
  \bibinfo {author} {\bibfnamefont {P.}~\bibnamefont {Delaney}}, \bibinfo
  {author} {\bibfnamefont {F.}~\bibnamefont {Jelezko}}, \bibinfo {author}
  {\bibfnamefont {J.}~\bibnamefont {Wrachtrup}}, \ and\ \bibinfo {author}
  {\bibfnamefont {L.~C.~L.}\ \bibnamefont {Hollenberg}},\ }\href {\doibase
  10.1016/j.physrep.2013.02.001} {\bibfield  {journal} {\bibinfo  {journal}
  {Phys. Rep.}\ }\textbf {\bibinfo {volume} {528}},\ \bibinfo {pages} {1}
  (\bibinfo {year} {2013})}\BibitemShut {NoStop}%
\bibitem [{\citenamefont {Wolters}\ \emph {et~al.}(2013)\citenamefont
  {Wolters}, \citenamefont {Strau\ss{}}, \citenamefont {Schoenfeld},\ and\
  \citenamefont {Benson}}]{Wolters13}%
  \BibitemOpen
  \bibfield  {author} {\bibinfo {author} {\bibfnamefont {J.}~\bibnamefont
  {Wolters}}, \bibinfo {author} {\bibfnamefont {M.}~\bibnamefont {Strau\ss{}}},
  \bibinfo {author} {\bibfnamefont {R.~S.}\ \bibnamefont {Schoenfeld}}, \ and\
  \bibinfo {author} {\bibfnamefont {O.}~\bibnamefont {Benson}},\ }\href
  {\doibase 10.1103/PhysRevA.88.020101} {\bibfield  {journal} {\bibinfo
  {journal} {Phys. Rev. A}\ }\textbf {\bibinfo {volume} {88}},\ \bibinfo
  {pages} {020101} (\bibinfo {year} {2013})}\BibitemShut {NoStop}%
\bibitem [{\citenamefont {Talkner}\ \emph {et~al.}(2007)\citenamefont
  {Talkner}, \citenamefont {Lutz},\ and\ \citenamefont {H\"anggi}}]{Talkner07}%
  \BibitemOpen
  \bibfield  {author} {\bibinfo {author} {\bibfnamefont {P.}~\bibnamefont
  {Talkner}}, \bibinfo {author} {\bibfnamefont {E.}~\bibnamefont {Lutz}}, \
  and\ \bibinfo {author} {\bibfnamefont {P.}~\bibnamefont {H\"anggi}},\ }\href
  {\doibase 10.1103/PhysRevE.75.050102} {\bibfield  {journal} {\bibinfo
  {journal} {Phys. Rev. E}\ }\textbf {\bibinfo {volume} {75}},\ \bibinfo
  {pages} {050102} (\bibinfo {year} {2007})}\BibitemShut {NoStop}%
\bibitem [{\citenamefont {Rastegin}\ and\ \citenamefont {\ifmmode~\dot{Z}\else
  \.{Z}\fi{}yczkowski}(2014)}]{Rastegin14}%
  \BibitemOpen
  \bibfield  {author} {\bibinfo {author} {\bibfnamefont {A.~E.}\ \bibnamefont
  {Rastegin}}\ and\ \bibinfo {author} {\bibfnamefont {K.}~\bibnamefont
  {\ifmmode~\dot{Z}\else \.{Z}\fi{}yczkowski}},\ }\href {\doibase
  10.1103/PhysRevE.89.012127} {\bibfield  {journal} {\bibinfo  {journal} {Phys.
  Rev. E}\ }\textbf {\bibinfo {volume} {89}},\ \bibinfo {pages} {012127}
  (\bibinfo {year} {2014})}\BibitemShut {NoStop}%
\bibitem [{\citenamefont {Cimini}\ \emph {et~al.}(2020)\citenamefont {Cimini},
  \citenamefont {Gherardini}, \citenamefont {Barbieri}, \citenamefont
  {Gianani}, \citenamefont {Sbroscia}, \citenamefont {Buffoni}, \citenamefont
  {Paternostro},\ and\ \citenamefont {Caruso}}]{Cimini20}%
  \BibitemOpen
  \bibfield  {author} {\bibinfo {author} {\bibfnamefont {V.}~\bibnamefont
  {Cimini}}, \bibinfo {author} {\bibfnamefont {S.}~\bibnamefont {Gherardini}},
  \bibinfo {author} {\bibfnamefont {M.}~\bibnamefont {Barbieri}}, \bibinfo
  {author} {\bibfnamefont {I.}~\bibnamefont {Gianani}}, \bibinfo {author}
  {\bibfnamefont {M.}~\bibnamefont {Sbroscia}}, \bibinfo {author}
  {\bibfnamefont {L.}~\bibnamefont {Buffoni}}, \bibinfo {author} {\bibfnamefont
  {M.}~\bibnamefont {Paternostro}}, \ and\ \bibinfo {author} {\bibfnamefont
  {F.}~\bibnamefont {Caruso}},\ }\href {\doibase 10.1038/s41534-020-00325-7}
  {\bibfield  {journal} {\bibinfo  {journal} {npj Quantum Information}\
  }\textbf {\bibinfo {volume} {6}},\ \bibinfo {pages} {96} (\bibinfo {year}
  {2020})}\BibitemShut {NoStop}%
\bibitem [{\citenamefont {Guarnieri}\ \emph {et~al.}(2017)\citenamefont
  {Guarnieri}, \citenamefont {Campbell}, \citenamefont {Goold}, \citenamefont
  {Pigeon}, \citenamefont {Vacchini},\ and\ \citenamefont
  {Paternostro}}]{Guarnieri17}%
  \BibitemOpen
  \bibfield  {author} {\bibinfo {author} {\bibfnamefont {G.}~\bibnamefont
  {Guarnieri}}, \bibinfo {author} {\bibfnamefont {S.}~\bibnamefont {Campbell}},
  \bibinfo {author} {\bibfnamefont {J.}~\bibnamefont {Goold}}, \bibinfo
  {author} {\bibfnamefont {S.}~\bibnamefont {Pigeon}}, \bibinfo {author}
  {\bibfnamefont {B.}~\bibnamefont {Vacchini}}, \ and\ \bibinfo {author}
  {\bibfnamefont {M.}~\bibnamefont {Paternostro}},\ }\href {\doibase
  10.1088/1367-2630/aa8cf1} {\bibfield  {journal} {\bibinfo  {journal} {New
  Journal of Physics}\ }\textbf {\bibinfo {volume} {19}},\ \bibinfo {pages}
  {103038} (\bibinfo {year} {2017})}\BibitemShut {NoStop}%
\bibitem [{\citenamefont {Gardas}\ and\ \citenamefont
  {Deffner}(2015)}]{Gardas15}%
  \BibitemOpen
  \bibfield  {author} {\bibinfo {author} {\bibfnamefont {B.}~\bibnamefont
  {Gardas}}\ and\ \bibinfo {author} {\bibfnamefont {S.}~\bibnamefont
  {Deffner}},\ }\href {\doibase 10.1103/PhysRevE.92.042126} {\bibfield
  {journal} {\bibinfo  {journal} {Phys. Rev. E}\ }\textbf {\bibinfo {volume}
  {92}},\ \bibinfo {pages} {042126} (\bibinfo {year} {2015})}\BibitemShut
  {NoStop}%
\bibitem [{\citenamefont {Gherardini}\ \emph {et~al.}(2021)\citenamefont
  {Gherardini}, \citenamefont {Belenchia}, \citenamefont {Paternostro},\ and\
  \citenamefont {Trombettoni}}]{Gherardini20x}%
  \BibitemOpen
  \bibfield  {author} {\bibinfo {author} {\bibfnamefont {S.}~\bibnamefont
  {Gherardini}}, \bibinfo {author} {\bibfnamefont {A.}~\bibnamefont
  {Belenchia}}, \bibinfo {author} {\bibfnamefont {M.}~\bibnamefont
  {Paternostro}}, \ and\ \bibinfo {author} {\bibfnamefont {A.}~\bibnamefont
  {Trombettoni}},\ }\href {\doibase 10.1103/PhysRevA.104.L050203} {\bibfield
  {journal} {\bibinfo  {journal} {Phys. Rev. A}\ }\textbf {\bibinfo {volume}
  {104}},\ \bibinfo {pages} {L050203} (\bibinfo {year} {2021})}\BibitemShut
  {NoStop}%
\bibitem [{\citenamefont {Micadei}\ \emph {et~al.}(2020)\citenamefont
  {Micadei}, \citenamefont {Landi},\ and\ \citenamefont {Lutz}}]{Micadei20}%
  \BibitemOpen
  \bibfield  {author} {\bibinfo {author} {\bibfnamefont {K.}~\bibnamefont
  {Micadei}}, \bibinfo {author} {\bibfnamefont {G.~T.}\ \bibnamefont {Landi}},
  \ and\ \bibinfo {author} {\bibfnamefont {E.}~\bibnamefont {Lutz}},\ }\href
  {\doibase 10.1103/PhysRevLett.124.090602} {\bibfield  {journal} {\bibinfo
  {journal} {Phys. Rev. Lett.}\ }\textbf {\bibinfo {volume} {124}},\ \bibinfo
  {pages} {090602} (\bibinfo {year} {2020})}\BibitemShut {NoStop}%
\bibitem [{\citenamefont {Sone}\ \emph {et~al.}(2020)\citenamefont {Sone},
  \citenamefont {Liu},\ and\ \citenamefont {Cappellaro}}]{Sone20}%
  \BibitemOpen
  \bibfield  {author} {\bibinfo {author} {\bibfnamefont {A.}~\bibnamefont
  {Sone}}, \bibinfo {author} {\bibfnamefont {Y.-X.}\ \bibnamefont {Liu}}, \
  and\ \bibinfo {author} {\bibfnamefont {P.}~\bibnamefont {Cappellaro}},\
  }\href {\doibase 10.1103/PhysRevLett.125.060602} {\bibfield  {journal}
  {\bibinfo  {journal} {Phys. Rev. Lett.}\ }\textbf {\bibinfo {volume} {125}},\
  \bibinfo {pages} {060602} (\bibinfo {year} {2020})}\BibitemShut {NoStop}%
\bibitem [{\citenamefont {Levy}\ and\ \citenamefont
  {Lostaglio}(2020)}]{Levy20}%
  \BibitemOpen
  \bibfield  {author} {\bibinfo {author} {\bibfnamefont {A.}~\bibnamefont
  {Levy}}\ and\ \bibinfo {author} {\bibfnamefont {M.}~\bibnamefont
  {Lostaglio}},\ }\href {\doibase 10.1103/PRXQuantum.1.010309} {\bibfield
  {journal} {\bibinfo  {journal} {PRX Quantum}\ }\textbf {\bibinfo {volume}
  {1}},\ \bibinfo {pages} {010309} (\bibinfo {year} {2020})}\BibitemShut
  {NoStop}%
\bibitem [{\citenamefont {Roulet}\ and\ \citenamefont
  {Bruder}(2018)}]{Roulet18}%
  \BibitemOpen
  \bibfield  {author} {\bibinfo {author} {\bibfnamefont {A.}~\bibnamefont
  {Roulet}}\ and\ \bibinfo {author} {\bibfnamefont {C.}~\bibnamefont
  {Bruder}},\ }\href {\doibase 10.1103/PhysRevLett.121.063601} {\bibfield
  {journal} {\bibinfo  {journal} {Phys. Rev. Lett.}\ }\textbf {\bibinfo
  {volume} {121}},\ \bibinfo {pages} {063601} (\bibinfo {year}
  {2018})}\BibitemShut {NoStop}%
\bibitem [{\citenamefont {Abobeih}\ \emph {et~al.}(2019)\citenamefont
  {Abobeih}, \citenamefont {Randall}, \citenamefont {Bradley}, \citenamefont
  {Bartling}, \citenamefont {Bakker}, \citenamefont {Degen}, \citenamefont
  {Markham}, \citenamefont {Twitchen},\ and\ \citenamefont
  {Taminiau}}]{Abobeih19}%
  \BibitemOpen
  \bibfield  {author} {\bibinfo {author} {\bibfnamefont {M.~H.}\ \bibnamefont
  {Abobeih}}, \bibinfo {author} {\bibfnamefont {J.}~\bibnamefont {Randall}},
  \bibinfo {author} {\bibfnamefont {C.~E.}\ \bibnamefont {Bradley}}, \bibinfo
  {author} {\bibfnamefont {H.~P.}\ \bibnamefont {Bartling}}, \bibinfo {author}
  {\bibfnamefont {M.~A.}\ \bibnamefont {Bakker}}, \bibinfo {author}
  {\bibfnamefont {M.~J.}\ \bibnamefont {Degen}}, \bibinfo {author}
  {\bibfnamefont {M.}~\bibnamefont {Markham}}, \bibinfo {author} {\bibfnamefont
  {D.~J.}\ \bibnamefont {Twitchen}}, \ and\ \bibinfo {author} {\bibfnamefont
  {T.~H.}\ \bibnamefont {Taminiau}},\ }\href {\doibase
  10.1038/s41586-019-1834-7} {\bibfield  {journal} {\bibinfo  {journal}
  {Nature}\ }\textbf {\bibinfo {volume} {576}},\ \bibinfo {pages} {411}
  (\bibinfo {year} {2019})}\BibitemShut {NoStop}%
\bibitem [{\citenamefont {Poggiali}\ \emph {et~al.}(2017)\citenamefont
  {Poggiali}, \citenamefont {Cappellaro},\ and\ \citenamefont
  {Fabbri}}]{Poggiali17}%
  \BibitemOpen
  \bibfield  {author} {\bibinfo {author} {\bibfnamefont {F.}~\bibnamefont
  {Poggiali}}, \bibinfo {author} {\bibfnamefont {P.}~\bibnamefont
  {Cappellaro}}, \ and\ \bibinfo {author} {\bibfnamefont {N.}~\bibnamefont
  {Fabbri}},\ }\href {http://link.aps.org/doi/10.1103/PhysRevB.95.195308}
  {\bibfield  {journal} {\bibinfo  {journal} {Phys. Rev. B}\ }\textbf {\bibinfo
  {volume} {95}},\ \bibinfo {pages} {195308} (\bibinfo {year}
  {2017})}\BibitemShut {NoStop}%
\bibitem [{\citenamefont {Bar-Gill}\ \emph {et~al.}(2013)\citenamefont
  {Bar-Gill}, \citenamefont {Pham}, \citenamefont {Jarmola}, \citenamefont
  {Budker},\ and\ \citenamefont {Walsworth}}]{Bar-Gill13}%
  \BibitemOpen
  \bibfield  {author} {\bibinfo {author} {\bibfnamefont {N.}~\bibnamefont
  {Bar-Gill}}, \bibinfo {author} {\bibfnamefont {L.~M.}\ \bibnamefont {Pham}},
  \bibinfo {author} {\bibfnamefont {A.}~\bibnamefont {Jarmola}}, \bibinfo
  {author} {\bibfnamefont {D.}~\bibnamefont {Budker}}, \ and\ \bibinfo {author}
  {\bibfnamefont {R.~L.}\ \bibnamefont {Walsworth}},\ }\href {\doibase
  10.1038/ncomms2771} {\bibfield  {journal} {\bibinfo  {journal} {Nat.
  Commun.}\ }\textbf {\bibinfo {volume} {4}},\ \bibinfo {pages} {1743}
  (\bibinfo {year} {2013})}\BibitemShut {NoStop}%
\bibitem [{\citenamefont {Balasubramanian}\ \emph {et~al.}(2009)\citenamefont
  {Balasubramanian}, \citenamefont {Neumann}, \citenamefont {Twitchen},
  \citenamefont {Markham}, \citenamefont {Kolesov}, \citenamefont {Mizuochi},
  \citenamefont {Isoya}, \citenamefont {Achard}, \citenamefont {Beck},
  \citenamefont {Tissler}, \citenamefont {Jacques}, \citenamefont {Hemmer},
  \citenamefont {Jelezko},\ and\ \citenamefont
  {Wrachtrup}}]{Balasubramanian09}%
  \BibitemOpen
  \bibfield  {author} {\bibinfo {author} {\bibfnamefont {G.}~\bibnamefont
  {Balasubramanian}}, \bibinfo {author} {\bibfnamefont {P.}~\bibnamefont
  {Neumann}}, \bibinfo {author} {\bibfnamefont {D.}~\bibnamefont {Twitchen}},
  \bibinfo {author} {\bibfnamefont {M.}~\bibnamefont {Markham}}, \bibinfo
  {author} {\bibfnamefont {R.}~\bibnamefont {Kolesov}}, \bibinfo {author}
  {\bibfnamefont {N.}~\bibnamefont {Mizuochi}}, \bibinfo {author}
  {\bibfnamefont {J.}~\bibnamefont {Isoya}}, \bibinfo {author} {\bibfnamefont
  {J.}~\bibnamefont {Achard}}, \bibinfo {author} {\bibfnamefont
  {J.}~\bibnamefont {Beck}}, \bibinfo {author} {\bibfnamefont {J.}~\bibnamefont
  {Tissler}}, \bibinfo {author} {\bibfnamefont {V.}~\bibnamefont {Jacques}},
  \bibinfo {author} {\bibfnamefont {P.~R.}\ \bibnamefont {Hemmer}}, \bibinfo
  {author} {\bibfnamefont {F.}~\bibnamefont {Jelezko}}, \ and\ \bibinfo
  {author} {\bibfnamefont {J.}~\bibnamefont {Wrachtrup}},\ }\href {\doibase
  10.1038/nmat2420} {\bibfield  {journal} {\bibinfo  {journal} {Nature
  Materials}\ }\textbf {\bibinfo {volume} {8}},\ \bibinfo {pages} {383}
  (\bibinfo {year} {2009})}\BibitemShut {NoStop}%
\bibitem [{\citenamefont {Giachetti}\ \emph {et~al.}(2020)\citenamefont
  {Giachetti}, \citenamefont {Gherardini}, \citenamefont {Trombettoni},\ and\
  \citenamefont {Ruffo}}]{Giachetti20}%
  \BibitemOpen
  \bibfield  {author} {\bibinfo {author} {\bibfnamefont {G.}~\bibnamefont
  {Giachetti}}, \bibinfo {author} {\bibfnamefont {S.}~\bibnamefont
  {Gherardini}}, \bibinfo {author} {\bibfnamefont {A.}~\bibnamefont
  {Trombettoni}}, \ and\ \bibinfo {author} {\bibfnamefont {S.}~\bibnamefont
  {Ruffo}},\ }\href {\doibase 10.3390/condmat5010017} {\bibfield  {journal}
  {\bibinfo  {journal} {Condensed Matter}\ }\textbf {\bibinfo {volume} {5}},\
  \bibinfo {pages} {17} (\bibinfo {year} {2020})}\BibitemShut {NoStop}%
\bibitem [{\citenamefont {Jarzynski}\ and\ \citenamefont
  {W\'ojcik}(2004)}]{Jarzynski04}%
  \BibitemOpen
  \bibfield  {author} {\bibinfo {author} {\bibfnamefont {C.}~\bibnamefont
  {Jarzynski}}\ and\ \bibinfo {author} {\bibfnamefont {D.~K.}\ \bibnamefont
  {W\'ojcik}},\ }\href {\doibase 10.1103/PhysRevLett.92.230602} {\bibfield
  {journal} {\bibinfo  {journal} {Phys. Rev. Lett.}\ }\textbf {\bibinfo
  {volume} {92}},\ \bibinfo {pages} {230602} (\bibinfo {year}
  {2004})}\BibitemShut {NoStop}%
\end{thebibliography}%

\newpage

\begin{widetext}
\begin{center}
\textbf{\large Supplemental Material}
\end{center}

\subsection{Nitrogen nuclear spin initialization}

The electronic spin of the NV center is intrinsically coupled to the nuclear spin of the nitrogen atom. The NV center we work with is coupled to a \textsuperscript{14}N atom with nuclear spin $I=1$. For that reason each of the electronic spin levels is split into three hyperfine sub-levels for the nuclear spin projections $m_I=0,\pm1$. In order to obtain a genuine three-level system of the electronic spin, we polarize the nuclear spin in the $m_I=-1$ projection. This is done with a sequence of selective microwave and radio-frequency $\pi$-pulses, as illustrated schematically in Figure~\ref{fig:nuc_initialization}.
We assume to have as initial state a completely mixed state involving the nine hyperfine sublevels, although this polarizing protocol is adequate for any initial state.
The NV center electronic spin is initialized into the $m_S=0$ state by means of a long laser pulse.
The external magnetic field is far away from the excited-state level anti-crossing (ESLAC)~\cite{Steiner10, Poggiali17}, hence the nuclear spin is unaffected by the interaction with laser pulses.
From now on we adopt the notation $\ket{m_S, m_I}$ to describe a state of the joint system. In the first step, the population of $\ket{0,1}$ is transferred to $\ket{1,1}$ by a controlled-NOT operation on the electronic spin (selective MW $\pi$-pulse) and subsequently transferred to $\ket{1,0}$ by a controlled-NOT operation on the nuclear spin (selective RF $\pi$-pulse). Then a long laser pulse pumps the population to $\ket{0,0}$ while leaving unchanged the population that was already in the $m_S=0$ level.
Analogously, in a second step the resulting population in $\ket{0,0}$ is transferred to $\ket{0,-1}$ passing through $\ket{1,0}$ and $\ket{1,-1}$.
Among the possibilities of initializing a different $m_I$ using different controlled-NOT operations, we have chosen the described one as it gave the best polarization result. We have measured an initialization fidelity of $79\%$, measured with a standard electron spin resonance (ESR) experiment, Figure~\ref{fig:nuc_init_ESR}.

As a part of the nuclear spin remains in the undesired $m_I = 0,+1$ a normalization of the measured fluorescence has to be done that is different from the standard references of $m_S=0$ and $m_S=\pm1$ where the nuclear spin is not relevant. For that reason spin-lock measurements after the initialization gates for the Hamiltonian eigenstates (see \textbf{Appendix~B} of the main text) have been done. In this way the correct function of the initialization gates is proven and the fluorescence references for the correct normalization are obtained.

Note that in the presence of a relatively weak magnetic bias field ($B\simeq100$~MHz$/\gamma_e$), as used in our experiments,  the nitrogen nuclear spin lifetime is expected to be of the order of milliseconds~\cite{Bar-Gill13,Balasubramanian09},  much longer than a single experimental realization ($\sim 10$~$\mu$s). 

\begin{figure}[b]
\includegraphics[width=0.75\columnwidth]{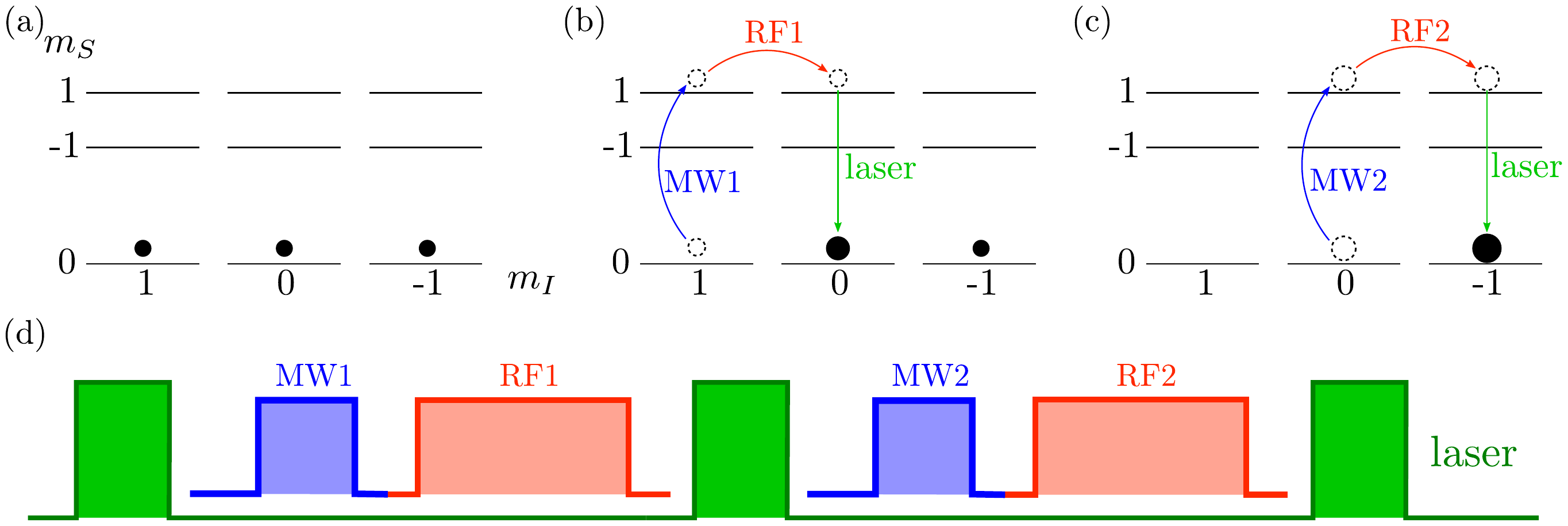}
\caption{\textbf{Scheme for nuclear spin initialization of the \textsuperscript{14}N nucleus}. \textbf{(a)} Initially the population is distributed among the $m_I=0,\pm1$. \textbf{(b, c)} With selective MW and RF pulses the population is first transferred from $m_I=+1$ to $m_I=0$ and then to $m_I=-1$ to initialize/polarize the nuclear spin. The laser is used to pump population in $m_S=1$ back to $m_S=0$. \textbf{(d)} Pulse protocol for the initialization process.}
\label{fig:nuc_initialization}
\end{figure}

\begin{figure}
\includegraphics[width=0.5\columnwidth]{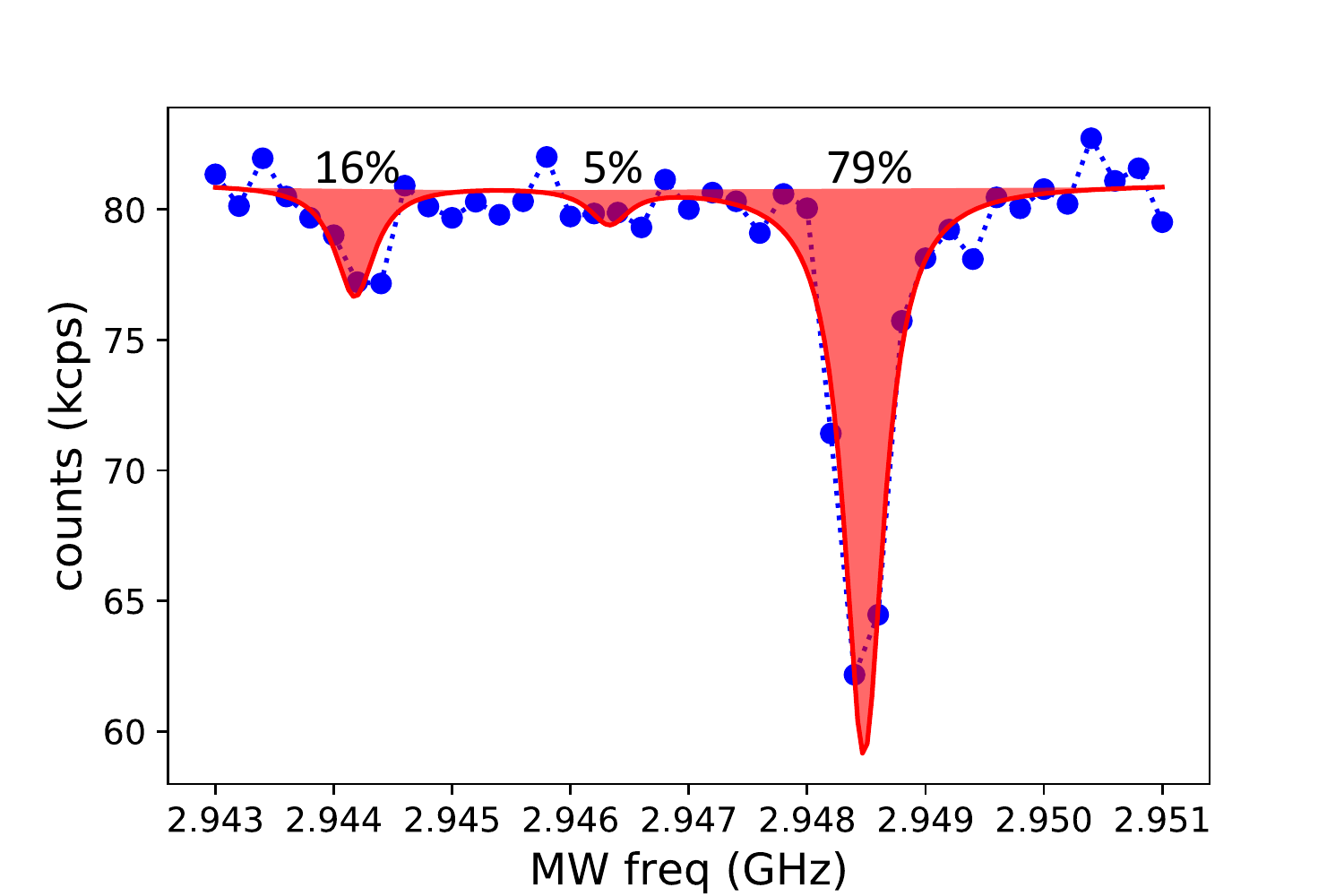}
\caption{\textbf{Nitrogen nuclear spin polarization}. Results from an electron spin resonance (ESR) experiment performed immediately after the nuclear spin polarization protocol.
}
\label{fig:nuc_init_ESR}
\end{figure}

\subsection{Lindbladian dissipation super-operator}
\label{subsec:Lindbladian_superoperator}

As also described in the \textbf{Appendix~A} of the main text, the interaction between the NV center and a green laser pulse can modeled as a POVM followed by a dissipation operator, and, overall, the dynamics of the open quantum system can be described by a master equation in Lindblad form. In this section, we will explicitly provide the expression of this Lindbladian dissipation operator.

For this purpose, from now on we will express operators in their matrix representation (with respect to the $S_z$-basis), such that
\begin{equation}
\ket{-1}\!\!\bra{-1} = \begin{pmatrix}
0 & 0 & 0 \\
0 & 0 & 0 \\
0 & 0 & 1
\end{pmatrix} \; ; \;
\ket{0}\!\!\bra{0} = \begin{pmatrix}
0 & 0 & 0 \\
0 & 1 & 0 \\
0 & 0 & 0
\end{pmatrix} \; ; \;
\ket{+1}\!\!\bra{+1} = \begin{pmatrix}
1 & 0 & 0 \\
0 & 0 & 0 \\
0 & 0 & 0
\end{pmatrix}.
\end{equation}
In addition, we will use indistinguishably the term matrix or operator, unless otherwise specified, and
we will adopt the formalism described in Ref.\,\cite{Havel03} for superoperators represented by $N^2\times N^2$ matrices, where $N=3$ is the dimension
of the quantum system's Hilbert space.

In this formalism, the Lindbladian dissipation super-operator (defined in \textbf{Appendix~A}) is written as
\begin{equation} \label{eq:3LS_Lindblad_superoperator}
\boldsymbol{\mathcal{L}}^{(t_\mathrm{L})} =\begin{pmatrix}
e^{-t_\mathrm{L} \Gamma} & 0 & 0 & 0 & 0 & 0 & 0 & 0 & 0 \\
0 & e^{-t_\mathrm{L} \Gamma/2} & 0 & 0 & 0 & 0 & 0 & 0 & 0 \\
0 & 0 & e^{-t_\mathrm{L} \Gamma} & 0 & 0 & 0 & 0 & 0 & 0 \\
0 & 0 & 0 & e^{-t_\mathrm{L} \Gamma/2} & 0 & 0 & 0 & 0 & 0 \\
1-e^{-t_\mathrm{L} \Gamma} & 0 & 0 & 0 & 1 & 0 & 0 & 0 & 1-e^{-t_\mathrm{L} \Gamma} \\
0 & 0 & 0 & 0 & 0 & e^{-t_\mathrm{L} \Gamma/2} & 0 & 0 & 0 \\
0 & 0 & 0 & 0 & 0 & 0 & e^{-t_\mathrm{L} \Gamma} & 0 & 0 \\
0 & 0 & 0 & 0 & 0 & 0 & 0 & e^{-t_\mathrm{L} \Gamma/2} & 0 \\
0 & 0 & 0 & 0 & 0 & 0 & 0 & 0 & e^{-t_\mathrm{L} \Gamma}
\end{pmatrix}
\end{equation}
where $t_\mathrm{L}$ denotes the laser duration, and $\Gamma$ is an effective decay rate. Using laser pulses with $t_\mathrm{L} = 41$~ns,
we characterized that the effective decay rate equals $\Gamma\sim 12.2$~MHz.

\subsection{Analytic \texorpdfstring{$\gamma$}{gamma} in the case of \texorpdfstring{$\mathcal{H}=\mathcal{H}_\mathrm{NV}$}{H\_NV}}
\label{subsec:gamma_case_of_Hnv}

As explained in the main text, the effect of applying a single laser pulse can be described by the superoperator
\begin{equation}
\boldsymbol{\mathcal{A}} \equiv \sum_{j=1}^4 \boldsymbol{\mathcal{D}}_j \boldsymbol{m}_{j}\,.
\end{equation}
This superoperator can be computed explicitly from the definitions of $\boldsymbol{\mathcal{D}}_j$ and $\boldsymbol{m}_j$ [see also \textbf{Appendix~A} in main text]. In this regard, by introducing the auxiliary matrices

\begin{small}
\begin{equation*}\begingroup 
\setlength\arraycolsep{0.3em}
\boldsymbol{A}_1 \equiv \begin{pmatrix}
1 & 0 & 0 & 0 & 0 & 0 & 0 & 0 & 0 \\
0 & 0 & 0 & 0 & 0 & 0 & 0 & 0 & 0 \\
0 & 0 & 0 & 0 & 0 & 0 & 0 & 0 & 0 \\
0 & 0 & 0 & 0 & 0 & 0 & 0 & 0 & 0 \\
0 & 0 & 0 & 0 & 1 & 0 & 0 & 0 & 0 \\
0 & 0 & 0 & 0 & 0 & 0 & 0 & 0 & 0 \\
0 & 0 & 0 & 0 & 0 & 0 & 0 & 0 & 0 \\
0 & 0 & 0 & 0 & 0 & 0 & 0 & 0 & 0 \\
0 & 0 & 0 & 0 & 0 & 0 & 0 & 0 & 1
\end{pmatrix} ; \;\; \boldsymbol{A}_2 \equiv \begin{pmatrix}
0 & 0 & 0 & 0 & 0 & 0 & 0 & 0 & 0 \\
0 & 0 & 0 & 0 & 0 & 0 & 0 & 0 & 0 \\
0 & 0 & 0 & 0 & 0 & 0 & 0 & 0 & 0 \\
0 & 0 & 0 & 0 & 0 & 0 & 0 & 0 & 0 \\
1 & 0 & 0 & 0 & 1 & 0 & 0 & 0 & 1 \\
0 & 0 & 0 & 0 & 0 & 0 & 0 & 0 & 0 \\
0 & 0 & 0 & 0 & 0 & 0 & 0 & 0 & 0 \\
0 & 0 & 0 & 0 & 0 & 0 & 0 & 0 & 0 \\
0 & 0 & 0 & 0 & 0 & 0 & 0 & 0 & 0
\end{pmatrix} ; \;\; \boldsymbol{A}_3 \equiv \mathbb{1}_{9\times 9} - \boldsymbol{A}_1
\endgroup \,,
\end{equation*}
\end{small}

\noindent
it holds that
\begin{equation}\label{eq:mathcalA}
\boldsymbol{\mathcal{A}} = \mu \boldsymbol{A}_1 + (1 - \mu) \boldsymbol{A}_2 + \boldsymbol{A}_3 (1 - p_\mathrm{abs}),
\end{equation}
where
\begin{equation}
\mu \equiv  1 - (1 - e^{-t_\mathrm{L} \Gamma})p_\mathrm{abs}\,.
\end{equation}
It is no surprising that, if $ p_\mathrm{abs} =0$, then $\boldsymbol{\mathcal{A}}_{(p_\mathrm{abs} =0)} = \mathbb{1}_{9\times 9}$. On the other hand, if $ p_\mathrm{abs} =1$, then the POVM is actually a quantum projective measurement in the $S_z$-basis. This is the reason why $\boldsymbol{\mathcal{A}}_{(p_\mathrm{abs} =1)} = e^{-t_\mathrm{L} \Gamma} \boldsymbol{A}_1 + (1 - e^{-t_\mathrm{L} \Gamma}) \boldsymbol{A}_2$ is
practically equal to the Lindbladian dissipation super-operator $\boldsymbol{\mathcal{L}}$~[Eq.~\eqref{eq:3LS_Lindblad_superoperator}], but without the terms involving coherences that is a signature of have applied a quantum projective measurement. Moreover, another special case occurs when $e^{-t_\mathrm{L} \Gamma}=1$, namely the case without dissipation, where Eq.~\eqref{eq:mathcalA} can be rewritten as $\boldsymbol{\mathcal{A}} = \boldsymbol{A}_1 p_\mathrm{abs} +  \mathbb{1}_{9\times 9} (1 - p_\mathrm{abs})$ that identifies the mean effect of applying a quantum projective measurement of $S_z$ with probability $p_\mathrm{abs}$.

On the other hand, by taking $\mathcal{H}=\mathcal{H}_\mathrm{NV}$, the unitary evolution of the system is described by the following superoperator:
\begin{align*}
\boldsymbol{U}(\tau) &\equiv \exp(-i \tau ( \mathcal{H}\otimes \mathbb{1}_{3\times 3}  - \mathbb{1}_{3\times 3} \otimes \mathcal{H} )) \notag \\
&=\begin{pmatrix}
1 & 0 & 0 & 0 & 0 & 0 & 0 & 0 & 0 \\
0 & e_{+} & 0 & 0 & 0 & 0 & 0 & 0 & 0 \\
0 & 0 & \frac{e_{+}}{e_{-}} & 0 & 0 & 0 & 0 & 0 & 0 \\
0 & 0 & 0 & \frac{1}{e_{+}} & 0 & 0 & 0 & 0 & 0 \\
0 & 0 & 0 & 0 & 1 & 0 & 0 & 0 & 0 \\
0 & 0 & 0 & 0 & 0 & \frac{1}{e_{-}} & 0 & 0 & 0 \\
0 & 0 & 0 & 0 & 0 & 0 & \frac{e_{-}}{e_{+}} & 0 & 0 \\
0 & 0 & 0 & 0 & 0 & 0 & 0 & e_{-} & 0 \\
0 & 0 & 0 & 0 & 0 & 0 & 0 & 0 & 1
\end{pmatrix},
\end{align*}
where $e_{\pm} \equiv e^{-i\tau E_{\pm1}}$. Therefore, $\boldsymbol{\mathcal{B}} \equiv \boldsymbol{\mathcal{A}}\boldsymbol{U}$ can be written as
\begin{equation} \label{eq:mathcalB}
\boldsymbol{\mathcal{B}} = \mu \boldsymbol{A}_1 + (1 - \mu) \boldsymbol{A}_2 + \boldsymbol{U}_1(\tau) (1 - p_\mathrm{abs})
\end{equation}
with $\boldsymbol{U}_1(\tau) \equiv \boldsymbol{U}(\tau) -\boldsymbol{A}_1$. Writing $\boldsymbol{\mathcal{B}}$ as in Eq.~\eqref{eq:mathcalB} is very useful, because the following properties hold:
\begin{align*}
\boldsymbol{A}_i^2 &= \boldsymbol{A}_i \;\;\mathrm{for}\;i\in\{1,2,3\} \\
\boldsymbol{A}_1 \boldsymbol{A}_3 &= \boldsymbol{A}_3 \boldsymbol{A}_1 = \boldsymbol{A}_2 \boldsymbol{A}_3 = \boldsymbol{A}_3 \boldsymbol{A}_2 = \mathbb{0}_{9\times 9} \\
\boldsymbol{A}_1 \boldsymbol{A}_2 &= \boldsymbol{A}_2 \boldsymbol{A}_1 = \boldsymbol{A}_2\\
(\boldsymbol{U}_1(\tau))^n &= \boldsymbol{U}_1(n\tau) \\
\boldsymbol{A}_i \boldsymbol{U}_1(\tau) &= \boldsymbol{U}_1(\tau) \boldsymbol{A}_i = \mathbb{0}_{9\times 9} \;\;\mathrm{for}\;i\in\{1,2\},
\end{align*}
thus implying that
\begin{equation} \label{eq:BnL_for_Hnv}
\boldsymbol{\mathcal{B}}^{N_\mathrm{L}} = \mu^{N_\mathrm{L}} \boldsymbol{A}_1 + (1 - \mu^{N_\mathrm{L}}) \boldsymbol{A}_2 + \boldsymbol{U}_1(N_\mathrm{L}\tau) (1 - p_\mathrm{abs})^{N_\mathrm{L}}\,.
\end{equation}

Moreover, being in the case of $\mathcal{H}=\mathcal{H}_\mathrm{NV}$, a thermal state $\rho^{\mathrm{th}}$ can be written according to the superoperators formalism \cite{Havel03} in the following way:
\begin{equation*}
\col(\rho^{\mathrm{th}}) = \frac{1}{Z}\col\left(\begin{bmatrix}
e^{-\beta E_{+1}} & 0 & 0 \\
0 & 1 & 0 \\
0 & 0 & e^{-\beta E_{-1}}
\end{bmatrix}\right) = \frac{1}{Z}\begin{pmatrix}
e^{-\beta E_{+1}}\\
0\\
0\\
0\\
1\\
0\\
0\\
0\\
e^{-\beta E_{-1}}
\end{pmatrix}
\end{equation*}
where $Z \equiv \sum_{k=-1}^1e^{-\beta E_i}$. Hence, as final calculation,
by exploiting that $\boldsymbol{A}_1^\dagger \col(\rho^{\mathrm{th}}) = (\boldsymbol{U}_1(N_\mathrm{L}\tau))^\dagger \col(\rho^{\mathrm{th}}) = \col(\rho^{\mathrm{th}})$, and $\boldsymbol{A}_2^\dagger \col(\rho^{\mathrm{th}})= \col(\mathbb{1}_{3\times 3})/Z$,
the use of Eq.~\eqref{eq:BnL_for_Hnv} allows us to obtain an analytic expression for $\gamma$, namely
\begin{align}
\gamma &= \Tr_{3\times3}[ (\boldsymbol{\mathcal{B}}^\dagger)^{N_\mathrm{L}} \,\col[\rho^{\mathrm{th}}]] \notag\\
 &= \mu^{N_\mathrm{L}}  + \frac{3}{Z}( 1-\mu^{N_\mathrm{L}})\,.
\end{align}
Let us observe that, apart the special cases mentioned before,
\begin{equation*}
\lim_{N_\mathrm{L} \rightarrow \infty} \mu^{N_\mathrm{L}} = 0\,,
\end{equation*}
meaning that
\begin{equation}
\lim_{N_\mathrm{L} \rightarrow \infty} \gamma = \frac{3}{Z}.
\end{equation}

\subsection{A method to extract $\eta^*$ from experimental data}

Let us consider a 3LS initialized in the thermal state $\rho^{\mathrm{th}} \equiv \exp(-\beta H)/Z_{\beta}$ with $Z_{\beta} \equiv {\rm Tr}[\exp(-\beta H)]$. By diagonalizing the system Hamiltonian as $H=\sum_{i}E_{i}|E_{i}\rangle\!\langle E_{i}|$, the initial thermal initial state is equal to
\begin{equation*}
\rho^{\mathrm{th}} = \sum_{i}\frac{e^{-\beta E_i}}{Z_{\beta}}|E_{i}\rangle\!\langle E_{i}| = \sum_{i}P_{i}|E_{i}\rangle\!\langle E_{i}|
\end{equation*}
where $P_i \equiv \exp(-\beta E_i)/Z_{\beta}$.

Then, let us introduce the mean value of the exponentiated energy variation $\Delta E$, i.e.,
\begin{equation}
\langle \exp(-\eta\Delta E)\rangle = \sum_{i,j}P_{i}P_{j|i}e^{-\eta(E_j - E_i)},
\end{equation}
which strictly depends on the energy scaling factor $\eta$. Note that the average $\langle \exp(-\eta\Delta E)\rangle$ is defined by the sets $\{P_i\}$ and $\{P_{j|i}\}$, corresponding respectively to the probabilities to measure the initial energies of the system and the conditional probabilities to get $E_j$ at the end of the procedure after have measured $E_i$.

In the SSE regime, as described in the main text, the conditional probabilities $P_{j|i}$ does not depend upon the initially measured energies $E_i$. This entails that
\begin{equation}
P_{j|1}=P_{j|2}=P_{j|3}\equiv\widetilde{P}_j
\end{equation}
for any $j$ such that
\begin{equation}\label{eq:SSE_G}
\langle \exp(-\eta\Delta E)\rangle = \sum_{i,j}P_{i}\widetilde{P}_{j}e^{-\eta(E_j - E_i)}\,,
\end{equation}
whereby the quantum state $\rho_{\rm fin}$ after the application of the 2nd energy projective measurement of the TPM measurement protocol is a mixed state, not necessarily thermal but completely described by the probabilities $\widetilde{P}_j$, namely
\begin{equation*}
\rho_{\rm fin}=\sum_{j}\widetilde{P}_{j}|E_{j}\rangle\!\langle E_{j}|.
\end{equation*}

Now, to make our derivation compatible with the experimental setup and in particular with the case of $\mathcal{H} = \mathcal{H}_\mathrm{mw}$, we assume that the energy values are symmetric around zero, and, for the sake of brevity, we call $E_1=-\overline{E}$, $E_2=0$ and $E_3 = \overline{E}$, with $\overline{E}$ constant value (in the experiment $\overline{E}=\hbar \omega/2$). In this way, by means of the substitution
\begin{equation}
x \equiv e^{\eta^* \overline{E}} \Longleftrightarrow \eta^* = \frac{1}{\overline{E}}\ln x \,,
\end{equation}
the equation
\begin{equation*}
  \sum_{i,j}P_{i}\widetilde{P}_{j}e^{-\eta^*(E_j - E_i)} = 1
\end{equation*}
can be rewritten as the following polynomial equation:
\begin{equation}\label{eq:receipe}
(x-1)\left(P_{1}\widetilde{P}_{3}\,x^{3} + (P_{1}\widetilde{P}_{2} + P_{1}\widetilde{P}_{3} + P_{2}\widetilde{P}_{3})x^{2}
- (P_{2}\widetilde{P}_{1} + P_{3}\widetilde{P}_{1} + P_{3}\widetilde{P}_{2})x - P_{3}\widetilde{P}_{1}\right)=0
\end{equation}
Clearly, Eq.\,(\ref{eq:receipe}) contains the trivial solution $x=1$, i.e., $\eta^*=0$, while solving the third-order algebraic equation
\begin{equation}\label{eq:receipe_2}
P_{1}\widetilde{P}_{3}\,x^{3} + (P_{1}\widetilde{P}_{2} + P_{1}\widetilde{P}_{3} + P_{2}\widetilde{P}_{3})x^{2}
- (P_{2}\widetilde{P}_{1} + P_{3}\widetilde{P}_{1} + P_{3}\widetilde{P}_{2})x = P_{3}\widetilde{P}_{1}
\end{equation}
provides us the other value of $\eta^* \neq 0$ that obeys the fluctuation relation $\mathcal{G}(\eta^*)=1$. In this regard, it is worth noting that, by applying the well-known Routh-Hurwitz criterion to the polynomial (\ref{eq:receipe_2}), we can also prove that just only root of Eq.\,(\ref{eq:receipe_2}) has positive real part. Indeed, according to the Routh-Hurwitz criterion, we recall that to each variation (permanence) of the sign of the coefficients of the first column of the Routh table corresponds to a root of the polynomial with a positive (negative) real part. In our case, there are always $2$ sign-permanences and only $1$ variation, for any possible value of the probabilities $P_{i}$ and $\widetilde{P}_{j}$. Being $\eta^*\propto \ln x$, only the unique solution $x\neq 0$ with positive real part is physical, thus providing us the (unique) non-trivial energy scaling factor $\eta^*$ such that $\mathcal{G}(\eta^*)=1$.

\subsubsection*{Non-equilibrium steady-state condition fulfilled by $\eta^*$}

Here, let us provide some more insights on the interpretation of $\eta^*$, and the condition that it has to fulfill by imposing the validity of the equality $\mathcal{G}(\eta^*)=1$.

For this purpose, each probability $\widetilde{P}_j$ is decomposed in the product of two contributions: One is thermal and is associated to the inverse temperature $\beta_{\rm fin}$, while the other is a correction term that accounts for the (geometric) distance $\lambda$ concerning $\rho_{\rm fin}$ from being thermal~\cite{Giachetti20}. Specifically, given the set $\{E_j\}$ of the system energies after the application of the measurement protocol, $\widetilde{P}_j$ can be written as
\begin{equation}\label{eq:tilde_P_j}
\widetilde{P}_{1}=\frac{e^{-\beta_{\rm fin}E_1}e^{\lambda(E_2 - E_3)^2}}{Z_{\beta_{\rm fin}}(\lambda)};\,\,\,\,\,\,
\widetilde{P}_{2}=\frac{e^{-\beta_{\rm fin}E_2}e^{\lambda(E_3 - E_1)^2}}{Z_{\beta_{\rm fin}}(\lambda)};\,\,\,\,\,\,
\widetilde{P}_{3}=\frac{e^{-\beta_{\rm fin}E_3}e^{\lambda(E_1 - E_2)^2}}{Z_{\beta_{\rm fin}}(\lambda)}\,,
\end{equation}
where
\begin{equation*}
Z_{\beta_{\rm fin}}(\lambda)=\exp(-\beta_{\rm fin}A + 2\lambda B),\,\,\,\,\text{with}\,\,\,\,A=\sum_{j}E_j\,\,\,\,\text{and}\,\,\,\,B=\sum_{j}E_{j}^{2}-E_{1}E_{2}-E_{2}E_{3}-E_{3}E_{1},
\end{equation*}
denotes the corresponding discrete partition function. It is worth noting that the probabilities $\widetilde{P}_{1}$, $\widetilde{P}_{2}$ and $\widetilde{P}_{3}=1-\widetilde{P}_{1}-\widetilde{P}_{2}$ are written as a function of two free-parameters. Thus, known the energies $E_j$ and experimentally obtained the values of the probabilities $\widetilde{P}_{j}$ with $j=1,2,3$, $\beta_{\rm fin}$ and $\lambda$ can be derived by means of standard non-linear regression techniques.

Now, let us consider that $\mathcal{H} = \mathcal{H}_\mathrm{mw}$, which is the more involved one among the cases we have analyzed so far. In accordance with the experimental setup, the energies can be assumed symmetric around zero, as even shown above. In this way, by substituting the relations of Eq.~(\ref{eq:tilde_P_j}) into Eq.~(\ref{eq:SSE_G}) and imposing $\mathcal{G}(\eta^*)=1$, after simple calculations we can end up in the following equality that corresponds to a \textit{non-equilibrium steady-state condition} for the analysed open 3LS:
\begin{equation}\label{eq:NESS_cond}
Z_{\beta}Z_{\beta_{\rm fin}} + C(\lambda)\left(3 - Z_{\beta + \beta_{\rm fin}}\right) = C(\lambda)\left(Z_{(\beta - \beta_{\rm fin})-2\eta^*}+Z_{\beta_{\rm fin}+\eta^*}\right) + C(\lambda)^{4}Z_{\beta - \eta^*}
\end{equation}
where
\begin{eqnarray*}
&C(\lambda)=e^{\lambda\overline{E}^2};\,\,\,\,\,\,Z_{\beta - \eta^*}=\displaystyle{\sum_{k=-1}^{1}e^{-k\overline{E}(\beta - \eta^*)}};&\nonumber \\
&Z_{\beta_{\rm fin}+\eta^*}=\displaystyle{\sum_{k=-1}^{1}e^{-k\overline{E}(\beta_{\rm fin}+\eta^*)}};\,\,\,\,\,\,Z_{(\beta - \beta_{\rm fin})-2\eta^*}=\displaystyle{\sum_{k=-1}^{1}e^{-k\overline{E}((\beta-\beta_{\rm fin})-2\eta^*)}}.&
\end{eqnarray*}
Let us observe that $\eta^*=0$ is always solution of Eq.~(\ref{eq:NESS_cond}) for any value of $\lambda$, while for $\lambda=0$ the non-trivial solution $\eta^*$ of Eq.~(\ref{eq:NESS_cond}) is $\eta^*=\beta-\beta_{\rm fin}$ in perfect agreement with the Jarzynski-W\'ojcik relation~\cite{Jarzynski04}. Instead, for a value of $\lambda\neq 0$, Eq.~(\ref{eq:NESS_cond}) is numerically solved as a function of $\eta$, whereby the non-trivial solution $\eta^*$ is provided by the real and positive value of $\eta$ obeying the energy exchange fluctuation relation $\langle\exp(-\eta^*\Delta E)\rangle = 1$.

\end{widetext}


\end{document}